\documentclass[11pt, oneside]{article}   	

\usepackage{jheppub}

\usepackage{amsmath}
\usepackage{amssymb}
\usepackage{amsfonts}
\usepackage{amsthm}
\usepackage{amsbsy}
\usepackage{array}

\usepackage{subcaption} 
\usepackage{hyperref}

\usepackage{float}

\usepackage{mathtools}
\usepackage{xcolor}

\usepackage{empheq}

\usepackage{graphicx}

\usepackage{pepper}

\def\bea{\begin{eqnarray}} \def\eea{\end{eqnarray}}

\newcommand*\widefbox[1]{\fbox{\hspace{2em}#1\hspace{2em}}}

\newcommand{\p}{\partial}

\newcommand{\f}{\frac}

\newcommand{\ve}{\varepsilon}

\title{Correlation Functions and Trace Anomalies in Weakly Relevant Flows}

\author[a]{Denis Karateev}
\author[b,c]{and Biswajit Sahoo}

\affiliation[a]{
	D\'epartment de Physique Th\'eorique, Universit\'e de Gen\`eve,\\
	24 quai Ernest-Ansermet, 1211 Gen\`eve 4, Switzerland}

\affiliation[b]{Fields and Strings Laboratory, Institute of Physics\\ École Polytechnique Fédéral de Lausanne (EPFL)
	\\ Route de la Sorge, CH-1015 Lausanne, Switzerland}

\affiliation[c]{
	Department of Mathematics, King’s College London,\\ Strand, London WC2R 2LS, United
	Kingdom}

\abstract{
	We study abstract weakly relevant flows in a general number of dimensions. They arguably provide the simplest example of renormalization group (RG) flows between two non-trivial fixed points.
	We compute several two-point correlation functions in position space valid along the whole RG flow. This is done by using conformal perturbation theory together with the solution of the Callan-Symanzik equation.
	From the explicit expressions of the two-point functions of conserved currents and the stress-tensor we extract the change in the central charges between the UV and IR fixed points. This immediately gives us $\Delta c$, the change of the $c-$trace anomaly between the UV and IR fixed points in 4d. We also discuss three-point functions.
	
	We couple weakly relevant flows to non-dynamical dilaton and graviton background fields in 4d. We compute the three-dilaton vertex in terms of the scalar two-point function and extract the value of $\Delta a$, the change of the $a$-trace anomaly between the UV and IR fixed points. We also compute the graviton-graviton-dilaton vertex in terms of the three-point function of two stress-tensors and a scalar, and extract the value of $\Delta c$.
	
	The $\Delta c$ values obtained with the two different  methods agree.
	
}

\dedicated{Dedicated to the memory of Jim Simons for his enormous support of fundamental research and the establishment of the Simons Foundation, from which both authors have received invaluable support and research funding. }

\begin{document}
\maketitle

\section{Introduction}

Quantum field theories (QFTs) are defined as renormalization group (RG) flows between UV and IR fixed points. We assume that these fixed points are described by conformal field theories (CFTs). The typical problem one needs to solve is to compute the  spectrum of the IR CFT (anomalous dimensions, central charges and the OPE coefficients) in terms  of the UV description of the QFT. A more general problem is to compute correlation functions valid along the whole RG flow. The IR spectrum can then be extracted by taking the IR limit of two- and three-point functions.

The most famous QFT, where such computations can be done analytically is the Wilson-Fisher (WF) theory in $d=4-\epsilon $ dimensions  \cite{Wilson:1971dc}.
A lot of work has been done on computing anomalous dimensions in the WF theory, see \cite{Kleinert:1991rg,Batkovich:2016jus}. For some recent interesting analytic and numerical results on the spectrum of the WF theory and related theories see \cite{Brezin:1973igb,Petkou:1995vu,JACK1990647,Cappelli:1991ke,Rychkov:2015naa,Hogervorst:2015akt,Rychkov:2018vya,Alday:2017zzv,Carmi:2020ekr,Fei:2014yja,Fei:2015kta,Stergiou:2015roa,Giombi:2021cnr}. 
Computation of the IR CFT OPE coefficients is a much harder task than computation of anomalous dimensions, much less work has been done in this direction. For some recent works in this direction see \cite{Gopakumar:2016wkt,Codello:2017hhh,Pagani:2020ejb}. 

One can define an abstract class of QFTs with non-trivial UV and IR fixed points, called \textit{weakly relevant flows}, where perturbative analytic calculations can also be performed. It is defined in any number of dimensions $d$ by the following action
\begin{equation}
	\label{eq:theory_definition}
	A =  A_\text{UV CFT} + \lambda_0 m^{d-\Delta} \int d^d x\, \cO(x).
\end{equation}
Here $\lambda_0$ is the coupling constant and $m$ is a mass scale. The operator $\cO(x)$ is assumed to be the only relevant scalar operator in the UV CFT. Its scaling dimension is $\Delta = d-\delta$, where $0<\delta\ll1$. In other words, the operator $\cO(x)$ is weakly relevant. Some aspects of this model have been discussed in 2d in \cite{Cappelli:1989yu} and in general $d$ in \cite{Klebanov:2011gs,Komargodski:2011xv,Komargodski:2016auf}. In $3d$, this model serves as the dual QFT for certain inflationary slow-roll models and has been extensively studied in \cite{Bzowski:2012ih,McFadden:2013ria}. One assumes that $\lambda_0$ is small, thus, one can use it as a perturbative parameter. Since we perturb around a non-trivial UV CFT, we refer to this type of perturbation theory as the conformal perturbation theory. Conformal perturbation theory was first introduced in $d=2$ in \cite{Zamolodchikov:1987ti,Zamolodchikov:1989hfa}. Some interesting works using it in two and higher dimensions in various contexts include \cite{Gaberdiel:2008fn,Gorbenko:2018ncu,Gorbenko:2018dtm,Rutter:2018aog,Burrington:2023vei}. Some other related works include \cite{Behan:2017mwi,Behan:2017dwr,Antunes:2022vtb,Behan:2023ile,Antunes:2024hrt,Cuomo:2024vfk}.

This paper contains two parts. The first part consists of sections \ref{sec:conformal_perturbation_theory} - \ref{sec:stress-tensor}. It is dedicated to computing two- and three-point functions in position space along the whole RG flow in weakly relevant flows in a general number of dimensions. In particular, we compute two-point functions of scalars, conserved currents and the stress-tensor. These expressions allow to extract the scalar IR anomalous dimension and the difference of the UV and IR central charges for the current and the stress-tensor. We also initiate the study of three-point functions.  The second part of this paper consists of section \ref{sec:background_field_method}. There we use non-dynamical background fields (dilatons and gravitons) to derive the change of the $a$- and $c$-trace anomalies between the UV and IR fixed points in 4d.

In order to compute correlation functions in position space valid along the whole RG flow we use the strategy of \cite{Brezin:1973igb,Cappelli:1989yu}. It consists of two steps. 
First, one uses conformal perturbation theory in order to obtain a perturbative result. This result, however, breaks down far from the UV fixed point. In order to obtain the result, which is valid along the whole RG flow, one needs to re-sum the leading perturbative contributions to all orders. This is done in the second step by using the renormalization group techniques, more precisely by writing down and solving the Callan-Symanzik equation.

The rest of the paper is organized as follows. In subsection \ref{sec:conventions} we provide our conventions and we summarize our results in subsection \ref{sec:summary}. Section \ref{sec:conformal_perturbation_theory} is dedicated to conformal perturbation theory. In section \ref{sec:Callan-Symanzik_correlators} we derive and solve the Callan-Symanzik equation for scalar two- and three-point functions in position space. In section \ref{sec:conserved_currents} we discuss conserved currents. In section \ref{sec:stress-tensor} we discuss the stress-tensor. 
Section \ref{sec:background_field_method} is dedicated to the background field method approach.
We discuss open problems and future applications in section \ref{sec:discussion}. Various additional details are provided in appendices. For clarity we framed key expressions in sections \ref{sec:conformal_perturbation_theory} to \ref{sec:background_field_method}.

\subsection{Conventions}
\label{sec:conventions}
We work in $d$ dimensional Euclidean space. The surface area of the $(d-1)$-dimensional sphere embedded in $d$ dimensions is given by
\begin{equation}
	\Omega_{d-1} \equiv \frac{2\pi^{\frac{d}{2}}}{\Gamma\left(\frac{d}{2}\right)}.
\end{equation}
The connected part of scalar correlation functions of the local operator $\cO(x)$ in position space are computed from the action $A$ using the standard path-integral expression
\begin{equation}
	\label{eq:connected_correlation_functions}
	\langle \cO(x_1)\ldots \cO(x_n) \rangle_\text{QFT} =\frac{ \int [d\Phi] \cO(x_1)\ldots \cO(x_n) e^{-A[\Phi]} }{ \int [d\Phi]\ e^{-A[\Phi]}},
\end{equation}
where $[\Phi]$ are independent UV CFT degrees of freedom. 

Our convention for generic two- and three-point conformal correlators is
\begin{align}
	\label{eq:2pt_CFT}
	\langle\cO_{\Delta_1}(x_1)\cO_{\Delta_2}(x_2)\rangle_{\text{CFT}} &= \frac{\delta_{\Delta_1\Delta_2}}{\left(x_{12}^2\right)^{\Delta_1}},\\
	\label{eq:3pt_CFT}
	\langle\cO_{\Delta_1}(x_1)\cO_{\Delta_2}(x_2)\cO_{\Delta_3}(x_3)\rangle_{\text{CFT}} &= \frac{C_{\cO_1\cO_2\cO_3}}{\left(x_{12}^2\right)^\frac{\Delta_1+\Delta_2-\Delta_3}{2}\left(x_{13}^2\right)^\frac{\Delta_1-\Delta_2+\Delta_3}{2}\left(x_{23}^2\right)^\frac{-\Delta_1+\Delta_2+\Delta_3}{2}}.
\end{align}
Here $C_{\cO_1\cO_2\cO_3}$ is the OPE coefficient for three different scalar operators and
\begin{equation}
	x_{ij}^\mu \equiv x_i^\mu - x_j^\mu.
\end{equation}
For later purposes it is also convenient to define
\begin{equation}
	\Delta_{ij} \equiv \Delta_i-\Delta_j.
\end{equation}
When writing scaling dimensions and OPE coefficients, most of the time we will refer to the UV CFT. When we discuss the IR CFT, we will put an explicit label IR to the scaling dimensions in order to distinguish them from the UV CFT ones.

Due to Poincare\'e invariance all the QFT scalar two-point correlation functions depend on a single variable defined as
\begin{equation}
	\label{eq:definition_r}
	r \equiv \left(x_{12}^2\right)^\frac{1}{2}.
\end{equation}
Analogously, all the scalar three-point functions depend only on the three variables $x_{12}^2$, $x_{13}^2$ and $x_{23}^2$. For the later purposes another basis of variables is much more convenient. It consists of one dimensionful variable $\rho$ and two dimensionless variables $y_1$ and $y_2$. We define them as
\begin{equation}
	\label{eq:definition_r_y1_y2_app}
	\rho \equiv (x_{12}^2 x_{13}^2 x_{23}^2)^\frac{1}{6},\qquad
	y_1 \equiv \frac{x^2_{12}}{x_{23}^2},\qquad
	y_2 \equiv \frac{x_{13}^2}{x_{23}^2}.
\end{equation}
Notice that the CFT three-point function \eqref{eq:3pt_CFT} with all scalars being identical depends only on $\rho$.
We can invert the above relations and get
\begin{equation}
	\label{eq:change_variables}
	x_{12}^2 = \rho^2 \times y_1^{2/3}y_2^{-1/3},\qquad
	x_{13}^2 = \rho^2 \times y_1^{-1/3}y_2^{2/3},\qquad
	x_{23}^2 = \rho^2 \times y_1^{-1/3}y_2^{-1/3}.
\end{equation}
The following relation is often useful
\begin{equation}
	\label{eq:relation_dot_product}
	x_{ij}\cdot x_{ik} = \frac{1}{2}\left(x_{ij}^2+x_{ik}^2-x_{kj}^2\right).
\end{equation}

\subsection{Summary of the results}
\label{sec:summary}

The action \eqref{eq:theory_definition} is super-renormalizable as long as $d/2<\Delta<d$. All the quantities we consider in this paper are finite and no regularization is required.  Renormalization techniques are used here simply as an efficient way to re-sum perturbative corrections to all orders. We find easier to summarize our results below in terms of the renormalized quantities. These can be, however, straightforwardly rewritten in terms of the original (bare) quantities. Instead of the scalar operator $\cO(x)$ we will work with the renormalized operator $ \mathbf{O}(x)$, see \eqref{eq:bold_O} and \eqref{eq:renormalization_Z} for its precise definition. Instead of the bare coupling $\lambda_0$ we will work with the renormalized coupling $\lambda=\lambda(\mu)$. Here $\mu<m$ is some energy scale at which the value of $\lambda$ is defined. The $\beta$-function for the renormalized coupling reads as
\begin{equation}
	\beta(\lambda)= \delta\lambda \left(-1+\frac{\lambda}{\lambda_\star} \right) + O(\lambda^3),
\end{equation}
where the non-negative real parameter $\lambda_\star$ is given by
\begin{equation}
	\lambda_{\star} = \frac{2\delta}{C_{\cO\cO\cO}\Omega_{d-1}}+O(\delta^2).
\end{equation}
We assume that $C_{\cO\cO\cO}>0$.
The $\beta$-function has a non-trivial perturbative IR fixed point with the effective couplings $\lambda_\star$. The renormalized coupling $\lambda$ is related to $\lambda_0$ via the following relation
\begin{equation}
	\lambda(\mu) = \lambda_0\, (m\mu^{-1})^\delta\left(1-\frac{\lambda_0}{\lambda_\star}
	(m\mu^{-1})^{\delta}\right) + O(\lambda_0^3).
\end{equation}

By solving the Callan-Symanzik equation for two-point functions of scalar operators we obtain the following quantities
\begin{equation}
	\label{eq:running_coupling}
	\chi(s,\lambda) \equiv \frac{1}{1+\frac{\lambda}{\lambda_\star}\,(s^\delta-1)},\qquad
	\Lambda(s,\lambda)\equiv
	\lambda s^\delta\times\chi(s,\lambda),
\end{equation}
where $s$ is a dimensionless variable.
The $\Lambda(s,\lambda)$ quantity is often called the running coupling. These two objects provide nice interpolating functions between the UV $(s\rightarrow 0)$ and the IR $(s\rightarrow \infty)$ fixed points.

\paragraph{Results (part 1).}
The scalar two-point function valid along the whole RG flow reads as
\begin{equation}
	\label{eq:two-point_function_QFT}
	\langle \mathbf{O}(x_1) \mathbf{O}(x_2) \rangle_\text{QFT} = \frac{1}{r^{2\Delta}}
	\times
	\Big(\chi\left(\mu r,\lambda\right)\Big)^4.
\end{equation}
The normalization condition we use to define the renormalized operator $\mathbf{O}(x)$ is the requirement that the above two-point function at the distance $r = \mu^{-1}$ simply reduces to the UV CFT expression \eqref{eq:2pt_CFT}. We can see that this is indeed the case since $\chi(1,\lambda)=1$. Our result \eqref{eq:two-point_function_QFT} agrees with the one obtained in \cite{Bzowski:2012ih} by using a different method. For completeness, we review this method with some modifications in appendix \ref{App:2-point_function_independent_derivation}.

The two-point function of conserved currents reads as
\begin{equation}
	\label{eq:JJ_intro}
	\< J^\mu(x_1) J^\nu(x_2) \>_\text{QFT} = \frac{1}{r^{2(d-1)}}\times\left(h_1(\mu r,\lambda)\delta^{\mu\nu}+h_2(\mu r,\lambda)\frac{x_{12}^\mu x_{12}^\nu}{r^2}\right),
\end{equation}
where the two scalar dimensionless functions are given by
\begin{equation}
	\label{solution_JJ}
	\begin{aligned}
		h_1(\mu r,\lambda)  &= C_J - (\Delta-1)E_{JJ}C_{JJ\cO}\times\Lambda(\mu r,\lambda),\\
		h_2(\mu r,\lambda) &= -2C_J +  (\Delta+d-2)E_{JJ}C_{JJ\cO}\times\Lambda(\mu r,\lambda).
	\end{aligned}
\end{equation}
Here $C_{JJ\cO}$ is the OPE coefficient in the UV CFT three-point function of two conserved currents and the scalar operator $\cO$. For its definition see \eqref{eq:CFT_JJO}. The coefficient $E_{JJ}$ is given in \eqref{eq:EJJ}. In the IR we must recover conformal invariance. According to the condition \eqref{eq:JJ_limits} the IR central charge can be extracted as
\begin{equation}
	C_J^{IR} = h_1(\infty,\lambda) = -\frac{1}{2}\,h_2(\infty,\lambda).
\end{equation}
Plugging here our solution \eqref{solution_JJ} we obtain the change of the current central charge along the RG flow
\begin{equation}
	\label{eq:deltaCJ}
	C_J^{UV}-C_J^{IR}= \frac{4(d-1)}{d}\frac{C_{JJ\cO}}{C_{\cO\cO\cO}}\delta+O(\delta^2).
\end{equation}
The solution \eqref{solution_JJ} breaks conservation along the RG flow at the order $O(\lambda^2)$. We carefully discuss this issue in the end of section \ref{sec:JJ_two-point_functions} and propose a possible way to restore conservation at all orders. Despite this issue, our result \eqref{eq:JJ_intro} together with \eqref{solution_JJ} obeys conservation and enjoys conformal invariance in the IR limit at the order $O(\delta). $ This is the reason why \eqref{eq:deltaCJ} still holds.

We obtain an analogous to \eqref{eq:JJ_intro} result for the case of the stress-tensor in section \ref{sec:stress-tensor}. More precisely see equations
\eqref{eq:decomposition_TT_QFT}, \eqref{eq:tensor_structures}, \eqref{eq:relations_tracelessness}, \eqref{eq:coefficient_E} and \eqref{eq:functions_h_CPT_2}. Using these we can compute the change of the stress-tensor central charge along the RG flow. It reads
\begin{equation}
	C_T^{UV}-C_T^{IR}= \frac{8(d+1)}{d(d+2)^2}\f{C_{TT\mathcal{O}}}{C_{\mathcal{O}\mathcal{O}\mathcal{O}}}\delta \ +\ O(\delta^2).
\end{equation}
Here $C_{TT\cO}$ is the OPE coefficient in the UV CFT three-point function of two stress-tensors and the scalar operator $\cO$. For its definition see appendix \ref{sec:Osborn_Petkou}.
In 4d CFTs one can define the $c$-anomaly. In the standard conventions \cite{Osborn:1993cr}, see also section 2.1 in  \cite{Karateev:2022jdb} for a concise review, it is related to the stress-tensor central charge as $c\equiv \f{\pi^2}{640}C_T$. Using the above result we can then evaluate the change of the $c$-anomaly along the RG flow in 4d. It reads
\begin{equation}
	\label{eq:Delta_c}
	d=4:\qquad	\Delta c\equiv c_{UV} - c_{IR}
	=\f{\delta\, \pi^2}{2304} \f{C_{TT\mathcal{O}}}{C_{\mathcal{O}\mathcal{O}\mathcal{O}}}\  \ +\ O(\delta^2).
\end{equation}

We have also computed several renormalized three-point functions. When solving the three-point Callan-Symanzik equation we have used as the boundary condition only the three-point function computed at the trivial order in conformal perturbation theory. The scalar three-point function along the whole RG flow is then given by \eqref{eq:3pt_solution_intermed}. It can be compactly written as
\begin{equation}
	\label{eq:OOO_intro}
	\langle \mathbf{O}(x_1) \mathbf{O}(x_2)  \mathbf{O}(x_3) \rangle_\text{QFT} =
	\langle O(x_1) O(x_2)  O(x_3) \rangle_\text{UV CFT}
	\times \Big(\chi\left(\mu \rho,\lambda\right)\Big)^6.
\end{equation}
The three-point function of two conserved currents and one scalar operator is given by
\begin{multline}
	\label{eq:JJO_intro}
	\langle J^\mu(x_1) J^\nu(x_2)  \mathbf{O}(x_3) \rangle_\text{QFT} = \\
	\langle J^\mu(x_1) J^\nu(x_2)  O(x_3) \rangle_\text{UV CFT}
	\times \Big(\chi\left(\mu \rho,\lambda\right)\Big)^2 + H^{\mu\nu}(x_1,x_2,x_3),
\end{multline}
where $H^{\mu\nu}$ is a function which we will discuss shortly. The factor $\chi(\mu r,\lambda)$ explicitly breaks conservation in the right-hand side. This effect is similar to the one already encountered at the level of the two-point function \eqref{eq:JJ_intro}. One can construct the function $H^{\mu\nu}$ in such a way that the non-conservation effects are pushed to higher orders in $\lambda$. We did not attempt to do it in this paper.
Analogously to \eqref{eq:JJO_intro} we can write the three-point function of two stress-tensors and a scalar operator. It reads
\begin{multline}
	\label{eq:three-point_function}
	\langle T^{\mu\nu}(x_1) T^{\rho\sigma}(x_2) \mathbf{O}(x_3)\rangle_\text{QFT} =\\
	\langle T^{\mu\nu}(x_1) T^{\rho\sigma}(x_2)\cO(x_3)\rangle_\text{UV CFT}\times \Big(\chi\left(\mu \rho,\lambda\right)\Big)^2  + H^{\mu\nu\rho\sigma}(x_1,x_2,x_3).
\end{multline}
The CFT three-point function is defined in appendix \ref{sec:Osborn_Petkou}.

\paragraph{Results (part 2).}
In the second part of the paper given by section \ref{sec:background_field_method} we couple weakly relevant flows to non-dynamical dilaton and graviton background fields in 4d. We compute the three-dilaton vertex in terms of the scalar two-point function \eqref{eq:two-point_function_QFT} and extract the value of $\Delta a\equiv a_{UV} - a_{IR}$. We also compute the graviton-graviton-dilaton vertex in terms of the the three-point function \eqref{eq:three-point_function}, and extract the value of $\Delta c$. In order to circumvent our ignorance about the object  $H^{\mu\nu\rho\sigma}$ in \eqref{eq:three-point_function}, for computing $\Delta c$ we use only particular tensor structures in the graviton-graviton-dilaton vertex which are not sensitive to it.  The values of $\Delta a $ and $\Delta c$ are extracted by comparing the vertices obtained here with the general prediction given in \cite{Komargodski:2011vj} and \cite{Karateev:2023mrb}.
Our results read
\begin{equation}
	\label{eq:change_trace_anomalies}
	d=4:\qquad
	\Delta a=\frac{\delta^3}{2304\pi^2C_{\cO\cO\cO}^2} +O\left(\delta^4\right),\qquad
	\Delta c
	=\f{\delta\,\pi^2}{2304} \f{C_{TT\mathcal{O}}}{C_{\mathcal{O}\mathcal{O}\mathcal{O}}} \ +\ O(\delta^2).
\end{equation}
The $\Delta a$ value agrees with the result of section 2 in \cite{Komargodski:2011xv}.\footnote{Section 2 of \cite{Komargodski:2011xv} contains two independent derivations of $\Delta a$ in weakly relevant flows. The derivation of section 2.1 is based on the background field method as in this work. However, since the form of the two-point function \eqref{eq:two-point_function_QFT} was not known at the time, the author had to rely on an ad hoc regularization procedure. The rigorous derivation in section 2.2 is based on the expression of the partition function of weakly relevant flows on a sphere derived in \cite{Klebanov:2011gs}.} The $\Delta c$ value agrees with  \eqref{eq:Delta_c}.

\section{Conformal perturbation theory}
\label{sec:conformal_perturbation_theory}

Correlation functions in weakly relevant flows are computed by using the definition \eqref{eq:connected_correlation_functions} together with the action \eqref{eq:theory_definition}. In practice we expand the right-hand side of  \eqref{eq:connected_correlation_functions} around small values of $\lambda_0$. At the leading order we can then write the following expression for the scalar two-point function
 \begin{multline}
 	\label{eq:to_compute_2}
	\langle \cO(x_1) \cO(x_2) \rangle_\text{QFT} = \langle \cO(x_1) \cO(x_2) \rangle_\text{UV CFT} \\- \lambda_0 m^{d-\Delta} \int d^dx_3
	\langle \cO(x_1) \cO(x_2) \cO(x_3)\rangle_\text{UV CFT} + O(\lambda_0^2),
\end{multline}
and the following expression for the scalar three-point function
\begin{multline}
	\label{eq:to_compute_3}
	\langle \cO(x_1) \cO(x_2) \cO(x_3) \rangle_\text{QFT} = \langle \cO(x_1) \cO(x_2) \cO(x_3)\rangle_\text{UV CFT} \\- \lambda_0 m^{d-\Delta} \int d^dx_4
	\langle \cO(x_1) \cO(x_2) \cO(x_3) \cO(x_4)\rangle_\text{UV CFT} + O(\lambda_0^2).
\end{multline}
In this section we will discuss how to evaluate the integral in the expressions \eqref{eq:to_compute_2}. In this paper we will only use \eqref{eq:to_compute_3} at the trivial order of conformal perturbation theory, namely we will ignore the integral over the CFT four-point function.

The most general CFT three-point function is given in equation \eqref{eq:3pt_CFT}. Our goal is to evaluate the following integral
\begin{multline}
	\label{eq:integral_3pt_subsection}
	\int d^dx_3\ \langle\cO_{\Delta_1}(x_1)\cO_{\Delta_2}(x_2)\cO_{\Delta_3}(x_3)\rangle_{\text{UV CFT}}=\\
	\frac{C_{\cO_1\cO_2\cO_3}}{\left(x_{12}^2\right)^\frac{\Delta_1+\Delta_2-\Delta_3}{2}}\int d^dx_3\
	\frac{1}{\left(x_{13}^2\right)^\frac{\Delta_1-\Delta_2+\Delta_3}{2}\left(x_{23}^2\right)^\frac{-\Delta_1+\Delta_2+\Delta_3}{2}}.
\end{multline}
Let us start by discussing its divergences.  The divergences appearing when $x_3=x_1$ or $x_3=x_2$ are called the UV divergences. Let us consider the situation when $x_3=x_1$, then the leading part of the above integral becomes
\begin{equation}
 \int d^dx_3\ \frac{1}{\left(x_{13}^2\right)^\frac{\Delta_1-\Delta_2+\Delta_3}{2}} = \Omega_{d-1}
  \int_0^\infty d|x_{13}|\ \frac{|x_{13}|^{d-1}}{|x_{13}|^{\Delta_1-\Delta_2+\Delta_3}}.
\end{equation}
The UV divergence comes from the integration limit $|x_{13}|=0$. It is absent if
\begin{equation}
	\label{eq:condition_13}
	\Delta_1-\Delta_2+\Delta_3<d.
\end{equation} 
We should also consider the $x_3=x_2$ case. It will lead to an analogous to \eqref{eq:condition_13} condition. Combining the two we conclude that the UV divergences are absent if the following conditions are satisfied
\begin{equation}
	\label{eq:UV_divergence_constraint_2}
	\Delta_3<d,\qquad
	\Delta_1-\Delta_2+\Delta_3<d.
\end{equation}
The divergence appearing when $x_3=\infty$ is called the IR divergence. In the $x_3=\infty$ limit the leading part of the integral \eqref{eq:integral_3pt_subsection} reads as
\begin{equation}
	\int d^d x_3 \frac{1}{\left(x_3^2\right)^{\Delta_3}} = \Omega_{d-1}	\int_0^\infty d|x_3| \frac{|x_3|^{d-1}}{|x_3|^{2\Delta_3}}.
\end{equation}
The IR divergence comes from the integration limit $|x_{3}|=\infty$. It is absent if
\begin{equation}
	\label{eq:IR_divergence_constraint_2}
	\Delta_3>d/2.
\end{equation}

The integral appearing in \eqref{eq:to_compute_2} needs to be evaluated for the case $\Delta_i=\Delta=d-\delta$. This case satisfies both bounds \eqref{eq:UV_divergence_constraint_2} and \eqref{eq:IR_divergence_constraint_2}, and, thus, is free of UV and IR divergences. 

The integral \eqref{eq:integral_3pt_subsection} can be straightforwardly evaluated in the range when it is free of UV and IR divergences. We compute this integral in appendix \ref{app:integral_3pt}. Here let us simply provide the final answer
\begin{equation}
	\label{eq:integral_3pt}
	\int d^dx_3\ \langle\cO_{\Delta_1}(x_1)\cO_{\Delta_2}(x_2)\cO_{\Delta_3}(x_3)\rangle =
	\frac{C_{\cO_1\cO_2\cO_3}}{\left(x_{12}^2\right)^\frac{\Delta_1+\Delta_2+\Delta_3-d}{2}} N(\Delta_1, \Delta_2, \Delta_3),
\end{equation}
where $N(\Delta_1, \Delta_2, \Delta_3)$ reads as
\begin{equation}
	\label{eq:N}
	N(\Delta_1, \Delta_2, \Delta_3) = \pi^{\f{d}{2}}\,
	\f{\Gamma\left( \f{\Delta_1-\Delta_2-\Delta_3+d}{2}\right) \Gamma\left( \f{\Delta_2-\Delta_3-\Delta_1+d}{2}\right)}{\Gamma\left(\f{\Delta_2+\Delta_3-\Delta_1}{2}\right)\Gamma\left(\f{\Delta_3+\Delta_1-\Delta_2}{2}\right)}\times
	\frac{\Gamma\left(\Delta_3-\f{d}{2}\right)}{\Gamma\left( d-\Delta_3\right)}.
\end{equation}

Using this we can obtain the  explicit expression for the two-point function \eqref{eq:to_compute_2}. It reads
\begin{empheq}[box=\widefbox]{equation}
	\label{eq:G_CPT_2}
\langle\mathcal{O}(x_1)\mathcal{O}(x_2)\rangle_{\text{QFT}} = \f{1}{r^{2\Delta}}
\left(1-\lambda_0 (m r)^{d-\Delta} C_{\cO\cO\cO}N(\Delta, \Delta, \Delta)
+ O\big(\lambda_0^2\big) \right).
\end{empheq}
For the later purposes it is important to define the following quantity
\begin{equation}
	\label{eq:definition_B}
	B(\delta) \equiv \delta \times N(d-\delta,\,d-\delta,\, d-\delta).
\end{equation}
This function remains finite in the limit $\delta\rightarrow 0$, more precisely
\begin{equation}
	\label{eq:expansion_B}
	B(\delta) = 2\Omega_{d-1} + O(\delta^2).
\end{equation}

\section{Scalar correlators along the RG flow}
\label{sec:Callan-Symanzik_correlators}

In the previous section we computed the scalar two-point function using conformal perturbation at the leading order. The result is given in equation \eqref{eq:G_CPT_2}. In the UV limit $r\rightarrow 0$ we recover conformally invariant two-point function as expected. However, the IR limit $r\rightarrow \infty$ is ill-defined, because the perturbative expansion breaks down due to the large value of the second term. In order to be able to take the IR limit one has to either compute \eqref{eq:G_CPT_2} to all orders of conformal perturbation theory or to at least re-sum the leading order perturbative corrections to all orders. In this section we perform the latter re-summation by using the renormalization group techniques. By solving the Callan-Symanzik equation for the scalar two-point function we will obtained the re-summed version of \eqref{eq:G_CPT_2} valid at all energies. We will also solve the Callan-Symanzik equation for the scalar three-point function.

\subsection{Renormalization}
Our theory is defined via the action \eqref{eq:theory_definition}. We can provide an alternative description of this theory by integrating out the degrees of freedom up to some energy scale $\mu$. Our action will take the following form then 
\begin{equation}
	\label{eq:theory_alternative}
	A =  A_\text{EFT} + \lambda(\mu) \mu^{d-\Delta} \int d^d x\, \mathbf{O}(x).
\end{equation}
Here $A_\text{EFT}$ stands for the effective action at the scale $\mu$ obtained from the original action by integrating out the high energy degrees of freedom up to scale $\mu$, and $\lambda(\mu)$ is an effective coupling usually called the renormalized coupling. The bold operator is defined as
\begin{equation}
	\label{eq:bold_O}
	\mathbf{O}(x) \equiv Z(\mu)^{-\frac{1}{2}} \cO(x),
\end{equation}
where $Z(\mu)$ is a numerical factor, whose choice depends on the choice of $\mu$. The operator \eqref{eq:bold_O} is usually called the renormalized operator. At the energy scale $\mu$, the correlation functions of $\mathbf{O}(x)$ take the same form as the conformal correlators given in \eqref{eq:2pt_CFT} and \eqref{eq:3pt_CFT}, with the conformal dimension $\Delta = d - \delta$ and the OPE coefficient $C_{\mathcal{O}\mathcal{O}\mathcal{O}}$. This will be implemented through the renormalization conditions below.

In this section we will study the renormalized QFT scalar two- and three-point functions
\begin{align}
	\label{eq:renormalized_2pt}
	F_2(r,\mu,\lambda) &\equiv \langle\mathbf{O}(x_1)\mathbf{O}(x_2)\rangle_{\text{QFT}},\\
	\label{eq:renormalized_3pt}
	F_3(\rho,y_1,y_2,\mu,\lambda) &\equiv \langle\mathbf{O}(x_1)\mathbf{O}(x_2)\mathbf{O}(x_3)\rangle_{\text{QFT}}.
\end{align}
Recall, that $r$, $\rho$, $y_1$ and $y_2$ are defined in \eqref{eq:definition_r} and \eqref{eq:definition_r_y1_y2_app}.
Due to \eqref{eq:bold_O} the renormalized correlators can be expressed in terms of the original ones as
\begin{align}
	\label{eq:2pt_practical}
	F_2(r,\mu,\lambda) &=
	Z(\mu)^{-1}\langle\cO(x_1)\cO(x_2)\rangle_{\text{QFT}}
	,\\
	\label{eq:3pt_practical}
	F_3(\rho,y_1,y_2,\mu,\lambda) &=
	Z(\mu)^{-\frac{3}{2}}\langle\cO(x_1)\cO(x_2)\cO(x_3)\rangle_{\text{QFT}}.
\end{align}

The numerical factor $Z(\mu)$ allows to pick a convenient normalization of (renormalised) operators at the scale $\mu$. The most natural choice is given by the following (renormalization) condition
\begin{equation}
	\label{eq:renormalization_condition}
	F_2(r=\mu^{-1},\,\mu,\lambda) = \langle\cO(x_1)\cO(x_2)\rangle_{\text{CFT}}\Big|_{r=\mu^{-1}} = \mu^{2\Delta}.
\end{equation}
Here we have used the convention \eqref{eq:2pt_CFT}.
This condition enforces our renormalized correlator to be conformally invariant at the distance $r=\mu^{-1}$.
Plugging our perturbative result \eqref{eq:G_CPT_2} for the two-point function in \eqref{eq:2pt_practical}, and requiring \eqref{eq:renormalization_condition} we obtain the following result for the numerical factor $Z(\mu)$,
\begin{equation}
	\label{eq:Zmu_bare}
	Z(\mu)=1-(m \mu^{-1})^{d-\Delta}\lambda_0\,C_{\cO\cO\cO} N(\Delta,\Delta,\Delta)
	+ O\big(\lambda_0^2\big).
\end{equation}

The $\beta$-function for the effective (renormalized) coupling is defined as
\begin{equation}
	\label{eq:beta-function_def}
	\beta(\lambda) \equiv \mu \frac{d\lambda(\mu)}{d\mu}.
\end{equation}
It shows how the value of the renormalized coupling changes with the change of the scale choice $\mu$. Following the strategy of  \cite{Brezin:1973igb,Cappelli:1989yu} it can be derived by studying the trace of the stress-tensor obtained using the two actions \eqref{eq:theory_definition} and \eqref{eq:theory_alternative}. Since the stress-tensor cannot be renormalized (as a conserved operator) the two results must agree. One can then show that
\begin{equation}
T_{\text{QFT}}{}^\mu_{\ \mu}(x)=(d-\Delta) \lambda_0m^{d-\Delta} \cO(x)=-\beta(\lambda)\mu^{d-\Delta}\mathbf{O}(x). \label{eq:traceT_relation}
\end{equation}
See appendix \ref{app:trace_ST} for the detailed derivation of this statement.
Plugging here \eqref{eq:Zmu_bare} we obtain
\be
\beta(\lambda) =-\delta (m\mu^{-1})^{\delta}\lambda_0 +\frac{1}{2}\delta(m \mu^{-1})^{2\delta}\lambda_0^2C_{\cO\cO\cO}\,  N(\Delta,\Delta,\Delta)
+ O\big(\lambda_0^3\big).\label{eq:Beta_bare}
\ee

At extremely high energies we should recover the original UV CFT. As a result we must require the following condition
\begin{equation}
	\lambda(\infty) = 0.
\end{equation}
The relation between the original (bare) coupling $\lambda_0$ and the effective (renormalized) coupling $\lambda(\mu)$ can be obtained by solving the equation \eqref{eq:beta-function_def} together with the above boundary condition. We can write then
\begin{equation}
	\int_{0}^{\lambda(\mu)} d\lambda = \int_\infty^\mu \frac{d\mu}{\mu} \beta(\lambda).
\end{equation}
Notice, that the $\beta$-function in the form \eqref{eq:Beta_bare} is a function of $\mu$ only. Performing the integral we obtain
\begin{equation}
	\label{eq:lambda(mu)}
	\lambda(\mu) = \lambda_0 (m\mu^{-1})^\delta\left(1-\frac{1}{4}\lambda_0\,C_{\cO\cO\cO} N(\Delta,\Delta,\Delta)
	(m\mu^{-1})^{\delta}+O(\lambda_0^2)\right).
\end{equation}
We can invert this relation, the result reads
\begin{equation}
	\label{eq:relation_lambda0_lambda}
	\lambda_0 = \left(\mu m^{-1}\right)^\delta \lambda(\mu)\left(1
	+\frac{1}{4}\lambda(\mu) C_{\cO\cO\cO} N(\Delta,\Delta,\Delta)+
	O(\lambda^2)\right).
\end{equation}
Plugging \eqref{eq:relation_lambda0_lambda} in \eqref{eq:Zmu_bare} and \eqref{eq:Beta_bare} we obtain the final expressions
\begin{align}
	\label{eq:renormalization_Z}
	Z(\mu) &= 1 - \lambda(\mu)C_{\cO\cO\cO}N(\Delta,\Delta,\Delta)
	+ O\big(\lambda^2\big),\\
	\label{eq:beta_function}
	\beta(\lambda)&= \delta\,\lambda \left(-1+\f{1}{4}\lambda C_{\cO\cO\cO} N(\Delta,\Delta,\Delta)+O(\lambda^2)\right).
\end{align}

Finally, let us also introduce the $\gamma$-function defined as
\begin{equation}
	\label{eq:gamma_def}
	\gamma(\lambda) \equiv \frac{1}{2}\, \frac{d \log Z(\mu)}{d\log\mu}.
\end{equation}
Applying this definition to \eqref{eq:Zmu_bare} we get
\begin{equation}
	\gamma(\lambda) = \frac{\delta}{2}\, \lambda_0 \left(m\mu^{-1}\right)^\delta C_{\cO\cO\cO}N(\Delta,\Delta,\Delta)
+O(\lambda_0^2).
\end{equation}
Notice that since \eqref{eq:Zmu_bare} has only been computed up to $\lambda_0$ order, the $\gamma$-function can only be evaluated up to the order $O(\lambda_0)$.
Plugging here the relation \eqref{eq:relation_lambda0_lambda} we finally conclude that
\begin{equation}
	\label{eq:anomalous_dimension}
	\gamma(\lambda) = \frac{\delta}{2}\lambda\, C_{\cO\cO\cO} N(\Delta,\Delta,\Delta)+O(\lambda^2).
\end{equation}

\subsubsection{IR fixed point}
\label{sec:IR_fixed_point}
\begin{figure}[t]
	\centering
	\includegraphics[scale=0.99]{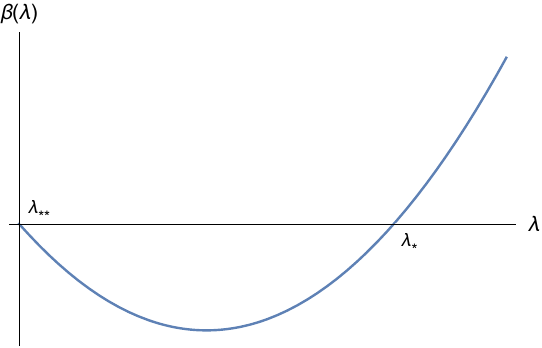}
	\caption{The $\beta$-function in weakly relevant flows. }
	\label{fig_beta-function}
\end{figure}

The $\beta$-function \eqref{eq:beta_function} has two zeros
\begin{equation}
	\label{eq:zero_beta}
	\beta(\lambda) = 0.
\end{equation}
They are achieved with the following values of the renormalized coupling
\begin{equation}
	\label{eq:lambda_star}
	\lambda_{\star} = \frac{4}{C_{\cO\cO\cO}N(\Delta,\Delta,\Delta)},\qquad
	\lambda_{\star\star} =  0.
\end{equation}
Using these, the $\beta$-function \eqref{eq:beta_function} can be rewritten as
\begin{equation}
	\label{eq:beta_rewritten}
	\beta(\lambda)= \delta\lambda \left(-1+\frac{\lambda}{\lambda_\star} \right) + O(\lambda^3).
\end{equation}
The first value $\lambda_{\star} $ corresponds to the IR fixed point. The second value $\lambda_{\star\star} $ corresponds to the UV fixed point. The easiest way to see these statements is by using the running coupling constant which we will introduce and compute in the next subsection. The $\beta$-function \eqref{eq:beta_rewritten} has the form depicted in figure \ref{fig_beta-function}.
Finally, using the definition of $B(\delta)$ given in
\eqref{eq:definition_B} and its expansion \eqref{eq:expansion_B} we can write
\begin{equation}
	\label{eq:lambda_star_final}
	\boxed{
	\lambda_{\star} = \frac{2\delta}{C_{\cO\cO\cO}\Omega_{d-1}}+O(\delta^2).
}
\end{equation}
Plugging \eqref{eq:lambda_star_final} into \eqref{eq:beta_rewritten} we get
\begin{equation}
	\label{eq:beta_rewritten_2}
	\beta(\lambda)=  -\delta\lambda+\frac{1}{2}C_{\cO\cO\cO}\Omega_{d-1}\lambda^2 + O(\lambda^3,\, \lambda^2 \delta).
\end{equation}
In order to be able to trust the IR solution \eqref{eq:lambda_star} of \eqref{eq:zero_beta} we must require that the two terms in \eqref{eq:beta_rewritten_2} are roughly of the same order. This can be achieved by demanding that $\delta\ll1$, which makes the a priori leading term in \eqref{eq:beta_rewritten_2} accidentally small.

\subsubsection{Renormalized correlators}
In the case of the two-point function we can plug \eqref{eq:G_CPT_2} together with \eqref{eq:renormalization_Z} and \eqref{eq:relation_lambda0_lambda} into \eqref{eq:2pt_practical}. The result reads
\begin{equation}
	F^{\text{pert}}_2(r,\mu,\lambda) = \frac{1}{r^{2\Delta}}\,\left(1+\lambda\, C_{\cO\cO\cO} N(\Delta,\Delta,\Delta)(1-(r \mu)^\delta) + O (\lambda^2)\right).
\end{equation}
Recall the relations \eqref{eq:definition_B} and \eqref{eq:expansion_B}. At the leading order in $\delta$ we can rewrite the above expression as
\begin{equation}
	\label{eq:F_QFT_2pt_pert}
	\boxed{
	F^{\text{pert}}_2(r,\mu,\lambda) = \frac{1}{r^{2\Delta}}\,\left(1-2\lambda\,C_{\cO\cO\cO} \Omega_{d-1}\log(r \mu) + O (\lambda^2, \lambda\delta)\right).
}
\end{equation}
It is interesting to notice that the renormalized scalar two-point function \eqref{eq:F_QFT_2pt_pert} remains finite in the limit $\delta\rightarrow 0$, whereas the bare one given by \eqref{eq:G_CPT_2} diverges due to the presence of the $1/\delta$ pole.
Analogously, at the trivial order of conformal perturbation theory we get
\begin{equation}
	\label{eq:F_QFT_3pt_pert}
	\boxed{
	F^{\text{pert}}_3(\rho,y_1,y_2,\mu,\lambda)=
	\frac{1}{\rho^{3\Delta}}\times
	\Bigg(C_{\cO\cO\cO}+O(\lambda)\Bigg).
}
\end{equation}

\subsection{Derivation of the Callan-Symanzik equation}
\label{sec:Callan-Symanzik}
Using dimensional analysis we can represent the renormalized two-point function \eqref{eq:renormalized_2pt} as follows
\begin{equation}
	\label{eq:form_2}
	F_2(r,\mu,\lambda) = \frac{f_2(r\mu,\lambda)}{r^{2\Delta}}.
\end{equation}
Comparing this with equation \eqref{eq:F_QFT_2pt_pert} we obtain the perturbative expression for the function $f_2$. It reads as
\begin{equation}
	\label{eq:f_QFT_2pt_pert}
	f^\text{pert}_2(r\mu,\lambda) = 1-2\lambda\,C_{\cO\cO\cO} \Omega_{d-1}\log(r \mu) + O (\lambda^2, \delta).
\end{equation}
Analogously, we can represent the renormalized three-point function \eqref{eq:renormalized_3pt} as follows
\begin{equation}
	\label{eq:form_3}
	F_3(\rho,\mu,\lambda) = \frac{f_3(\rho\mu,y_1,y_2,\lambda)}{\rho^{3\Delta}}.
\end{equation}
Notice that the variables $y_1$ and $y_2$ are dimensionless. Comparing with \eqref{eq:F_QFT_3pt_pert} we simply conclude that
\begin{equation}
	\label{eq:f_QFT_3pt_pert}
	f^\text{pert}_3(\rho\mu,y_1,y_2,\lambda)=
	C_{\cO\cO\cO}+O(\lambda).
\end{equation}

Consider the following renormalized $n$-point function
\begin{equation}
	\label{eq:n-point_function_renormalized}
	F_n(x_1,\ldots, x_n, \mu,\lambda) \equiv Z(\mu)^{-\frac{n}{2}}\langle\cO(x_1)\ldots\cO(x_n)\rangle_{\text{QFT}}.
\end{equation}
The original (unrenormalized) $n$-point function does not depend on the scale $\mu$. As a result we have the following condition
\begin{equation}
	\mu\frac{d}{d\mu} \langle\cO(x_1)\ldots\cO(x_n)\rangle_{\text{QFT}} =0.
\end{equation}
Plugging here \eqref{eq:n-point_function_renormalized} we obtain
\begin{multline}
	\frac{n}{2}\,Z(\mu)^{\frac{n}{2}-1} \left(\mu \frac{dZ(\mu)}{d\mu}\right) 
	F_n(x_1,\ldots, x_n,\mu,\lambda)+\\
	Z(\mu)^\frac{n}{2}\left(\mu\frac{\partial}{\partial\mu}+\mu\frac{d\lambda}{d\mu}\frac{\partial}{\partial\lambda}\right)F_n(x_1,\ldots, x_n,\mu,\lambda) = 0.
\end{multline}
Using the definitions of the $\beta$-function and the anomalous dimension $\gamma$ given by \eqref{eq:beta-function_def} and \eqref{eq:gamma_def}, and stripping off the factor $Z(\mu)^\frac{n}{2}$ we obtain the Callan-Symanzik equation
\begin{equation}
	\label{eq:CS_equation_n}
	\left( \mu\frac{\partial}{\partial \mu}+\beta(\lambda)\frac{\partial}{\partial\lambda}+n\gamma(\lambda)\right)  F_n(x_1,\ldots, x_n,\mu,\lambda)=0.
\end{equation}

Let us focus on the case of $n=2$ and $n=3$ point functions. Plugging \eqref{eq:form_2} into the Callan-Symanzik equation \eqref{eq:CS_equation_n} we obtain the following equation
\begin{equation}
	\label{eq:CS_2}
	s\equiv r\mu:\qquad
	\left( s\frac{\partial}{\partial s}+\beta(\lambda)\frac{\partial}{\partial\lambda}+2\gamma(\lambda)\right) f_2(s,\lambda)=0.
\end{equation}
Analogously, plugging \eqref{eq:form_3} into the Callan-Symanzik equation \eqref{eq:CS_equation_n} we obtain 
\begin{equation}
	\label{eq:CS_3}
	\sigma\equiv \rho\mu:\qquad
	\left( \sigma\frac{\partial}{\partial \sigma}+\beta(\lambda)\frac{\partial}{\partial\lambda}+3\gamma(\lambda)\right)   f_3(\sigma,y_1,y_2,\lambda)=0.
\end{equation}

\subsection{Solution of the Callan-Symanzik equation: two-point function}
\label{sec:Callan-Symanzik_2pt}
Let us now obtain the most general solution of the Callan-Symanzik equation \eqref{eq:CS_2}.

 We denote the characteristics of the equation \eqref{eq:CS_2}, also known as the running coupling, by  $\Lambda (s,\lambda)$. It is a dimensionless function, which obeys the characteristic equation
\begin{equation}
	\label{eq:characteristic}
	s\,\frac{\partial}{\partial s} \Lambda(s,\lambda) = - \beta\big(\Lambda(s,\lambda)\big).
\end{equation}
The function $\Lambda (s,\lambda)$ provides an effective description of the interaction strength in a QFT depending on the position on the RG flow. In order to solve the above equation we must provide a boundary condition. We require that at some point of the flow $s_0$ the characteristic $\Lambda (s,\lambda)$ simply reduces to $\lambda$, namely
\begin{equation}
	\label{eq:boundary_condition}
	\Lambda (s=s_0,\lambda) = \lambda.
\end{equation}

We can solve \eqref{eq:characteristic} with the boundary condition \eqref{eq:boundary_condition} by using the explicit form of the $\beta$-function given in \eqref{eq:beta_rewritten}. The result reads as
\begin{equation}
	\label{eq:lambda-bar_sol}
	\Lambda(s,\lambda)
	=	\lambda (s/s_0)^\delta\times\frac{1}{1+\frac{\lambda}{\lambda_\star}\,\big((s/s_0)^\delta-1\big)},
\end{equation}
where the constant $\lambda_\star$ is given in \eqref{eq:lambda_star}. The IR ($s\rightarrow \infty$) and the UV ($s\rightarrow 0$) limits of the running coupling read
\begin{equation}
	\Lambda(\infty,\lambda) = \lambda_\star,\qquad
	\Lambda(0,\lambda) = 0. 
\end{equation}
This is in a perfect agreement with the statements made in subsection \ref{sec:IR_fixed_point}.

The standard solution of the Callan-Symanzik equation \eqref{eq:CS_2} reads
\begin{equation}
	\label{eq:solution_2}
	f_2(s,\lambda) =   f_2^\text{boundary}\big(s_0,\Lambda(s,\lambda)\big) \times
	\exp\left(-2 \int_{s_0}^s \frac{ds'}{s'}\gamma\big(\Lambda(s',\lambda)\big)\right),
\end{equation}
where $f_2^\text{boundary}(s,\lambda)$ is the boundary condition. In practice, we use the perturbative result \eqref{eq:f_QFT_2pt_pert} as the boundary condition. We verify the validity of the solution \eqref{eq:solution_2} in appendix \ref{app:CS_details} for completeness.

Let us now compute the exponent 
\begin{equation}
	\label{eq:exponent}
	\exp\left(-n \int_{s_0}^s \frac{ds'}{s'}\gamma\big(\Lambda(s',\lambda)\big)\right)
\end{equation}
appearing in the solution \eqref{eq:solution_2} with $n=2$. The anomalous dimension has been obtained in \eqref{eq:anomalous_dimension}, it reads
\begin{equation}
	\label{eq:anomalous_dimension_leading}
	\gamma(\lambda) = \frac{2\delta}{\lambda_\star}\,\lambda(\mu)	+ O(\lambda^2).
\end{equation}
This approximation is enough since the $O(\lambda^2)$ will be sub-sub-leading in the evaluation of the above exponent. Plugging \eqref{eq:anomalous_dimension_leading} into \eqref{eq:exponent} and using \eqref{eq:lambda-bar_sol} one obtains
\begin{equation}
	\label{eq:exp_solution}
	\exp\left(-n \int_{s_0}^s \frac{ds'}{s'}\gamma\big(\Lambda(s',\lambda)\big)\right)= \left(\frac{1}{1+\frac{\lambda}{\lambda_\star}\,\big((s/s_0)^\delta-1\big)}\right)^{2n}.
\end{equation}

Using \eqref{eq:solution_2} together with \eqref{eq:form_2} and \eqref{eq:f_QFT_2pt_pert} we simply get
\begin{equation}
	F_2(r,\mu,\lambda) = \frac{1}{r^{2\Delta}}
	\times\left( 1-2\Lambda(\mu r,\lambda)\,C_{\cO\cO\cO} \Omega_{d-1}\log(s_0)\right)
	\left(\frac{1}{1+\frac{\lambda}{\lambda_\star}\,\big((\mu r/s_0)^\delta-1\big)}\right)^4.
\end{equation}
We can now fix the value of $s_0$ by recalling the renormalization condition \eqref{eq:renormalization_condition} on the two-point function. The above result agrees with \eqref{eq:renormalization_condition} only after setting
\begin{equation}
	s_0=1.
\end{equation}
Taking this into account we obtain the final solution
\begin{equation}
	\label{eq:2pt_solution}
	\boxed{F_2(r,\mu,\lambda) = \frac{1}{r^{2\Delta}}
		\times
		\left(\frac{1}{1+\frac{\lambda}{\lambda_\star}\,\big((\mu r)^\delta-1\big)}\right)^4.\ }
\end{equation}

\paragraph{IR limit}
In the IR limit $r\rightarrow \infty$ for the two-point function we expect to recover conformal symmetry, namely
\begin{equation}
	\label{eq:expectations}
	\lim_{r\rightarrow \infty}\mu^{2\gamma_\mathbf{O}} F_2(r,\mu,\lambda) = \frac{N_{IR}^2}{r^{2\Delta_{IR}}},
\end{equation}
where the IR anomalous dimension for the operator $\mathbf{O}$ reads as
\begin{equation}
	\gamma_\mathbf{O} \equiv\Delta_{IR}-\Delta.
\end{equation}
Taking the limit inside our explicit expressions \eqref{eq:2pt_solution} we get
\begin{equation}
	\lim_{r\rightarrow \infty}\mu^{4\delta} F_2(r,\mu,\lambda) =  \frac{1}{r^{2(\Delta+2\delta)}} \times \left(\frac{\lambda_\star}{ \lambda}\right)^4.
\end{equation}
Comparing these with our expectation \eqref{eq:expectations} we get
\begin{equation}
	\label{eq:IR_parameters}
	\Delta_{IR}=\Delta+2\delta,\qquad
	\gamma_\mathbf{O} = 2\delta,\qquad
	N_{IR} = \left(\frac{\lambda_\star}{\lambda}\right)^2.
\end{equation}
The second entry in the above expressions is expected from \eqref{eq:anomalous_dimension_leading}, since $\gamma_\mathbf{O}= \gamma(\lambda_\star)$.

\paragraph{Unrenormalized two-point function}
Using \eqref{eq:2pt_solution} we can also obtain the two-point function of un-renormalized operators. It reads
\begin{equation}
	\langle\cO(x_1)\cO(x_2)\rangle_{\text{QFT}} = Z(\mu) F_2(r,\mu,\lambda)
	= \frac{1 - \frac{4\lambda}{\lambda_\star}}{r^{2\Delta}}
	\times
	\left(\frac{\lambda_\star}{\lambda_\star+\lambda\,\left((r\mu)^\delta-1\right)}\right)^4.
\end{equation}
Here we have used \eqref{eq:2pt_practical}, \eqref{eq:renormalization_Z} and \eqref{eq:lambda_star}. We can now plug here the expression for $\lambda(\mu)$ in terms of $\lambda_0$ given by \eqref{eq:lambda(mu)}. The result will be independent on $\mu$ up to $O(\lambda_0^2)$ order. Using this independence we can set $\mu$ to any value we like. One convenient choice is to set $\mu=m$. Taking this into account we can write
\begin{equation}
	\label{eq:bare_OO}
	\boxed{
	\langle\cO(x_1)\cO(x_2)\rangle_{\text{QFT}}
	=  \frac{1 - \frac{4\lambda_0}{\lambda_\star}}{r^{2\Delta}}
	\times
	\left(\frac{\lambda_\star}{\lambda_\star+\lambda_0\,\left((rm)^\delta-1\right)}\right)^4 +O(\lambda_0^2).
}
\end{equation}
In the UV limit we get
\begin{equation}
	\lim_{r\rightarrow 0} \langle\cO(x_1)\cO(x_2)\rangle_{\text{QFT}}
	= \frac{1}{r^{2\Delta}}+O(\lambda_0^2).
\end{equation}
In agreement with our expectations at the order in $\lambda_0$ we are working at.

\paragraph{Further discussion} In order to develop a better understanding of our result \eqref{eq:2pt_solution} let us expand it around small coupling $\lambda$. We get then
\begin{equation}
	\label{eq:expansions_our_result}
	F_2(r,\mu,\lambda) = \frac{1}{r^{2(d-\delta)}}\sum_{n=0}^\infty c_{nn}\lambda^n q(\mu r)^n,
\end{equation}
where the coefficient $c_{nn}$ reads as
\begin{equation}
	\label{eq:coefficients_cnn}
	c_{nn}\equiv \frac{(n+3)!}{3! n!}(-1)^n 
\end{equation}
and the function $q(\mu r)$ is given by
\begin{equation}
	\label{eq:definition_q}
	q(\mu r)\equiv \frac{(\mu r)^\delta-1}{\lambda_\star}=
	\f{C_{\cO\cO\cO}\Omega_{d-1}}{2}
	\frac{(\mu r)^\delta-1}{\delta} = \f{C_{\cO\cO\cO}\Omega_{d-1}}{2}
	\log(\mu r) + O(\delta).
\end{equation}

From conformal perturbation theory it is clear, however, that the most general expression of the two-point function has the following series representation
\begin{equation}
\label{eq:expansions_expectation}
F_2(r,\mu,\lambda) = \frac{1}{r^{2(d-\delta)}}\sum_{n=0}^\infty \sum_{m=0}^n c_{nm}\lambda^n \left(\log(\mu r)\right)^m +O(\delta),
\end{equation}
where $n$ is the order of conformal perturbation theory and $m\leq n$ for each n. Comparing \eqref{eq:expansions_our_result} with the expectation \eqref{eq:expansions_expectation} and taking into account that $q(\mu r)\sim \log(\mu r)$, we conclude that
our result \eqref{eq:2pt_solution} sums only the ``leading-log'' contributions $\lambda^n \log^n$ and completely misses contributions of the form $\lambda^n \log^m$ with $n\geq 2$ and $m<n$.

\subsection{Solution of the Callan-Symanzik equation: three-point function}
\label{eq:CS_3pt}
Let us now obtain the most general solution of the 3-point Callan-Symanzik equation \eqref{eq:CS_3}. The discussion here remains almost identical to the previous section.

The characteristic equation for \eqref{eq:CS_3} reads as
\begin{equation}
	\label{eq:characteristic_3}
	\sigma\,\frac{\partial}{\partial \sigma} \Lambda(\sigma,\lambda) = - \beta\big(\Lambda(\sigma,\lambda)\big),
\end{equation}
together with the boundary condition
\begin{equation}
	\label{eq:boundary_condition_2}
	\Lambda (\sigma=\sigma_0,\lambda) = \lambda.
\end{equation}
The most general solution of  \eqref{eq:CS_3} is then given by
\begin{equation}
	\label{eq:solution_CS_3}
	f_3(\sigma,y_1,y_2,\lambda) =   f_3^\text{boundary}\big(\sigma_0,y_1,y_2,\Lambda(\sigma,\lambda)\big) \times
	\exp\left(-3 \int_{\sigma_0}^\sigma \frac{d\sigma'}{\sigma'}\gamma\big(\Lambda(\sigma',\lambda)\big)\right),
\end{equation}
where $f_3^\text{boundary}(\sigma,y_1,y_2,\lambda)$ is the boundary condition.  In practice, we will use the perturbative result \eqref{eq:f_QFT_3pt_pert} as this boundary condition. Using the results of the previous section we can write
\begin{equation}
	\label{eq:3pt_solution_intermed}
	\boxed{
    F_3(\rho,y_1,y_2,\mu,\lambda) = \frac{1}{\rho^{3\Delta}}
	\times C_{\cO\cO\cO}\times
	\left(
	\frac{1}{1+\frac{\lambda}{\lambda_\star}\,\big((\mu\rho/\sigma_0)^\delta-1\big)}
	\right)^6.
}
\end{equation}
Without loss of generality we can set $\sigma_0=1$.

\section{Conserved currents}
\label{sec:conserved_currents}

Let us assume that our UV CFT has a U(1) global symmetry and the associated Abelian current is $J^\mu(x)$. The goal of this section is to compute the following QFT correlation functions along the RG flow
\begin{equation}
	\label{eq:correlators_JJO}
	\< J^\mu(x_1) J^\nu(x_2) \>_\text{QFT},\qquad
	\< J^\mu(x_1) J^\nu(x_2) \cO(x_3)\>_\text{QFT}.
\end{equation}

\subsection{Two-point function}
\label{sec:JJ_two-point_functions}
The CFT two-point function has the following simple form
\begin{equation}
	\label{eq:CFT_JJ}
	\< J^\mu(x_1) J^\nu(x_2) \>_\text{CFT} = \frac{C_J}{r^{2(d-1)}}\times\left(\delta^{\mu\nu}-2\,\frac{x_{12}^\mu x_{12}^\nu}{r^2}\right),
\end{equation}
where $C_J>0$ is called the current central charge. Its value is part of the CFT data. Using dimensional analysis one can conclude that the most general current two-point function in weakly relevant flows has the following form
\begin{equation}
	\label{eq:QFT_JJ}
	\< J^\mu(x_1) J^\nu(x_2) \>_\text{QFT} = \frac{1}{r^{2(d-1)}}\times\left(h_1(mr,\lambda_0)\delta^{\mu\nu}+h_2(mr,\lambda_0)\frac{x_{12}^\mu x_{12}^\nu}{r^2}\right),
\end{equation}
where $h_i(mr,\lambda_0)$ are dimensionless functions. In what follows we will compute these functions using conformal perturbation theory. In order to reproduce the UV and IR fixed points these functions must obey the following asymptotic conditions
\begin{equation}
	\label{eq:JJ_limits}
	\begin{aligned}
		&\lim_{r\rightarrow 0} h_1(mr,\lambda_0) = C_J^{UV},\quad\,
		\lim_{r\rightarrow 0} h_2(mr,\lambda_0) = -2C_J^{UV},\\
		&\lim_{r\rightarrow \infty} h_1(mr,\lambda_0) = C_J^{IR},\quad
		\lim_{r\rightarrow \infty} h_2(mr,\lambda_0) = -2C_J^{IR}.
	\end{aligned}
\end{equation}
Conservation of the current $\partial_{\mu} J^\mu(x)=0$ implies the following differential equation
\begin{equation}
	\label{eq:conservation_JJ}
	t\frac{\partial}{\partial t}\left(h_1(t,\lambda_0) + h_2(t,\lambda_0) \right)= (d-1)\times\left(2h_1(t,\lambda_0)+h_2(t,\lambda_0)\right),
\end{equation}
where $t\equiv mr$.

\paragraph{Conformal perturbation theory}
Analogously to \eqref{eq:to_compute_2} using conformal perturbation theory we can write
 \begin{multline}
	\label{eq:JJ_QFT}
	\langle J^\mu(x_1) J^\nu(x_2) \rangle_\text{QFT} = \langle J^\mu(x_1) J^\nu(x_2) \rangle_\text{UV CFT} \\- \lambda_0 m^{d-\Delta} \int d^dx_3
	\langle J^\mu(x_1) J^\nu(x_2) \cO(x_3)\rangle_\text{UV CFT} + O(\lambda_0^2).
\end{multline}
The index-free form of the UV CFT correlator entering the integral in \eqref{eq:JJ_QFT}  reads as
\begin{multline}
	\label{eq:CFT_JJO}
	\<J^\mu(x_1)J^\nu(x_2)\cO(x_3)\>_\text{UV CFT} =   \frac{C_{JJ\cO}}{\left(x_{12}^2\right)^{d-1-\frac{\Delta}{2}}\left(x_{13}^2\right)^{\frac{\Delta}{2}}\left(x_{23}^2\right)^{\frac{\Delta}{2}}}\times\\
\left(
 -\left( \delta^{\mu\nu} - 2\, \frac{x^\mu_{12} x^\nu_{12}}{x_{12}^2} \right)+\frac{\Delta}{d-1-\Delta}x_{12}^2
\left( \frac{x^\mu_{12}}{x_{12}^2}-\frac{x^\mu_{13}}{x_{13}^2}\right)
\left( \frac{x^\nu_{12}}{x_{12}^2}+\frac{x^\nu_{23}}{x_{23}^2}\right)
\right).
\end{multline}
Here $C_{JJ\cO}$ is the OPE coefficient of two currents and one scalar. Both \eqref{eq:CFT_JJ} and \eqref{eq:CFT_JJO} obey the conservation condition $\partial_{\mu} J^\mu(x)=0$. As a result the left-hand side of \eqref{eq:JJ_QFT} also obeys it.

In appendix \ref{app:integral_3pt} we obtain the following result
\begin{equation}
	\label{eq:definition_I}
\left(x_{12}^2\right)^{a_1+a_2-d/2}\int d^d x_3 \f{1}{\left(x_{13}^2\right)^{a_1}\left(x_{23}^2\right)^{a_2} } = P_2(a_1,a_2),
\end{equation}
where the numerical coefficient $P_2$ is given by \eqref{eq:P2}. Let us now plug \eqref{eq:CFT_JJ}, \eqref{eq:QFT_JJ} and \eqref{eq:CFT_JJO} into \eqref{eq:JJ_QFT}. Contracting the obtained expression with $\delta^{\mu\nu}$ and with $x_{12}^\mu x_{12}^\nu$ respectively, and using \eqref{eq:relation_dot_product} and \eqref{eq:definition_I} we obtain
\begin{multline}
	d \, h_1^\text{pert}(mr,\lambda_0)+h_2^\text{pert}(mr,\lambda_0) = (d-2)C_J\\ - \lambda_0 C_{JJ\cO} (m r)^{d-\Delta}\,
	\left(\frac{\Delta}{2(d-1-\Delta)}A - (d-2)P_2(\Delta/2,\Delta/2)\right) +O(\lambda_0^2),
\end{multline}
and
\begin{multline}
	h_1^\text{pert}(mr,\lambda_0)+h_2^\text{pert}(mr,\lambda_0) = -C_J\\ -\lambda_0 C_{JJ\cO} (m r)^{d-\Delta}\,
	\left(\frac{\Delta}{4(d-1-\Delta)}A - \frac{2-2d+\Delta}{2(d-1-\Delta)}P_2(\Delta/2,\Delta/2)\right) +O(\lambda_0^2),
\end{multline}
where the coefficient $A$ reads as
\begin{multline}
	 A\equiv P_2(\Delta/2-1,\Delta/2+1)+P_2(\Delta/2+1,\Delta/2-1)+P_2(\Delta/2+1,\Delta/2+1)\\
	-2P_2(\Delta/2+1,\Delta/2)-2P_2(\Delta/2,\Delta/2+1).
\end{multline}

Plugging here the explicit values of $P_2$ we get
\begin{equation}
	\label{eq:solution_JJ}
	\begin{aligned}
		h^\text{pert}_1(mr,\lambda_0)  &= C_J - \lambda_0 (m r)^{d-\Delta} (\Delta-1)E_{JJ}C_{JJ\cO}+O(\lambda_0^2),\\
		h^\text{pert}_2(mr,\lambda_0) &= -2C_J + \lambda_0 (m r)^{d-\Delta} (\Delta+d-2)E_{JJ}C_{JJ\cO}+O(\lambda_0^2),
	\end{aligned}
\end{equation}
where the coefficient $E_{JJ}$ is given by
\begin{equation}
	\label{eq:EJJ}
	E_{JJ} \equiv \frac{2^{\Delta-d}\pi^\frac{d+1}{2}\Delta}{(\Delta-d+1)} \frac{\Gamma(\Delta-d/2)}{\Gamma(\Delta/2+1)^2}
	\frac{\Gamma\left(\frac{d+2-\Delta}{2}\right)}{\Gamma\left(\frac{d+1-\Delta}{2}\right)}\, \overset{\delta\rightarrow 0}{=}\,  \frac{2\Omega_{d-1}}{d}+O(\delta).
\end{equation}
The solution \eqref{eq:solution_JJ} obeys the conservation condition \eqref{eq:conservation_JJ}.

\paragraph{Renormalized coupling}
In terms of the renormalized coupling $\lambda$ the decomposition \eqref{eq:QFT_JJ} can be written as
\begin{equation}
	\label{eq:QFT_JJ_renorm}
	\boxed{\< J^\mu(x_1) J^\nu(x_2) \>_\text{QFT} = \frac{1}{r^{2(d-1)}}\times\left(h_1(r\mu,\lambda)\delta^{\mu\nu}+h_2(r\mu,\lambda)\frac{x_{12}^\mu x_{12}^\nu}{r^2}\right),\ }
\end{equation}
where the perturbative expressions for the functions $h_i(r\mu,\lambda)$ are obtained from \eqref{eq:solution_JJ} using the leading part of the relation \eqref{eq:relation_lambda0_lambda}. We simply get
\begin{empheq}[box=\widefbox]{multline}
	\label{eq:solution_JJ_renorm}
	h^\text{pert}_1(r\mu,\lambda)  = C_J - \lambda\, (r\mu)^{d-\Delta} (\Delta-1)E_{JJ}C_{JJ\cO}+O(\lambda^2),\\
		h^\text{pert}_2(r\mu,\lambda) = -2C_J + \lambda\, (r\mu)^{d-\Delta} (\Delta+d-2)E_{JJ}C_{JJ\cO}+O(\lambda^2). \hspace{1.5cm}
\end{empheq}

\paragraph{Callan-Symanzik equation}
The two-point function of conserved currents does not depend on the renormalization scale $\mu$ as can be seen from \eqref{eq:QFT_JJ}. Thus, we can write
\begin{equation}
	\label{eq:CS}
	\frac{d}{d \mu} \langle J^\alpha(x_1) J^\beta(x_2) \rangle_\text{QFT} = 0.
\end{equation}
Plugging here the decomposition \eqref{eq:QFT_JJ_renorm}, we obtain a pair of the Callan-Symanzik equations
\begin{equation}
	\label{eq:CS_JJ_pair}
	s\equiv r\mu:\qquad
	\left( s\frac{\partial}{\partial s}+\beta(\lambda)\frac{\partial}{\partial\lambda}\right)  h_i(s,\lambda) = 0.
\end{equation}
The goal now is to solve the pair of equations \eqref{eq:CS_JJ_pair} together with the conservation equation \eqref{eq:conservation_JJ}, which can be simply rewritten as
\begin{equation}
	\label{eq:conservation_JJ_CS}
	s\frac{\partial}{\partial s}\left(h_1(s,\lambda) + h_2(s,\lambda)\right) = (d-1)\,\left(2h_1(s,\lambda)+h_2(s,\lambda)\right).
\end{equation}

Using the most general solution of the Callan-Symanzik equation for the scalar two-point function \eqref{eq:solution_2}, we can write
\begin{equation}
	\label{eq:solution_JJ_formal}
	h_i(s,\lambda) =   h_i^\text{boundary}\big(1,\Lambda(s,\lambda)\big).
\end{equation}
Notice, that this expression is so simple since conserved currents have zero anomalous dimension.
Using the perturbative result \eqref{eq:solution_JJ_renorm} as the boundary condition we conclude that
\begin{empheq}[box=\fbox]{multline}
	\label{eq:solution_JJ_explicit}
		\hspace{2cm}h_1(s,\lambda)  = C_J - (\Delta-1)E_{JJ}C_{JJ\cO}\times\Lambda(s,\lambda),\\
		h_2(s,\lambda) = -2C_J +  (\Delta+d-2)E_{JJ}C_{JJ\cO}\times\Lambda(s,\lambda).\hspace{3cm}
\end{empheq}
Plugging this solution into the conservation equation \eqref{eq:conservation_JJ_CS} we can see, however, that it is satisfied only up to the  order $O(\lambda)$. In the reminder of this section we carefully discuss the issue of conservation.

\paragraph{Series representation}
The result \eqref{eq:solution_JJ_explicit} can be compactly written as
\begin{equation}
	\label{eq:result_compact_h}
	h_i(s,\lambda)  = a_i + b_i\Lambda(s,\lambda),
\end{equation}
where the coefficients $a_i$ and $b_i$ read as
\begin{equation}
	a_2=-2a_1=-2C_J,\qquad
	b_1=- (\Delta-1)E_{JJ}C_{JJ\cO},\qquad
	b_2=(\Delta+d-2)E_{JJ}C_{JJ\cO}.
\end{equation}
Let us expand this expression around small values of $\lambda$ in the same way as was done in the end of section \ref{sec:Callan-Symanzik_2pt}.  The expansion of the ``running coupling''  \eqref{eq:running_coupling} reads as
\begin{equation}
	\label{eq:expansion_Lambda}
	\Lambda(s,\lambda) = \lambda s^\delta\times \sum_{n=0}^\infty (-1)^n\, \lambda^{n} q(s)^n = \lambda\times \sum_{n=0}^{\infty} c_n \lambda^n (\log s)^n+O(\delta),
\end{equation}
where $n$ denotes the order in conformal perturbation theory, the function $q(s)$ was defined in \eqref{eq:definition_q} and the coefficient $c_n$ reads as
\begin{equation}
c_n= (-1)^n \left(\f{C_{\cO\cO\cO}\Omega_{d-1}}{2}\right)^n.
\end{equation}

As we have already mentioned above, the solution \eqref{eq:result_compact_h} obeys the conservation condition \eqref{eq:conservation_JJ_CS} only up to $O(\lambda)$ order. We can better understand the source of the conservation breaking by plugging \eqref{eq:result_compact_h} together with \eqref{eq:expansion_Lambda} into the conservation condition \eqref{eq:conservation_JJ_CS}. The left-hand side (LHS) and right-hand side (RHS) of the conservation  condition \eqref{eq:conservation_JJ_CS} become
\begin{equation} \label{eq:conservation_condition_series}
	\begin{aligned}
		\text{LHS} &= \lambda (d-1) \f{2\Omega_{d-1}}{d} C_{JJ\cO}\sum_{n=1}^\infty n c_n \lambda^{n} (\log s)^{n-1} + O(\delta),\\
		\text{RHS} &=\delta\times  \lambda(d-1)\f{2\Omega_{d-1}}{d} C_{JJ\cO}\,\times \sum_{n=0}^\infty  c_n\lambda^{n} (\log s)^n +O(\delta^2).
	\end{aligned}
\end{equation}
The conservation condition \eqref{eq:conservation_JJ_CS} demands the equality of LHS and RHS above order by order in $n$. However, the equality of LHS and RHS in \eqref{eq:conservation_condition_series} is clearly violated, since at order $\lambda^{n+1}$ the RHS contains only the $\delta\times  \log^{n}$ terms and misses the $\delta^0\times \log^{n-1}$ terms present in the LHS.

\paragraph{Modified solution}
Let us, nevertheless, propose a procedure which could be used to obtain a modified version of the solution \eqref{eq:solution_JJ_explicit}, which obeys conservation at all orders in $\lambda$.

One could plug the solution \eqref{eq:solution_JJ_formal} without specifying the boundary conditions into the conservation condition \eqref{eq:conservation_JJ_CS}. The conservation condition then translates into the constraint on the boundary conditions.
We solve this constraint in appendix \ref{app:conservation_CS} using the expression \eqref{eq:solution_JJ_explicit} for the ``running coupling''. The result reads
\begin{align}
	\nn
	h_1(s,\lambda)  &= C_J - (\Delta-1)E_{JJ}C_{JJ\cO}\times\Lambda(s,\lambda),\\
	\label{eq:JJ_solution_CS_alternative}
	h_2(s,\lambda)  &= -2C_J +
	(\Delta-1)E_{JJ}C_{JJ\cO}\\ &\times\left((\Lambda(s,\lambda)+\lambda_\star)+(\Lambda(s,\lambda)-\lambda_\star)\,{}_2F_1\left(1,1,\frac{1-d+\delta}{\delta},\frac{\Lambda(s,\lambda)}{\lambda_\star}\right)\right).
	\nn
\end{align}
It is straightforward to check explicitly that this solution obeys the conservation equation \eqref{eq:conservation_JJ_CS}. One could interpret the solution \eqref{eq:JJ_solution_CS_alternative} as an alternative to \eqref{eq:solution_JJ_explicit} re-summation. The solutions \eqref{eq:solution_JJ_explicit} and \eqref{eq:JJ_solution_CS_alternative} match up to $O(\lambda)$ terms. 

Unfortunately, the expression \eqref{eq:JJ_solution_CS_alternative} is not well behaved along the whole RG flow, for instance $h_2(s,\lambda)$ diverges at $s=\infty$, where $\Lambda(\infty,s)=\lambda_\star$. We hope, however, that if we worked at one higher order in conformal perturbation theory, which corrects the first line of \eqref{eq:JJ_solution_CS_alternative} by providing access to $\lambda^2 \log^1$ terms as discussed above, we would have been able to obtain a modified version of \eqref{eq:JJ_solution_CS_alternative}, which is conserved and finite in the IR limit $s=\infty$.

Let us also make another comment on our result \eqref{eq:JJ_solution_CS_alternative}. Recall the series representation of the hypergoemtric function
\begin{equation}
	{}_2F_1(a,b,c,z) = \sum_{n=0}^\infty \frac{(a)_n (b)_n}{(c)_n} \frac{z^n}{n!}
\end{equation}
valid for any complex $z$ in the disk $|z|<1$. We could blindly use this representation inside the second line of \eqref{eq:JJ_solution_CS_alternative} up to (and including) the order $\left(\frac{\Lambda}{\lambda_\star}\right)^N$, where $N\geq 1$. 
Then we can explicitly check that in this order of approximation \eqref{eq:JJ_solution_CS_alternative} satisfy the conservation condition \eqref{eq:conservation_JJ_CS} up to the order $O(\lambda^{N+1})$. Such a truncated expression will be valid in the IR limit $s\rightarrow \infty$ and provide the Callan-Symanzik solution with the improved conservation properties.

\subsection{Current central charge $C_J$}
Let us evaluate the functions \eqref{eq:solution_JJ_explicit} in the IR. Taking into account that $\lambda_\star\propto\delta$ we have
\begin{equation}
	\begin{aligned}
		h_1(\infty,\lambda) &=C_J-(d-1)E_{JJ}C_{JJ\cO}\lambda_\star+O(\delta^2),\\
		h_2(\infty,\lambda) &=-2C_J+2(d-1)E_{JJ}C_{JJ\cO}\lambda_\star+O(\delta^2).
	\end{aligned}
\end{equation}
In the IR we must recover conformal invariance, thus we must have
\begin{equation}
	h_1(\infty,\lambda) = C_J^{IR},\qquad
	h_2(\infty,\lambda) = -2C_J^{IR}.
\end{equation}
Comparing the above equations we conclude that
\begin{equation}
	\Delta C_J\equiv C_J-C_J^{IR} = (d-1)E_{JJ}C_{JJ\cO}\lambda_\star +O(\delta).
\end{equation}
Using \eqref{eq:EJJ} and \eqref{eq:expansion_B} we finally get
\begin{equation}
\boxed{	\Delta C_J= \frac{4(d-1)}{d}\frac{C_{JJ\cO}}{C_{\cO\cO\cO}}\delta+O(\delta^2).\, }
\end{equation}

\subsection{Three-point function}
\label{sec:JJO}
Let us now consider the three-point function in \eqref{eq:correlators_JJO}.
We can construct tensor structures from the following ingredients only
\begin{equation}
	\delta^{\mu\nu},\quad
	\frac{x_{12}^\mu}{\rho},\quad
	\frac{x_{23}^\mu}{\rho},\quad 
	\frac{x_{13}^\mu}{\rho}.
\end{equation}
Using them we can write the following decomposition into tensor structures
\begin{multline}
	\label{eq:JJO_tensor_decomposition}
	\< J^\mu(x_1) J^\nu(x_2) \cO(x_3)\>_\text{QFT} = \frac{1}{\rho^{2(d-1)+\Delta}} \times
	\bigg(
	h_1(m\rho,y_1,y_2,\lambda_0) \delta^{\mu\nu} +
	h_2(m\rho,y_1,y_2,\lambda_0) \frac{x_{12}^\mu x_{12}^\nu}{\rho^2} \\
	+h_3(m\rho,y_1,y_2,\lambda_0) \frac{x_{12}^\mu x_{23}^\nu}{\rho^2} 
	+h_4(m\rho,y_1,y_2,\lambda_0) \frac{x_{13}^\mu x_{12}^\nu }{\rho^2} +
	h_5(m\rho,y_1,y_2,\lambda_0) \frac{x_{13}^\mu x_{23}^\nu}{\rho^2}
		\bigg).
\end{multline}
In writing these expressions we have used the fact that $x_{13}^\mu=x_{12}^\mu+x_{23}^\mu$, which eliminates some tensor structures.
The permutation symmetry of the two currents imply the following constraints
\begin{equation}
	i=1,2,5:\qquad h_i(m\rho,y_1,y_2,\lambda_0) = h_i\left(m\rho,y_1/y_2,1/y_2,\lambda_0\right),
	\end{equation}
together with
\begin{equation}
	h_3(m\rho,y_1,y_2,\lambda_0) = - h_4\left(m\rho,y_1/y_2,1/y_2,\lambda_0\right).
\end{equation}

At the trivial order of conformal perturbation theory we simply have
\begin{equation}
	\label{eq:JJO_QFT}
	\langle J^\mu(x_1) J^\nu(x_2) \cO(x_3)\rangle_\text{QFT} = \langle J^\mu(x_1) J^\nu(x_2) \cO(x_3)\rangle_\text{UV CFT} + O(\lambda_0).
\end{equation}
Using the explicit expression \eqref{eq:CFT_JJO} and comparing it with \eqref{eq:JJO_tensor_decomposition} we conclude that
\begin{equation}
	\label{eq:solution_h_JJ}
	\begin{aligned}
		h^\text{pert}_1(m\rho,y_1,y_2,\lambda_0) &= C_{JJ\cO} \left(\frac{y_1^2}{y_2}\right)^\frac{1-d+\Delta}{3} + (\rho m)^{d-\Delta} O(\lambda_0),\\
		h^\text{pert}_2(m\rho,y_1,y_2,\lambda_0) &= a\,C_{JJ\cO}\, \left(\frac{y_1^2}{y_2}\right)^\frac{-d+\Delta}{3} + (\rho m)^{d-\Delta} O(\lambda_0),\\
		h^\text{pert}_3(m\rho,y_1,y_2,\lambda_0) &= b\,C_{JJ\cO}\, y_1\left(\frac{y_1^2}{y_2}\right)^\frac{-d+\Delta}{3} + (\rho m)^{d-\Delta} O(\lambda_0),\\
		h^\text{pert}_4(m\rho,y_1,y_2,\lambda_0) &= -b\,C_{JJ\cO}\, \frac{y_1}{y_2}\left(\frac{y_1^2}{y_2}\right)^\frac{-d+\Delta}{3} + (\rho m)^{d-\Delta} O(\lambda_0),\\
		h^\text{pert}_5(m\rho,y_1,y_2,\lambda_0) &= -b\,C_{JJ\cO}\, \left(\frac{y_1^2}{y_2}\right)^{1+\frac{-d+\Delta}{3}} + (\rho m)^{d-\Delta} O(\lambda_0),
	\end{aligned}
\end{equation}
where we have defined
\begin{equation}
	a\equiv-2+b,\qquad
	b\equiv-\frac{\Delta}{d-1-\Delta}.
\end{equation}

\paragraph{Renormalized three-point function}
Let us consider now the renormalized three-point function
\begin{equation}
	\label{eq:JJO_QFT_in_terms_of _I_renormalized}
	\langle J^\mu(x_1) J^\nu(x_2) \mathbf{O}(x_3)\rangle_\text{QFT}.
\end{equation}
Analogously to \eqref{eq:JJO_tensor_decomposition} we can write the following decomposition
\begin{multline}
	\label{eq:JJO_tensor_decomposition_renorm}
	\< J^\mu(x_1) J^\nu(x_2) \mathbf{O}(x_3)\>_\text{QFT} = \frac{1}{\rho^{2(d-1)+\Delta}} \times
	\bigg(
	h_1(\mu\rho,y_1,y_2,\lambda) \delta^{\mu\nu} +
	h_2(\mu\rho,y_1,y_2,\lambda) \frac{x_{12}^\mu x_{12}^\nu}{\rho^2} \\
	+h_3(\mu\rho,y_1,y_2,\lambda) \frac{x_{12}^\mu x_{23}^\nu}{\rho^2} 
	+h_4(\mu\rho,y_1,y_2,\lambda) \frac{x_{13}^\mu x_{12}^\nu }{\rho^2} +
	h_5(\mu\rho,y_1,y_2,\lambda) \frac{x_{13}^\mu x_{23}^\nu}{\rho^2}
	\bigg).
\end{multline}

The renormalized three-point function at the trivial order of conformal perturbation theory simply reads
\begin{equation}
	\begin{aligned}
		\langle J^\mu(x_1) J^\nu(x_2) \mathbf{O}(x_3)\rangle_\text{QFT} &= Z(\mu)^{-\frac{1}{2}}
		\langle J^\mu(x_1) J^\nu(x_2) \cO(x_3)\rangle_\text{QFT}\\
		&= \langle J^\mu(x_1) J^\nu(x_2) \cO(x_3)\rangle_\text{UV CFT}  + O(\lambda_0).
	\end{aligned}
\end{equation}
Recall, that the relation between the bare coupling $\lambda_0$ and renormalized coupling $\lambda(\mu)$ is given by \eqref{eq:relation_lambda0_lambda}, and the expression for $Z(\mu)$ given in \eqref{eq:renormalization_Z}. 
From this one immediately concludes that $h^\text{pert}_i(\mu\rho,y_1,y_2,\lambda)=h^\text{pert}_i(m\rho,y_1,y_2,\lambda_0)$ at the trivial order of perturbation theory, where the latter functions are given in \eqref{eq:solution_h_JJ}.

\paragraph{Three-points}
The Callan-Symanzik equation for the renormalized correlator \eqref{eq:JJO_QFT_in_terms_of _I_renormalized} reads
\begin{equation}
	\left(\mu\frac{d}{d\mu}+\gamma(\lambda)\right)
	\langle J^\mu(x_1) J^\nu(x_2) \mathbf{O}(x_3)\rangle_\text{QFT} = 0.
\end{equation}
Plugging here the decomposition \eqref{eq:JJO_tensor_decomposition_renorm}, using the chain rule for the derivative and removing tensor structures we obtain the following five scalar Callan-Symanzik equations
\begin{equation}
	\left( \sigma\frac{\partial}{\partial \sigma}+\beta(\lambda)\frac{\partial}{\partial\lambda}+\gamma(\lambda)\right)  h_i(\sigma,y_1,y_2;\lambda) =0.
\end{equation}
The solution to these equations is given by
\begin{equation}
	\label{eq:solution_JJO_formal}
	h_i(\sigma,y_1,y_2;\lambda) =   h_i^\text{boundary}\big(1,y_1,y_2;\Lambda(s,\lambda)\big)\times
	\left(\frac{1}{1+\frac{\lambda}{\lambda_\star}\,(\sigma^\delta-1)}\right)^{2}.
\end{equation}
We can use \eqref{eq:solution_h_JJ} as the boundary conditions here. The solution \eqref{eq:solution_JJO_formal} combined with the decomposition \eqref{eq:JJO_tensor_decomposition_renorm} can be compactly written as
\begin{equation}
	\label{eq:JJO}
	\langle J^\mu(x_1) J^\nu(x_2)  \mathbf{O}(x_3) \rangle_\text{QFT} =
	\langle J^\mu(x_1) J^\nu(x_2)  O(x_3) \rangle_\text{UV CFT}
	\times \Big(\chi\left(\mu \rho,\lambda\right)\Big)^2,
\end{equation}
where the function $\chi(\mu\rho,\lambda)$ was defined in \eqref{eq:running_coupling}.

\section{Stress-tensor}
\label{sec:stress-tensor}

Let us now consider the QFT stress-tensor $T_\text{QFT}^{\mu\nu}(x)$. It can be decomposed into its trace given by equation  \eqref{eq:traceT_relation} and its CFT traceless symmetric part $ T^{\mu\nu}(x)$. We can write
\begin{equation}
	\label{eq:decomposition_stress-tensor}
	T_\text{QFT}^{\mu\nu}(x) = \lambda_0m^\delta \cO^{\mu\nu} (x)+T^{\mu\nu}(x),
\end{equation}
where $O^{\mu\nu}(x)$ is some local operator which obeys the condition $\cO^{\mu}_{\ \mu} (x)=\delta\times \cO(x)$.
In this section we will compute the following correlation functions in weakly relevant flows
\begin{equation}
	\langle T_\text{QFT}^{\mu\nu}(x_1) T_\text{QFT}^{\rho\sigma}(x_2) \rangle_\text{QFT},\qquad
	\langle T_\text{QFT}^{\mu\nu}(x_1) T_\text{QFT}^{\rho\sigma}(x_2) \cO(x_3)\rangle_\text{QFT}.
\end{equation}

\subsection{Two-point function}
The most general form of the stress-tensor two-point function in weakly relevant flows can be written as 
\begin{equation}
	\label{eq:decomposition_TT_QFT}
	\langle T_\text{QFT}^{\mu\nu}(x_1)T_\text{QFT}^{\rho\sigma}(x_2)\rangle_{\text{QFT}}= \frac{1}{\left(x_{12}^2\right)^d}\sum_{i=1}^5
	h_i(mr,\lambda_0) \mathbb{T}_i^{\mu\nu\rho\sigma}(x_{12}),
\end{equation}
where $h_i(mr,\lambda_0)$ are dimensionless functions and the basis of five tensor structures can be chosen as in \cite{Cardy:1988cwa}, namely
\begin{equation}
	\label{eq:tensor_structures}
	\begin{aligned}
		\mathbb{T}_1^{abcd}(x_{12})\equiv &\frac{x_{12}^a x_{12}^b x_{12}^c x_{12}^d}{r^4},\\
		\mathbb{T}_2^{abcd}(x_{12})\equiv &\frac{x_{12}^a x_{12}^b \delta^{cd}+x_{12}^c x_{12}^d \delta^{ab}}{r^2},\\
		\mathbb{T}_3^{abcd}(x_{12})\equiv&\frac{
			x_{12}^a x_{12}^c \delta^{bd}+
			x_{12}^b x_{12}^c \delta^{ad}+
			x_{12}^a x_{12}^d \delta^{bc}+
			x_{12}^b x_{12}^d \delta^{ac}}{r^2},\\
		\mathbb{T}_4^{abcd}(x_{12})\equiv&\delta^{ab}\delta^{cd}\\
		\mathbb{T}_5^{abcd}(x_{12})\equiv &
		\delta^{ac}\delta^{bd}+\delta^{bc}\delta^{ad}.
	\end{aligned}
\end{equation}

Let us split the two-point function using \eqref{eq:decomposition_stress-tensor} in the following way
\begin{multline}
	\langle T_\text{QFT}^{\mu\nu}(x_1)T_\text{QFT}^{\rho\sigma}(x_2)\rangle_{\text{QFT}} = 
	\langle T^{\mu\nu}(x_1)T^{\rho\sigma}(x_2)\rangle_{\text{QFT}} \\ +  \lambda_0m^\delta 
	\Big(\langle \cO^{\mu\nu}(x_1)T^{\rho\sigma}(x_2)\rangle_{\text{QFT}}+\langle T^{\mu\nu}(x_1)\cO^{\rho\sigma}(x_2)\rangle_{\text{QFT}}\Big) + O(\lambda_0^2).
\end{multline}
Using conformal perturbation theory, analogously to \eqref{eq:to_compute_2} we can write
\begin{multline}
	\label{eq:TT_CPT_intermediate}
	\langle T_\text{QFT}^{\mu\nu}(x_1)T_\text{QFT}^{\rho\sigma}(x_2)\rangle_{\text{QFT}}=
	\langle T^{\mu\nu}(x_1)T^{\rho\sigma}(x_2)\rangle_{\text{UV CFT}}\\
	-m^\delta\lambda_0\, \int d^{d}x_3 \langle T^{\mu\nu}(x_1)T^{\rho\sigma}(x_2)\cO(x_3)\rangle_{\text{UV CFT}} + O\big(\lambda_0^2\big).
\end{multline}
Here we used the fact that CFT two-point functions of two different operators vanish, in particular
\begin{equation}
	\langle \cO^{\mu\nu}(x_1)T^{\rho\sigma}(x_2)\rangle_{\text{UV CFT}} = 0.
\end{equation}
Since at the order $O(\lambda_0)$ the two-point function \eqref{eq:TT_CPT_intermediate} is traceless in both $(\mu\nu)$ and $(\rho\sigma)$ pairs of indices, there are two relations between the 5 functions $h_i(mr,\lambda_0)$ at the order $O(\lambda_0)$. They read as
\begin{equation}
	\label{eq:relations_tracelessness}
	\begin{aligned}
		h_3(mr,\lambda_0) &= -\frac{1}{4}\,\left(h_1(mr,\lambda_0)+d\,h_2(mr,\lambda_0)\right)+ O\big(\lambda_0^2\big),\\
		h_5(mr,\lambda_0) &= -\frac{1}{2}\,\left(h_2(mr,\lambda_0)+d\,h_4(mr,\lambda_0)\right)+ O\big(\lambda_0^2\big).
	\end{aligned}
\end{equation}

\paragraph{Conformal perturbation theory}
Comparing the decomposition \eqref{eq:decomposition_TT_QFT} and the stress-tensor two-point function \eqref{eq:TT_CPT_intermediate} computed in perturbation theory we can obtain the five functions $h_i(mr,\lambda_0)$. In practice this can be done by contracting both equations with the five tensor structures \eqref{eq:tensor_structures}. We will have five scalar equations which relate the functions $h_i(mr,\lambda_0)$ with the expressions given in terms of the integrals  \eqref{eq:integral_3pt_subsection}. In order to express everything in terms of the integrals of the type \eqref{eq:integral_3pt_subsection} we use relations of the form
\be 
(x_{13}\cdot x_{23})^2 &=\f{1}{4} (x_{13}^2+x_{23}^2-x_{12}^2)^2\nn\\
&= \f{1}{4}\big( x_{13}^4+x_{23}^4 +x_{12}^4 +2x_{13}^2 x_{23}^2 -2x_{13}^2 x_{12}^2 -2x_{23}^2 x_{12}^2\big).
\ee 
All these steps are completely analogous to the ones of section \ref{sec:JJ_two-point_functions}.
We are skipping all the details here and simply quote the final result. It reads
\begin{equation}
	\label{eq:functions_h_CPT}
	\begin{aligned}
		h^\text{pert}_1(mr,\lambda_0) &= 4 C_T^{UV}-\lambda_0 C_{TT\cO}^{UV} E_{TT} (d-2)(d+\Delta)(2+d+\Delta)\,(m r)^{d-\Delta}+ O\big(\lambda_0^2\big),\\
		h^\text{pert}_2(mr,\lambda_0) &= \lambda_0 C_{TT\cO}^{UV} E_{TT} (d-\Delta)(d+\Delta)\,(m r)^{d-\Delta}+ O\big(\lambda_0^2\big),\\
		h^\text{pert}_4(mr,\lambda_0) &= -\frac{1}{d}\, C_T^{UV}-\lambda_0 C_{TT\cO}^{UV} E_{TT} (2+d-\Delta^2)\,(m r)^{d-\Delta}+ O\big(\lambda_0^2\big),
	\end{aligned}
\end{equation}
where $E_{TT}$ is a numerical function given by
\begin{equation}
	\label{eq:coefficient_E}
	E_{TT}\equiv \frac{B(\delta)(2+d-\Delta)}{(d-2)(2d-\Delta)(2+2d-\Delta)\Delta(2+\Delta)}.
\end{equation}
Recall, the coefficient $B(\delta)$ was defined in \eqref{eq:definition_B}. This  result agrees with the unpublished notes of Guilherme Pimentel \cite{Pimentel:2015} communicated  to us by Hugh Osborn. The solution \eqref{eq:functions_h_CPT} aslo satisfies the relations (2.103) in \cite{Karateev:2019ymz} obtained by imposing conservation of the stress-tensor $\partial_\mu T^{\mu\nu}(x) = 0$.

\paragraph{Renormalized coupling}
Conserved operators like the stress-tensor do not renormalize. We can, however, rewrite the result \eqref{eq:functions_h_CPT} in terms of the renormalized coupling $\lambda$. By using the relation between $\lambda_0$ and $\lambda$ given in \eqref{eq:relation_lambda0_lambda} we obtain
\begin{empheq}[box=\fbox]{multline}
	\label{eq:functions_h_CPT_renormalized}
		h^\text{pert}_1(\mu r,\lambda) = 4 C_T^{UV}-\lambda\, C_{TT\cO}^{UV} E_{TT} (d-2)(d+\Delta)(2+d+\Delta)\,(\mu r)^{d-\Delta}+ O\big(\lambda^2\big),\\
		h^\text{pert}_2(\mu r,\lambda) = \lambda\, C_{TT\cO}^{UV} E_{TT}(d-\Delta)(d+\Delta)\,(\mu r)^{d-\Delta}+ O\big(\lambda^2\big),\hspace{3.6cm}\\
		h^\text{pert}_4(\mu r,\lambda) = -\frac{1}{d}\, C_T^{UV}-\lambda\, C_{TT\cO}^{UV} E_{TT}(2+d-\Delta^2)\,(\mu r)^{d-\Delta}+ O\big(\lambda^2\big).\hspace{2.1cm}
\end{empheq}

\paragraph{Callan–Symanzik equation}
Analogously to section \ref{sec:JJ_two-point_functions} we can write the Callan–Symanzik equation for the stress-tensor two-point function. Using its decomposition into tensor structures we can then rewrite it as five scalar equations
\begin{equation}
	\left( s\f{\p}{\p s}+\beta(\lambda)\f{\p}{\p\lambda}\right) h_i(s,\lambda) = 0,\qquad 
	i=1,\ldots,5,
\end{equation}
where $s\equiv \mu r$ as before. As explained in section \ref{sec:Callan-Symanzik}, the solution to these is given by
\begin{equation}
	h_i(s,\lambda) = h^\text{boundary}_i(s=1, \Lambda(s,\lambda)),
\end{equation}
where $ \Lambda(s,\lambda)$ is the running coupling given by \eqref{eq:lambda-bar_sol}. We use our perturbative result \eqref{eq:functions_h_CPT_renormalized} as the boundary condition. Writing out the result explicitly we get
\begin{equation}
	\label{eq:functions_h_CPT_2}
	\boxed{
	\begin{aligned}
		h_1(s,\lambda) &= 4 C_T^{UV}-C_{TT\cO}^{UV} E_{TT} (d-2)(d+\Delta)(2+d+\Delta)\times
		\Lambda(s,\lambda),\\
		h_2(s,\lambda) &= C_{TT\cO}^{UV} E_{TT}(d-\Delta)(d+\Delta)\times
		\Lambda(s,\lambda),\hspace{5.5cm}\\
		h_4(s,\lambda) &= -\frac{1}{d}\, C_T^{UV}-C_{TT\cO}^{UV} E_{TT}(2+d-\Delta^2)\times
		\Lambda(s,\lambda).
	\end{aligned}
}
\end{equation}

Analogous comments about conservation of the conserved currents made in section \ref{sec:JJ_two-point_functions} are applicable here.

\subsection{Stress-tensor central charge $C_T$}

The final solution for the two-point stress-tensor correlator is given by equations \eqref{eq:decomposition_TT_QFT}, \eqref{eq:tensor_structures} and \eqref{eq:functions_h_CPT_2}. Let us now consider the following contraction of this correlator
\begin{equation}
	\delta_{\mu\rho}\delta_{\nu\sigma}\langle T_{\text{QFT}}^{\mu\nu}(x_1)T_{\text{QFT}}^{\rho\sigma}(x_2)\rangle_{\text{QFT}}=
	\frac{1-d}{2r^{2d}}\left(h_1(r)+(d+2)(2h_2(r)+d\,h_4(r))\right).
\end{equation}
Here we have used the relations \eqref{eq:relations_tracelessness}. Plugging here the solution \eqref{eq:functions_h_CPT_2} we get
\begin{multline}
	\label{eq:contraction_TT_deltas}
	\delta_{\mu\rho}\delta_{\nu\sigma}\langle T_{\text{QFT}}^{\mu\nu}(x_1)T_{\text{QFT}}^{\rho\sigma}(x_2)\rangle_{\text{QFT}}=\frac{d-1}{2r^{2d}}
	\Big(
	(d-2)C_T^{UV} +\\
	C_{TT\cO}^{UV} E \left(4d^2(d+2)+2(d-2)(d+1)\Delta-(6+d(3+d))\Delta^2\right)
	\Big)	\times
	\Lambda(s,\lambda).
\end{multline}
Let us now take the IR limit of the above expression $r\rightarrow \infty$. The running coupling $\Lambda(s,\lambda)$ in this limit simply reduces to $\lambda_\star$. In the IR limit we can compare \eqref{eq:contraction_TT_deltas} with the IR CFT expression \eqref{eq:contraction_2pt}. We conclude then that
\begin{equation}
	\label{eq:difference_central_charges}
	C_T^{UV}-C_T^{IR} =
	-	\frac{\lambda_\star C_{TT\cO} E_{TT}}{d-2} \left(4d^2(d+2)+2(d-2)(d+1)\Delta-(6+3d+d^2)\Delta^2\right).
\end{equation}
Using the explicit values of $\lambda_\star$ and $E_{TT}$ given in equations \eqref{eq:lambda_star} and \eqref{eq:coefficient_E} and expanding in small $\delta$ we get the following simple answer
\begin{equation}
	\label{eq:CT_difference}
	C_T^{UV}-C_T^{IR}= \frac{8(d+1)}{d(d+2)^2}\f{C_{TT\mathcal{O}}}{C_{\mathcal{O}\mathcal{O}\mathcal{O}}}\delta \ +\ O(\delta^2).
\end{equation}
In particular focusing on $d=4$ we get 
\be
\label{eq:CPT_result}
\boxed{d=4:\qquad
	\Delta c=\f{\pi^2}{640}\big(C_T^{UV}-C_T^{IR}\big)=\f{\pi^2}{2304} \f{C_{TT\mathcal{O}}}{C_{\mathcal{O}\mathcal{O}\mathcal{O}}}\delta \ +\ O(\delta^2).
}
\ee
For an interesting discussion of $\Delta c$ in 4d see \cite{Cappelli:1990yc,Karateev:2020axc}.

\subsection{Three-point function}
Following the same logical steps as in section \ref{sec:JJO} it is straightforward to obtain the following expression
\begin{equation}
	\langle T_\text{QFT}^{\mu\nu}(x_1) T_\text{QFT}^{\rho\sigma}(x_2) \mathbf{O}(x_3)\rangle_\text{QFT} =\\
	\langle T^{\mu\nu}(x_1) T^{\rho\sigma}(x_2)\cO(x_3)\rangle_\text{UV CFT}\times \Big(\chi\left(\mu \rho,\lambda\right)\Big)^2,
\end{equation}
where the function $\chi(\mu\rho,\lambda)$ was defined in \eqref{eq:running_coupling}.

\section{Background field method and trace anomalies in 4d}
\label{sec:background_field_method}

In this section we will couple weakly relevant flows \eqref{eq:theory_definition} to non-dynamical background dilaton and graviton fields in 4d as was done in \cite{Komargodski:2011vj}, see also \cite{Karateev:2023mrb} for the recent discussion. In subsection \ref{sec:effective_action} we will compute the IR effective action of the background fields in terms of the correlation functions along the RG flow. In subsection \ref{sec:ddd_vertex} we will derive the three-dilaton effective vertex and extract the value of $\Delta a$. In subsection \ref{sec:HHD_vertex} we derive the graviton-graviton-dilaton vertex and extract the value of $\Delta c$. The values of $\Delta a $ and $\Delta c$ are extracted by comparing the vertices obtained here with the general predictions found in \cite{Karateev:2023mrb}.

\subsection{Effective action for the background fields}
\label{sec:effective_action}
We conformally couple our theory \eqref{eq:theory_definition} to a non-dynamical curved background with metric $g_{\mu\nu}(x)$. We also make the following replacement
\begin{equation}
	 \label{eq:compensation_UV}
	m \longrightarrow m\, \Omega(x),
\end{equation}
where $\Omega(x)$ is called the compensator field. It is introduced in order to restore Weyl invariance classically.\footnote{In this section we could have also worked with the renormalized action \eqref{eq:theory_alternative}. Because of \eqref{eq:lambda(mu)} at the level of renormalized quantities the replacement \eqref{eq:compensation_UV} translates into \cite{Goldberger:2007zk,DiVecchia:2017uqn}
\begin{equation*}
	\lambda(\mu) \longrightarrow \lambda\left(\Omega(x)^{-1} \mu\right).
\end{equation*}
}

We define the graviton field $h_{\mu\nu}(x)$ as a small fluctuation around the flat background, namely 
\begin{equation}
	\label{eq:metric}
	g_{\mu\nu}(x)=\delta_{\mu\nu}+2\kappa\, h_{\mu\nu}(x).
\end{equation}
The parameter $\kappa$ has mass dimension $-1$. It will be used as an expansion parameter. In such expansion it simply counts the total number of the graviton fields. We will work with traceless and transverse gravitons only, namely
\begin{equation}
	\eta^{\mu\nu}h_{\mu\nu}(x)=0,\qquad
	\p_\mu h^{\mu\nu}(x)=0. \label{eq:traceless_transverse_condition}
\end{equation}
The dilaton field $\varphi(x)$ enters through the compensator field $\Omega(x)$ via the following relation
\begin{equation}
	\label{eq:compensator}
	\Omega(x) = e^{-\tau(x)}= 1 - \frac{1}{\sqrt{2}f}\,\varphi(x).
\end{equation}
Here $f$ is a non-negative coupling constant of mass dimension $+1$ and $f^{-1}$ will be used as an expansion parameter which counts number of dilaton fields.

Let us denote the action describing the weakly relevant flow coupled to the compensator field $\Omega(x)$ and the background metric $g_{\mu\nu}(x)$  by $A'$. Looking at \eqref{eq:theory_definition} we conclude that
\begin{equation}
\label{eq:QFT_action_in_background}
	A'\left[g,\,\Omega\right] =A_{\text{UV CFT}}[g] + \lambda_0 m^\delta\int d^{d}x\sqrt{g}\  \Omega(x)^\delta \cO_{g}(x).
\end{equation}
The UV CFT on the curved background can be expanded as\footnote{Here we used the following definition of the stress-tensor
	\begin{equation*}
		T^{\mu\nu}(x)\equiv \f{2}{\sqrt{g}}\f{\delta A[g]}{\delta g_{\mu\nu}(x)}.
	\end{equation*}	
	From this it follows that
	\begin{equation*}
		\delta_{g}A[g]=\f{1}{2}\int d^dx\  \sqrt{g}\ \delta g_{\mu\nu}(x)T^{\mu\nu}(x).
	\end{equation*}	
}
\begin{equation}
	A_{\text{UV CFT}}[g]= A_{\text{UV CFT}}+\kappa\int d^dx\ h_{\mu\nu}(x)T^{\mu\nu}(x)+O(\kappa^2).
\end{equation}
Here $T^{\mu\nu}(x)$ is the stress-tensor of the UV CFT in flat space. Here $\sqrt{g}$ is the standard measure in the curved background. We also use the subscript $g$ for the scalar operator in order to emphasize that in the curved background it might depend on the metric and while expanding in power of $\kappa$, it matches with the flat space expression i.e. $\sqrt{g}\ \cO_{g}(x)=\cO(x)+O(\kappa^1)$. Combining these facts and expanding in power of $\kappa$ and $f^{-1}$, from \eqref{eq:QFT_action_in_background} we get
\begin{eqnarray}
	A'\left[g,\,\Omega\right] &=& A+\kappa\int d^dx\ h_{\mu\nu}(x)T^{\mu\nu}(x) -\f{\delta \lambda_0 m^\delta}{\sqrt{2}f} \int d^{d}x\,   \varphi(x)\cO(x)\nn\\
	&&-\f{\delta(1-\delta) \lambda_0 m^\delta}{4f^2} \int d^{d}x\,   \varphi(x)^2\cO(x)
	+O\left(\kappa^2, \kappa f^{-1}, f^{-3}\right), \label{eq:action_CPT}
\end{eqnarray}
where the flat space action for weakly relevant flow $A$ is given in \eqref{eq:theory_definition}.

Let us now take the action \eqref{eq:action_CPT} and integrate out all the high energy degrees of freedom along the short RG flow. This procedure will leaves us with an IR action which consists of the IR CFT in curved background, the IR EFT of background fields and possible interaction of dilaton with the IR CFT operators in curved background, as discussed in \cite{Karateev:2023mrb} for a generic RG flow. Here we only focus on deriving the IR EFT for background fields. The partition function of the theory \eqref{eq:action_CPT} reads as
\begin{equation}
	\label{eq:partition_function}
	Z[\varphi,\, h] \equiv \int [d\Phi]\ e^{-A'\left[g,\Omega\right]}.
\end{equation}
The effective action of the dilaton and graviton fields is defined then as
\begin{equation}
	\label{eq:effective_action}
	A_\text{EFT}[\varphi,\, h] \equiv - \log Z[\varphi,\, h] = - \log \int [d\Phi]\ e^{-A'\left[g,\Omega\right]}.
\end{equation}
It can be computed by plugging the equations \eqref{eq:action_CPT} and expanding in power of background fields.
By doing so, we obtain the following expression
\begin{equation}
	\label{eq:EFT_WRD}
	A_\text{EFT}[\varphi, h] = -\log\left(Z_{\text{QFT}}\right) - \log\Big(1+Y_{\cO\cO}+Y_{TT\cO}+\ldots\Big).
\end{equation}
Here $Y_{\cO\cO}$ and $Y_{TT\cO}$ denote the contribution of the two-point function $\<\cO\cO\>$ and the three-point function $\<TT\cO\>$. The ellipses represents the parts of the EFT action which do not contribute the three dilaton and graviton-graviton-dilaton vertices at the required order we would like to compute in subsection \ref{sec:ddd_vertex}  and subsection \ref{sec:HHD_vertex}. Below we provide the explicit expressions for the parts of $Y_{\cO\cO}$ and $Y_{TT\cO}$ which can contribute to  three dilaton and graviton-graviton-dilaton vertices (at leading order in small $\delta$) respectively
\begin{equation}
	\label{eq:YOO}
    Y_{\cO\cO} =
	\f{\left(\delta \lambda_0 m^{\delta}\right)^2}{4\sqrt{2}f^3}  \int d^dx_1\int d^dx_2\, \varphi(x_1)\varphi(x_2)^2\, 
	\langle  \mathcal{O}(x_1)\mathcal{O}(x_2)\rangle_{\text{QFT}}+\cdots,
\end{equation}
together with
\begin{multline}
	\label{eq:YTTO}
	Y_{TT\cO}=\f{\kappa^2}{2\sqrt{2}f} \delta\lambda_0 m^{\delta}\int d^dx_1  \int d^dx_2 \int d^dx_3
	\ h_{\mu\nu}(x_1)h_{\rho\sigma}(x_2) \varphi(x_3)\\
	\times \langle T^{\mu\nu}(x_1)T^{\rho\sigma}(x_2) \mathcal{O}(x_3)\rangle_{\text{QFT}}+\cdots.
\end{multline}

\subsection{Dilaton three-point vertex and the computation of $\Delta a$ in 4d} 
\label{sec:ddd_vertex}

The three-dilaton vertex is defined as
\begin{equation}
	\label{eq:vertex_ddd}
	(2\pi)^4 \delta^{(4)}(k_1+k_2+k_3) \times V_{(\varphi\varphi\varphi)}(k_1,k_2,k_3)
	\equiv (2\pi)^{12}\times\frac{i\ \delta^ 3A^L_\text{EFT}}{\delta \varphi(k_1)\delta \varphi(k_2)\delta \varphi(k_3)}\Bigg{|}_{h, \varphi=0},
\end{equation}
where $A^L_\text{EFT}$ is the momentum space EFT action in Lorentzian $d$ dimensional spacetime. Starting from the euclidean EFT action \eqref{eq:EFT_WRD} one needs to follow the following procedure to find the position space Lorenzian EFT action
\begin{equation}
\label{foot:Euclideanvs_Lorentzian}
	A^L_{\text{EFT}}=iA_{\text{EFT}}\Big{|}_{t_E=it_L},
\end{equation}
where $t_E$ denotes the Euclidean time and $t_L$ denotes the Lorentzian time. We need to perform this as, in \cite{Karateev:2023mrb}, we utilized the Lorentzian metric when deriving the vertices, which we are comparing to our expressions here.

Using the IR effective action \eqref{eq:EFT_WRD} we conclude that at the leading order in $\delta$ the relevant part of the effective action \eqref{eq:EFT_WRD} for this computation is simply given by 
\begin{equation}
	\label{eq:EFT_delA_computation}
	A_\text{EFT}[\varphi, h] = -Y_{\cO\cO}+\cdots
\end{equation}
Using the expression \eqref{eq:YOO} and keeping only terms with three dilatons we obtain the following result
\begin{multline}
	\label{eq:OO_QFT_section5_alternative}
	A_\text{EFT}[\varphi, h] = -\f{1}{4\sqrt{2}f^3}\,
	\delta^2 \left(m^{\delta}\lambda_0\right)^2  \int d^dx_1\int d^dx_2\ 
	\varphi(x_2)^2\varphi(x_1) \\
	\times    \langle\mathcal{O}(x_{12})\mathcal{O}(0)\rangle_{\text{QFT}} +\ \cdots
\end{multline}

Let us now perform the change of variables
\begin{equation}
	x^\mu \equiv x_2^\mu,\qquad
	y^\mu \equiv x_{12}^\mu.
\end{equation}
We can then Taylor expand the field $\varphi(x_1)=\varphi(x+y)$ around $x$ up to four powers in $y$. We will keep only terms at four derivative order which, as we will see shortly, are enough for the purposes of this section.
The relevant part of the action \eqref{eq:OO_QFT_section5_alternative} then becomes
\begin{equation}
	\label{eq:EFT_1}
	A_\text{EFT}[\varphi, h] = -\f{1}{4\sqrt{2}f^3} \ \int d^dx\ \varphi(x)^2\p_\mu\p_\nu\p_\rho\p_\sigma\varphi(x)\, L^{\mu\nu\rho\sigma}(\delta)+\cdots, 
\end{equation}
where we have defined
\begin{equation}
	\label{eq:integral_L1}
	L^{\mu\nu\rho\sigma}(\delta)\equiv \f{1}{24}
	\delta^2 \left(m^{\delta}\lambda_0\right)^2
	\int d^dy\  y^\mu y^\nu y^\rho y^\sigma\,
	\langle  \mathcal{O}(y)\mathcal{O}(0)\rangle_{\text{QFT}}.
\end{equation}
Using Lorentz covariance, the integral \eqref{eq:integral_L1} can be decomposed into tensor structures as follows
\begin{equation}
	\label{eq:decomposition_L1}
	L^{\mu\nu\rho\sigma}(\delta) =
	\left(\delta^{\mu\nu}\delta^{\rho\sigma}+\delta^{\mu\rho}\delta^{\nu\sigma}+\delta^{\mu\sigma}\delta^{\nu\rho}\right)\times L_1(\delta),
\end{equation}
where the scalar part reads as
\begin{equation}
	\label{eq:integral_L1_scalar}
	L(\delta)\equiv \f{\delta^2 \left(m^{\delta}\lambda_0\right)^2}{24d(d+2)}
	\int d^dy\ (y^2)^2 \,\langle  \mathcal{O}(y)\mathcal{O}(0)\rangle_{\text{QFT}}.
\end{equation}
Plugging \eqref{eq:decomposition_L1} together with \eqref{eq:integral_L1_scalar} into \eqref{eq:EFT_1} we get
\begin{equation}
	\label{eq:EFT_2pt}
	A_\text{EFT}[\varphi, h] = -\f{3L(\delta)}{4\sqrt{2}f^3} \, \int d^dx\ \varphi(x)^2\p^4\varphi(x)+\cdots .
\end{equation}
Using the prescription \eqref{foot:Euclideanvs_Lorentzian} in the above action, the Lorentzian action turns out
\begin{equation}
	\label{eq:EFT_2pt_L}
	A^L_\text{EFT}[\varphi, h] = \f{3L(\delta)}{4\sqrt{2}f^3} \, \int d^dx\ \varphi(x)^2\p^4\varphi(x)+\cdots .
\end{equation}

Fourier transforming \eqref{eq:EFT_2pt_L} and using the definition of the three-dilaton vertex \eqref{eq:vertex_ddd} we obtain the three-dilaton vertex
\begin{equation}
	\label{eq:vertexDDD}
	V_{(\varphi\varphi\varphi)}(k_1,k_2,k_3) =  \f{3iL(\delta)}{2\sqrt{2}f^3} \times \left((k_1^2)^2 +(k_2^2)^2+(k_3^2)^2\right)+\ldots.
\end{equation}

\paragraph{Integral evaluation in 4d}
Let us now evaluate the integral $L(\delta)$ defined in \eqref{eq:integral_L1_scalar}. Using the explicit expression of the scalar two-point function \eqref{eq:bare_OO} we get
\begin{equation}
	L(\delta)= \f{\delta^2 \left(m^{\delta}\lambda_0\right)^2}{24d(d+2)} \Omega_{d-1} \int_0^\infty dr r^{d+3}
	\frac{1 - \frac{4\lambda_0}{\lambda_\star}}{r^{2\Delta}}
	\left(\frac{\lambda_\star}{\lambda_\star+\lambda_0\,\left((rm)^\delta-1\right)}\right)^4.
\end{equation}
Focusing on $d=4$ dimensions we can trivially rewrite this integral as as
\begin{equation}
	L(\delta)= \f{\delta^2 \pi^2\left(m^{\delta}\lambda_0\right)^2}{288} \left(1 - \frac{4\lambda_0}{\lambda_\star}\right)
	\int_0^\infty dr 
	\frac{r^{2\delta-1}}{\left(1+\frac{\lambda_0}{\lambda_\star}\,\left((rm)^\delta-1\right)\right)^4}.
\end{equation}
Performing the integral we obtain
\begin{equation}
	L(\delta)= \f{\delta \pi^2\lambda_\star^2}{1728}
	\frac{1 - \frac{4\lambda_0}{\lambda_\star}}{\left(1-\frac{\lambda_0}{\lambda_\star}\right)^2}.
\end{equation}
Taking into account that $\lambda_0\ll1$ we get
\begin{equation}
	\label{eq:result_L}
	L(\delta)= \f{\delta \pi^2\lambda_\star^2}{1728}\left(1+O\left(\lambda_0\right)\right)=
	\f{\delta^3}{1728\pi^2C^2_{\cO\cO\cO}}\left(1+O\left(\lambda_0\right)\right).
\end{equation}
In the second equality we have used \eqref{eq:lambda_star_final}.

\paragraph{Computation of $\Delta a$}
In \cite{Karateev:2023mrb} we found that the three-point vertex of dilatons has the following form
\begin{multline}
	\label{eq:V3phi}
	V_{(\varphi\varphi\varphi)}(k_1,k_2,k_3)
	=  \frac{i\sqrt{2}}{f^3}\bigg(
	\Delta a\left(\left(k_1^2\right)^2+\left(k_2^2\right)^2+\left(k_3^2\right)^2\right)\\
	+2(18 r_1-\Delta a)\left(k_1^2 k_2^2+k_2^2 k_3^2 +k_3^2 k_1^2\right)+\ldots\bigg).
\end{multline}
Plugging \eqref{eq:result_L} into \eqref{eq:vertexDDD} and comparing the result with \eqref{eq:V3phi}, we conclude that
\begin{equation}
	\label{eq:result_delA_option_1}
	\boxed{\Delta a = \frac{3}{4} L(\delta)= \frac{\delta ^3}{2304 \pi ^2 C_{\cO\cO\cO}^2},\qquad r_1=\frac{\Delta a}{18}.\ }
\end{equation}

\subsection{Graviton-graviton-dilaton vertex and the computation of $\Delta c$ in 4d} 
\label{sec:HHD_vertex}
The graviton-graviton-dilaton vertex is defined as
\begin{multline}
	\label{eq:vertex_hhd}
	(2\pi)^4 \delta^{(4)}(k_1+k_2+k_3) \times V_{(h h\varphi)}(k_1,k_2,k_3;\ve_1,\ve_2)\\
	\equiv
	(2\pi)^{12}\times
	 \varepsilon_{\mu_1\nu_1}(k_1)\varepsilon_{\mu_2\nu_2}(k_2) \times\frac{i\delta^3 A^L_\text{EFT}}{\delta  h_{\mu_1\nu_1}(k_1)\delta  h_{\mu_2\nu_2}(k_2)\delta \varphi(k_3)}\Bigg{|}_{h,\varphi=0},
\end{multline}
where $A^L_\text{EFT}$ is the momentum space EFT action in Lorentzian $d$ dimensional spacetime following \eqref{foot:Euclideanvs_Lorentzian}.

The relevant part of the IR effective action describing the background fields for the computation of the vertex \eqref{eq:vertex_hhd} is given by
\begin{equation}
	\label{eq:EFT_relevant_part_deltaC}
	A_\text{EFT}[\varphi,\, h] = -Y_{TT\cO} + \ldots,
\end{equation}
where $Y_{TT\cO} $ is given in terms of the three-point function $\<TT\cO\>$ in \eqref{eq:YTTO}. Expanding this expression at large $f$, recall \eqref{eq:compensator}, and keeping only the term with a single dilaton we get
\begin{multline}
	\label{eq:EFT_action_CPT_3}
	A_\text{EFT}[\varphi,\, h] =- \f{\delta\kappa^2}{2\sqrt{2}f}\lambda_0m^\delta
	\int d^dx_1\int d^dx_2 \int d^dx_3\ \ h_{\mu\nu}(x_1)h_{\rho\sigma}(x_2) \varphi(x_3)\\
	\times \langle T^{\mu\nu}(x_1)T^{\rho\sigma}(x_2) \mathcal{O}(x_3)\rangle_{\text{QFT}}+\dots,
\end{multline}
Let us bring now this expression into a more convenient form. Due to translational invariance the following relation holds
\begin{equation}
	\cO (x_3) = e^{+P\cdot x_3} \cO (0) e^{-P\cdot x_3},
\end{equation}
where $P_\mu$ is the generator of translations. This allows to write
\begin{equation}
	\langle T^{\mu\nu}(x_1)T^{\rho\sigma}(x_2) \mathcal{O}(x_3)\rangle_{\text{QFT}} =
	\langle T^{\mu\nu}(x_{13})T^{\rho\sigma}(x_{23}) \mathcal{O}(0)\rangle_{\text{QFT}}. \label{eq:TTO_translation}
\end{equation}
Plugging this into \eqref{eq:EFT_action_CPT_3} and changing the integration variables according to
\begin{equation}
	\label{eq:coordinate_transformation}
	z_1^\mu \equiv x_{13}^\mu,\qquad
	z_2^\mu \equiv x_{23}^\mu,\qquad
	x^\mu \equiv x_3^\mu,
\end{equation}
we obtain the following expression
\begin{multline}
	\label{eq:EFT_action_CPT_4}
	A_\text{EFT}[\varphi,\, h] = - \f{\kappa^2}{2\sqrt{2}f}w
	\int d^dx\ \varphi(x)
	\int d^dz_1\int d^dz_2  \ h_{\mu\nu}(x+z_1)h_{\rho\sigma}(x+z_2)\\
	 \langle T^{\mu\nu}(z_1)T^{\rho\sigma}(z_2) \mathbf{O}(0)\rangle_{\text{QFT}}+\dots,
\end{multline}
where the coefficient $w$ is defined as
\begin{equation}
	\label{eq:def_w}
	w\equiv \delta\lambda_0m^\delta Z(\mu)^\frac{1}{2}.
\end{equation}
Above we have also used the definition of the renormalized scalar operator \eqref{eq:bold_O}.

The rest of this section is organised as follows. In \ref{sec:vertex_gravitons} we will compute the vertex \eqref{eq:vertex_hhd} using the effective action \eqref{eq:EFT_action_CPT_4} and the three-point function \eqref{eq:three-point_function}. From this vertex we will extract the value of $\Delta c$. There is a subtlety for doing this computation since in \eqref{eq:three-point_function} we do not know the object $H^{\mu\nu\rho\sigma}$. In order to bypass this issue we use the following trick. We set the object $H^{\mu\nu\rho\sigma}$ to zero. The effect of doing this propagates into the expression for the vertex, however, it remains contained in one of its three tensor structures. This tensor structure must be disregarded and the $\Delta c$ value must be extracted from the other two structures. We carefully justify this trick in subsection \ref{sec:trick}.

\subsubsection{Vertex and $\Delta c$ computation}
\label{sec:vertex_gravitons}
As we will see in the end of this section, in order to obtain $\Delta c$ we need to compute the graviton-graviton-dilaton vertex only at the four derivative order. We expand the effective action \eqref{eq:EFT_action_CPT_4} around $x$ for small values of $z_1$ and $z_2$ using the Taylor expansion of the graviton field
\begin{equation}
	h_{\mu\nu}(x+z_i) = \exp\Big(z_i\cdot \f{\p}{\p x}\Big)\ h_{\mu\nu}(x).
\end{equation}
We get then
\begin{equation}
	\label{eq:EFT_action_CPT_5}
	\begin{aligned}
		A_\text{EFT}[\varphi,\, h] = 
		&
		-\f{\kappa^2}{2\sqrt{2}f}\int d^dx\ \varphi(x) \Big(\\
		&\frac{1}{4}\left(\partial_{\alpha_1}\partial_{\alpha_2}h_{\mu\nu}(x)\right)
		\left(\partial_{\beta_1}\partial_{\beta_2}h_{\rho\sigma}(x)\right) g_1^{\alpha_1\alpha_2\beta_1\beta_2\mu\nu\rho\sigma}(\delta)\\
		+&\frac{1}{24}\left(\partial_{\alpha_1}\partial_{\alpha_2}\partial_{\beta_1}\partial_{\beta_2}h_{\mu\nu}(x)\right)
		h_{\rho\sigma}(x) g_2^{\alpha_1\alpha_2\beta_1\beta_2\mu\nu\rho\sigma}(\delta)\\
		+ &\frac{1}{24}h_{\mu\nu}(x)\left(\partial_{\alpha_1}\partial_{\alpha_2}\partial_{\beta_1}\partial_{\beta_2}h_{\rho\sigma}(x)\right) g_3^{\alpha_1\alpha_2\beta_1\beta_2\mu\nu\rho\sigma}(\delta)\\
		+ &\frac{1}{6}\left(\partial_{\alpha_1}h_{\mu\nu}(x)\right)\left(\partial_{\alpha_2}\partial_{\beta_1}\partial_{\beta_2}h_{\rho\sigma}(x)\right) g_4^{\alpha_1\alpha_2\beta_1\beta_2\mu\nu\rho\sigma}(\delta)\\
		+&\frac{1}{6}\left(\partial_{\alpha_1}\partial_{\alpha_2}\partial_{\beta_2} h_{\mu\nu}(x)\right)\left(\partial_{\beta_1}h_{\rho\sigma}(x)\right) g_5^{\alpha_1\alpha_2\beta_1\beta_2\mu\nu\rho\sigma}(\delta)\Big)+\cdots,
	\end{aligned}
\end{equation}
where we have defined
\begin{equation}
	\begin{aligned}
		g_1^{\alpha_1\alpha_2\beta_1\beta_2\mu\nu\rho\sigma}(\delta) &\equiv w
		\int d^dz_1\int d^dz_2  \left(z_1^{\alpha_1}z_1^{\alpha_2}z_2^{\beta_1}z_2^{\beta_2} 
		\langle T^{\mu\nu}(z_1)T^{\rho\sigma}(z_2) \mathbf{O}(0)\rangle_{\text{QFT}}\right),\\
		g_2^{\alpha_1\alpha_2\beta_1\beta_2\mu\nu\rho\sigma}(\delta) &\equiv w
		\int d^dz_1\int d^dz_2  \left(z_1^{\alpha_1}z_1^{\alpha_2}z_1^{\beta_1}z_1^{\beta_2} 
		\langle T^{\mu\nu}(z_1)T^{\rho\sigma}(z_2) \mathbf{O}(0)\rangle_{\text{QFT}}\right),\\
		g_3^{\alpha_1\alpha_2\beta_1\beta_2\mu\nu\rho\sigma}(\delta) &\equiv w
		\int d^dz_1\int d^dz_2  \left(z_2^{\alpha_1}z_2^{\alpha_2}z_2^{\beta_1}z_2^{\beta_2} 
		\langle T^{\mu\nu}(z_1)T^{\rho\sigma}(z_2) \mathbf{O}(0)\rangle_{\text{QFT}}\right),\\
		g_4^{\alpha_1\alpha_2\beta_1\beta_2\mu\nu\rho\sigma}(\delta) &\equiv w
		\int d^dz_1\int d^dz_2  \left(z_1^{\alpha_1}z_2^{\alpha_2}z_2^{\beta_1}z_2^{\beta_2} 
		\langle T^{\mu\nu}(z_1)T^{\rho\sigma}(z_2) \mathbf{O}(0)\rangle_{\text{QFT}}\right),\\
		g_5^{\alpha_1\alpha_2\beta_1\beta_2\mu\nu\rho\sigma}(\delta) &\equiv w
		\int d^dz_1\int d^dz_2  \left(z_1^{\alpha_1}z_1^{\alpha_2}z_2^{\beta_1}z_1^{\beta_2} 
		\langle T^{\mu\nu}(z_1)T^{\rho\sigma}(z_2) \mathbf{O}(0)\rangle_{\text{QFT}}\right).
	\end{aligned}
\end{equation}

From the definition of these integrals many symmetry properties among the indices follow. For example the function $g_1$ is symmetric under the independent exchanges $\alpha_1 \leftrightarrow \alpha_2$, $\beta_1 \leftrightarrow \beta_2$ and $\mu \leftrightarrow \nu$ indices. Also it is symmetric under the simultaneous exchange of the integration variables $z_1 \leftrightarrow z_2$ and the pairs of indices $(\mu\nu) \leftrightarrow (\rho\sigma)$. It is straightforward to find similar properties for all the other functions. Moreover, all the $g_i$ functions are traceless in both pairs of indices   $(\mu\nu)$ and $(\rho\sigma)$.

\paragraph{Decomposition into tensor structures}
Using Lorentz symmetry we can decompose the $g_i$ functions defined above into tensor structures as follows
\begin{equation}
	\label{eq:decomposition}
	g_i^{\alpha_1\alpha_2\beta_1\beta_2\mu\nu\rho\sigma}(\delta) =\sum_{I=1}^{N_i}g_{iI}(\delta) \, \mathbf{T}_I^{\alpha_1\alpha_2\beta_1\beta_2\mu\nu\rho\sigma},
\end{equation}
where $g_{iI}(\delta)$ are scalar functions and $\mathbf{T}$ are tensor structures built out of flat metric, namely
\begin{equation}
	\mathbf{T}_I^{\alpha_1\alpha_2\beta_1\beta_2\mu\nu\rho\sigma} = \delta^{\alpha_1\alpha_2}\delta^{\beta_1\beta_2}\delta^{\mu\rho}\delta^{\nu\sigma} + \text{symmetrizations}-\text{traces}.
\end{equation}
The number of independent tensor structures is denoted by $N_i$. Taking into account all the obvious symmetries discussed above one concludes that
\begin{equation}
	N_1 = 8,\qquad
	N_2=N_3=3,\qquad
	N_4=N_5=6.
\end{equation}

Let us now plug the definitions of the $g_i$ integrals together with the three-point function \eqref{eq:three-point_function} into the left-hand side of \eqref{eq:decomposition}. We then fully contract both sides of \eqref{eq:decomposition} with all possible combinations of Kronecker symbols. This allows to write explicitly all the scalar functions $g_{iI}(\delta)$  in terms of the basic scalar integrals
\begin{equation}
	\label{eq:integral_J}
	\bold{J}(a,b,c)\equiv 
	w\, C_{TT\cO}\times
	\int d^dz_1\int d^dz_2\ \f{\big(\chi\left(\mu \rho,\lambda\right)\big)^2}{\left(z^2_{12}\right)^\frac{a+\delta}{2} \left(z_1^2\right)^\frac{b-\delta}{2} \left(z_2^2\right)^\frac{c-\delta}{2}},
\end{equation}
where $\chi\left(s,\lambda\right)$ is defined in \eqref{eq:running_coupling} and $\rho$ is given by
\begin{equation}
	\rho= \left(z_1^2 z_2^2 (z_1-z_2)^2\right)^\frac{1}{6},
\end{equation}
which follows from \eqref{eq:definition_r_y1_y2_app} under the choice \eqref{eq:coordinate_transformation}.
Only the integrals subject to the following constraint
\begin{equation}
	\label{eq:condition}
	a+b+c=3d-4
\end{equation}
enter in the expressions for $g_{iI}(\delta)$.
This fact will be crucial for evaluating \eqref{eq:integral_J}. 

The steps performed here are logically simple, but extremely tedious in practice. We perform them in Mathematica. Here we also set the object $H^{\mu\nu\rho\sigma}$ appearing in \eqref{eq:three-point_function} to zero. We will discuss the consequences of this action in the next subsection.

\paragraph{Computation of the vertex}
We can now plug the decomposition \eqref{eq:decomposition} into the effective action \eqref{eq:EFT_action_CPT_5} to obtain its final form in position space. It will depend only on the fields and the integrals $\bold{J}(a,b,c)$ defined in \eqref{eq:integral_J}. We then compute the Fourier transformed effective action by using 
\begin{equation}
	\label{eq:Fourier_transform}
	\varphi(x) = \int \frac{d^4k}{(2\pi)^4}\ \exp(i x\cdot k) \varphi(k),\qquad
	h_{\mu\nu}(x) = \int \frac{d^4k}{(2\pi)^4}\ \exp(i x\cdot k) h_{\mu\nu}(k).
\end{equation}
The effective action in Fourier space can be then used together with the definition \eqref{eq:vertex_hhd} for evaluating the vertex. The result will be in the form
\begin{multline}
	\label{eq:hhphi_covariant}
	V_{(hh \varphi)}(k_1,k_2,k_3;\varepsilon_1,\varepsilon_2) =\\
	f_1(k_1,k_2)\times(\varepsilon_1. \ve_2)
	+f_2(k_1,k_2)\times(k_1.\varepsilon_2.k_1)(k_2.\ve_1.k_2)+f_3(k_1,k_2)\times(k_1.\ve_2.\ve_1.k_2),
\end{multline}
where the functions $f_i(k_1,k_2)$  depend on complicated linear combinations of integrals $\bold{J}(a,b,c)$. We obtain these functions also in Mathematica. In the above formula due to momentum conservation we have $k_3=-k_1-k_2$. We evaluate the integrals $\bold{J}(a,b,c)$ analytically in appendix \ref{app:double_integral_ebaluation}. Using the final result of this appendix given by equations \eqref{eq:result_integral_J} and \eqref{eq:h-integral_result}, we obtain
\begin{align}
	\nn
	f_1(k_1,k_2) &=\frac{i\pi^2\kappa^2\delta\, C_{TT\cO}}{288\sqrt{2}fC_{\cO\cO\cO}}\left(
	\frac{7}{12}(k_1.k_2)^2
	-\frac{17}{24}k_1^2 k_2^2
	+\frac{1}{48}(k_1^4+k_2^4)-\frac{5}{12}(k_1.k_2)(k_1^2+k_2^2)+\ldots\right),\\
	\label{eq:vertex_hhd_functions_WRF}
	f_2(k_1,k_2) &= \frac{i\pi^2\kappa^2\delta\, C_{TT\cO}}{288\sqrt{2}fC_{\cO\cO\cO}},\\
	\nn
	f_3(k_1,k_2) &= \frac{i\pi^2\kappa^2\delta\, C_{TT\cO}}{288\sqrt{2}fC_{\cO\cO\cO}}\left(
	-2(k_1.k_2)-0\times(k_1^2+k_2^2)+\ldots\right).
\end{align}

\paragraph{Computation of $\Delta c$}
\label{sec:result_delC}

In  in  \cite{Karateev:2023mrb} the following general prediction for the functions $f_i$ appearing in the graviton-graviton-dilaton vertex \eqref{eq:hhphi_covariant}  was given. It reads
	\begin{equation}
	\label{eq:vertex_hhd_functions}
	\begin{aligned}
		f_1(k_1,k_2) &= \frac{4i\kappa^2}{\sqrt{2}f}\,\Big(
		2 (-\Delta a+\Delta c+18 r_1)(k_1.k_2)^2
		+(2\Delta a-\Delta c+24 r_1)k_1^2 k_2^2\\
		&+12r_1(k_1^4+k_2^4)+ 42 r_1 (k_1.k_2)(k_1^2+k_2^2)+\ldots\Big),\\
		f_2(k_1,k_2) &= \frac{8i\kappa^2}{\sqrt{2}f}\,\left(-\Delta a+\Delta c+\ldots\right),\\
		f_3(k_1,k_2) &= \frac{8i\kappa^2}{\sqrt{2}f}\,\left(
		2(\Delta a-\Delta c-6r_1)(k_1.k_2)-6r_1(k_1^2+k_2^2)+\ldots\right).
	\end{aligned}
\end{equation}
Let us compare the result \eqref{eq:vertex_hhd_functions_WRF} with the general prediction \eqref{eq:vertex_hhd_functions}. Focusing on the functions $f_2(k_1,k_2)$ and $f_3(k_1,k_2)$ we can read off $\Delta c$ which is given by
\be
\label{eq:result_deltC}
\boxed{
	d=4:\qquad
	\Delta c=\f{\delta\pi^2}{2304} \f{C_{TT\mathcal{O}}}{C_{\mathcal{O}\mathcal{O}\mathcal{O}}}\ +\ O(\delta^2)
.\ }
\ee
In writing this we used the knowledge of  $\Delta a$ and $r_1$ from \eqref{eq:result_delA_option_1}, which states that those coefficients are simply zero at the leading order in $\delta$. The result \eqref{eq:result_deltC} is in a perfect agreement with \eqref{eq:CPT_result}.

If we now focus on the function $f_1(k_1,k_2)$ we can quickly realize that its form is incompatible with the general prediction \eqref{eq:vertex_hhd_functions}. This is not a computational mistake however. We will carefully explain the reason for this in the next subsection.

\subsubsection{Linearized gauge invariance and the IR regulator}
\label{sec:trick}

In order to perform the computations of the previous sections we used the three-point function \eqref{eq:three-point_function} and we set the $H^{\mu\nu\rho\sigma}$ object to zero. This action leads to breaking of the conservation for this three-point function. In what follows we will show that the effects due to breaking of conservation are contained only in the $f_1(k_1,k_2)$ function in the vertex \eqref{eq:hhphi_covariant}. As a consequence we should disregard the result of the computation for $f_1$, but we can safely keep the results for $f_2$ and $f_3$.

In order to show this let us define linearized gauge transformation as follows
\begin{equation}
	\delta h_{\mu\nu}(x)= -\f{1}{2\kappa}\left(\p_\mu\xi_\nu(x)+\p_\nu\xi_\mu(x)\right),
\end{equation} 
where $\xi_\mu(x)$ is some vector field. Let us investigate how the effective action \eqref{eq:EFT_action_CPT_3} which describes the three-point vertex of two gravitons and a dilaton, changes under these transformations. For convenience let us write here \eqref{eq:EFT_action_CPT_3} again
\begin{multline}
	\label{eq:EFT_action_CPT_6}
	A_\text{EFT}[\varphi,\, h] = - \f{\kappa^2}{2\sqrt{2}f}w
	\int d^dx_1\int d^dx_2 \int d^dx_3\ \ h_{\mu\nu}(x_1)h_{\rho\sigma}(x_2) \varphi(x_3)\\
	\times \langle T^{\mu\nu}(x_1)T^{\rho\sigma}(x_2) \mathbf{O}(x_3)\rangle_{\text{QFT}}+\dots .
\end{multline}
Here we used \eqref{eq:three-point_function}. After applying linearized gauge transformations we get
\begin{multline}
	\label{eq:symmetry_broken}
	\delta A_\text{EFT}[\varphi,h] = -\f{\kappa}{2\sqrt{2}f}w
	\int d^dx_1\int d^dx_2 \int d^dx_3\ \xi_\nu(x_1)h_{\rho\sigma}(x_2) \varphi(x_3)\\
	\times \left({\partial_{x_1}}_\mu\,f(\rho)\right) \, \langle T^{\mu\nu}(x_1)T^{\rho\sigma}(x_2) \mathcal{O}(x_3)\rangle_{\text{UV CFT}}+\dots,
\end{multline}
where we have defined $f(\rho) \equiv \chi(\mu\rho,\lambda)^2$ for simplicity of writing. We can write then
\be
\label{eq:derivative_f(z)}
{\partial_{x_1}}_\mu  f(\rho) = \f{\rho}{3}f'(\rho)\left( \f{x_{12\mu}}{x_{12}^2}+\f{x_{13\mu}}{x_{13}^2}\right) \neq 0.
\ee

\paragraph{Tracking linearized gauge invariance breaking (general expectation)}
The non-invariant part of the effective action under the linearized gauge transformation is given by \eqref{eq:symmetry_broken} and \eqref{eq:derivative_f(z)}. Let us now use \eqref{eq:TTO_translation} and move to the new variables defined in \eqref{eq:coordinate_transformation}. We can then rewrite \eqref{eq:symmetry_broken} as
\begin{multline}
	\delta A_\text{EFT}[\varphi,h] =- \f{\kappa}{2\sqrt{2}f}w
	\int d^dx\  \varphi(x)\int d^dz_1 \int d^dz_2\ \xi_\nu(x+z_1)\ h_{\rho\sigma}(x+z_2)\\ \times \f{\rho}{3}f'(\rho)\ \left( \f{z_{12\mu}}{z_{12}^2}+\f{z_{1\mu}}{z_{1}^2}\right)
	\times  \langle T^{\mu\nu}(z_1)T^{\rho\sigma}(z_2) \mathcal{O}(0)\rangle_{\text{UV CFT}}\ ,
\end{multline}

We can now proceed using the same strategy as in the previous section.
We first Taylor expand the fields $\xi_\nu(x+z_1)$ and $h_{\rho\sigma}(x+z_2)$ around $x$ for small values of $z_1$ and $z_2$. Keeping terms with four derivatives only we get
\begin{equation}
	\label{eq:variation_action}
	\begin{aligned}
		\delta A_\text{EFT}[\varphi,h] = 
		&
		-\f{\kappa}{2\sqrt{2}f}\int d^dx\ \varphi(x) \Big(\\
		&\frac{1}{120}\left(\partial_{\alpha_1}\partial_{\alpha_2}\partial_{\alpha_3}\partial_{\alpha_4}\partial_{\alpha_5}\xi_{\nu}(x)\right)
		h_{\rho\sigma}(x)\ q_1^{\alpha_1\alpha_2\alpha_3\alpha_4\alpha_5\nu\rho\sigma}(\delta)\\
		+&\frac{1}{24}\left(\partial_{\alpha_1}\partial_{\alpha_2}\partial_{\alpha_3}\partial_{\alpha_4}\xi_{\nu}(x)\right)
		\left(\partial_{\alpha_5}h_{\rho\sigma}(x)\right)\  q_2^{\alpha_1\alpha_2\alpha_3\alpha_4\alpha_5\nu\rho\sigma}(\delta)\\
		+ &\frac{1}{12}\left(\partial_{\alpha_1}\partial_{\alpha_2}\partial_{\alpha_3}\xi_{\nu}(x)\right)
		\left(\partial_{\alpha_4}\partial_{\alpha_5}h_{\rho\sigma}(x)\right) q_3^{\alpha_1\alpha_2\alpha_3\alpha_4\alpha_5\nu\rho\sigma}(\delta)\\
		+ &\frac{1}{12}\left(\partial_{\alpha_1}\partial_{\alpha_2}\xi_{\nu}(x)\right)
		\left(\partial_{\alpha_3}\partial_{\alpha_4}\partial_{\alpha_5}h_{\rho\sigma}(x)\right) q_4^{\alpha_1\alpha_2\alpha_3\alpha_4\alpha_5\nu\rho\sigma}(\delta)\\
		+ &\frac{1}{24}\left(\partial_{\alpha_1}\xi_{\nu}(x)\right)
		\left(\partial_{\alpha_2}\partial_{\alpha_3}\partial_{\alpha_4}\partial_{\alpha_5}h_{\rho\sigma}(x)\right)q_5^{\alpha_1\alpha_2\alpha_3\alpha_4\alpha_5\nu\rho\sigma}(\delta)\\
		+&\frac{1}{120}\xi_{\nu}(x)
		\left(\partial_{\alpha_1}\partial_{\alpha_2}\partial_{\alpha_3}\partial_{\alpha_4}\partial_{\alpha_5}h_{\rho\sigma}(x)\right) q_6^{\alpha_1\alpha_2\alpha_3\alpha_4\alpha_5\nu\rho\sigma}(\delta)\Big)+\cdots,
	\end{aligned}
\end{equation}
where we have defined
\begin{equation*}
	\resizebox{0.99\hsize}{!}{$
		\begin{aligned}
			q_1^{\alpha_1\alpha_2\alpha_3\alpha_4\alpha_5\nu\rho\sigma}(\delta)&\equiv w
			\int d^dz_1\ d^dz_2\  z_1^{\alpha_1}z_1^{\alpha_2}z_1^{\alpha_3}z_1^{\alpha_4}z_1^{\alpha_5}
			\langle T^{\mu\nu}(z_1)T^{\rho\sigma}(z_2) \mathcal{O}(0)\rangle_{\text{UV CFT}} \f{\rho}{3}f'(\rho) \left( \f{z_{12\mu}}{z_{12}^2}+\f{z_{1\mu}}{z_{1}^2}\right),\\
			q_2^{\alpha_1\alpha_2\alpha_3\alpha_4\alpha_5\nu\rho\sigma}(\delta)&\equiv w
			\int d^dz_1\ d^dz_2\  z_1^{\alpha_1}z_1^{\alpha_2}z_1^{\alpha_3}z_1^{\alpha_4}z_2^{\alpha_5}
			\langle T^{\mu\nu}(z_1)T^{\rho\sigma}(z_2) \mathcal{O}(0)\rangle_{\text{UV CFT}} \f{\rho}{3}f'(\rho) \left( \f{z_{12\mu}}{z_{12}^2}+\f{z_{1\mu}}{z_{1}^2}\right),\\
			q_3^{\alpha_1\alpha_2\alpha_3\alpha_4\alpha_5\nu\rho\sigma}(\delta)&\equiv w
			\int d^dz_1\ d^dz_2\  z_1^{\alpha_1}z_1^{\alpha_2}z_1^{\alpha_3}z_2^{\alpha_4}z_2^{\alpha_5}
			\langle T^{\mu\nu}(z_1)T^{\rho\sigma}(z_2) \mathcal{O}(0)\rangle_{\text{UV CFT}} \f{\rho}{3}f'(\rho) \left( \f{z_{12\mu}}{z_{12}^2}+\f{z_{1\mu}}{z_{1}^2}\right),\\
			q_4^{\alpha_1\alpha_2\alpha_3\alpha_4\alpha_5\nu\rho\sigma}(\delta) &\equiv w
			\int d^dz_1\ d^dz_2\  z_1^{\alpha_1}z_1^{\alpha_2}z_2^{\alpha_3}z_2^{\alpha_4}z_2^{\alpha_5} 
			\langle T^{\mu\nu}(z_1)T^{\rho\sigma}(z_2) \mathcal{O}(0)\rangle_{\text{UV CFT}} \f{\rho}{3}f'(\rho) \left( \f{z_{12\mu}}{z_{12}^2}+\f{z_{1\mu}}{z_{1}^2}\right),\\
			q_5^{\alpha_1\alpha_2\alpha_3\alpha_4\alpha_5\nu\rho\sigma}(\delta) &\equiv w
			\int d^dz_1\ d^dz_2\  z_1^{\alpha_1}z_2^{\alpha_2}z_2^{\alpha_3}z_2^{\alpha_4}z_2^{\alpha_5}
			\langle T^{\mu\nu}(z_1)T^{\rho\sigma}(z_2) \mathcal{O}(0)\rangle_{\text{UV CFT}} \f{\rho}{3}f'(\rho) \left( \f{z_{12\mu}}{z_{12}^2}+\f{z_{1\mu}}{z_{1}^2}\right),\\
			q_6^{\alpha_1\alpha_2\alpha_3\alpha_4\alpha_5\nu\rho\sigma}(\delta)&\equiv w
			\int d^dz_1\ d^dy_2\  z_2^{\alpha_1}z_2^{\alpha_2}z_2^{\alpha_3}z_2^{\alpha_4}z_2^{\alpha_5}
			\langle T^{\mu\nu}(z_1)T^{\rho\sigma}(z_2) \mathcal{O}(0)\rangle_{\text{UV CFT}} \f{\rho}{3}f'(\rho) \left( \f{z_{12\mu}}{z_{12}^2}+\f{z_{1\mu}}{z_{1}^2}\right).
		\end{aligned}
		$}
\end{equation*}
The six functions $q$ can be decomposed into tensor structures built out of the flat metric. The functions $q_1$ and $q_6$ have two linearly independent structures. The functions $q_2$ and $q_5$ have 5 linearly independent structures. The functions $q_3$ and $q_4$ have 8 linearly independent tensor structures. The coefficients of the depositions are given by linear combinations of the following integrals
\begin{equation}
	\label{eq:integral_K_definition}
	\bold{K}(a,b,c)\equiv w C_{TT\cO}\int d^dz_1\int d^dz_2\ \f{1}{\left(z^2_{12}\right)^\frac{a+\delta}{2} \left(z_1^2\right)^\frac{b-\delta}{2} \left(z_2^2\right)^\frac{c-\delta}{2}}\times \f{\rho}{3} f'(\rho)
\end{equation}
with $a+b+c=3d-4$. 

The integral \eqref{eq:integral_K_definition} can be computed by following the same steps as in appendix \ref{app:double_integral_ebaluation}. We get then
\begin{multline}
	\label{eq:integral_K1}
	\bold{K}(a,b,c)=  \Omega_{d-1}\Omega_{d-2}\,
	\ell(\delta) \frac{\sqrt{\pi } \Gamma \left(\frac{d-1}{2}\right) \Gamma \left(-\frac{a}{2}+\frac{2 d}{3}-\frac{2 \delta }{3}-\frac{2}{3}\right) }{2 \Gamma \left(\frac{1}{6} (3 a-d+4 \delta +4)\right) \Gamma \left(\frac{c}{2}-\frac{\delta }{3}+\frac{2}{3}-\frac{d}{6}\right) }\\
	\times \f{\Gamma \left(\frac{1}{6} (-3 c+4 d+2 \delta -4)\right) \Gamma \left(\frac{1}{6} (3 a+3 c-5 d+2 \delta +8)\right)}{\Gamma \left(-\frac{a}{2}+\frac{4 d}{3}-\frac{c}{2}-\frac{\delta }{3}-\frac{4}{3}\right)}\ ,
\end{multline}
where
\be
\label{eq:integral_K2}
\ell(\delta)\equiv w C_{TT\cO} \int_{0}^{\infty} d\rho\ \rho^{-d+3+\delta}\ \times \f{\rho}{3} f'(\rho)\ .
\ee
Focusing on $d=4$ we can evaluate this integral explicitly analogously to the step 3 given in appendix \ref{app:double_integral_ebaluation}. Skipping the details we get
\begin{equation}
	\ell(\delta)= -
	\f{\delta}{3}C_{TT\cO}\lambda_{\star}.
\end{equation}

Now let us plug the tensor structure decomposition of the functions $q_i$ into \eqref{eq:variation_action}. The result will depend on the integrals $\bold{K}(a,b,c)$. Performing the Fourier transform of the $\xi_\mu(x)$ field, the dilaton and the graviton field according to \eqref{eq:Fourier_transform} we obtain the effective action in Fourier space. 
Using this effective action we can compute the variation of the three-point vertex, which is given by the following expression
\begin{multline}
	\label{eq:variation_smart}
	(2\pi)^4 \delta^{(4)}(k_1+k_2+k_3) \times 
	\delta V_{(hh \varphi)}(k_1,k_2,k_3;\varepsilon_2) =\\ 2 \xi_\nu(k_1)\varepsilon_{2\rho\sigma}(k_2)\times
	\frac{i\,\delta^3 \left(\delta A^L_\text{EFT}\right)}{\delta  \xi_\nu(k_1)\delta  h_{\rho\sigma}(k_2)\delta \varphi(k_3)}\Bigg{|}_{h,\varphi=0}.
\end{multline}
The factor 2 is injected here since in the vertex computation \eqref{eq:vertex_hhd} we vary consequently with respect to two graviton fields and instead in \eqref{eq:variation_smart} we vary with respect to a single graviton.

Taking the variation with respect to one dilaton and one graviton, plugging there the explicit form of the integral $\bold{K}(a,b,c)$ given by \eqref{eq:integral_K1} and \eqref{eq:integral_K2}, and expanding at the leading order in $\delta$ we obtain the following result
\begin{equation}
	\label{eq:lin_gauge_condition_vertex_alternative}
	\delta V_{(hh \varphi)}(k_1,k_2,k_3;\varepsilon_2) =
	\frac{i\pi^2\kappa\delta\,C_{TT\cO}}{288\sqrt{2}fC_{\cO\cO\cO}} \left(+\frac{5i}{12}(k_1.k_2)^2\right)\times(k_1.\ve_2.\xi) +O(\delta^2).
\end{equation}

\paragraph{Testing linearized gauge invariance breaking in the solution} Let us now look at the obtained three-point vertex \eqref{eq:vertex_hhd_functions_WRF}. The linearized gauge transformation in momentum space reads as
\begin{equation}
	\label{eq:linearized transformation}
	\delta\varepsilon_{\mu\nu}(k) = -\frac{i}{2\kappa}\left(\xi_\mu(k) k_\nu + \xi_\nu(k) k_\mu\right),
\end{equation} 
where $ \xi_\mu(k)$ is some arbitrary vector field. Applying this transformation to \eqref{eq:hhphi_covariant} we get
\begin{multline}
	\label{eq:lin_gauge_condition_vertex}
	\delta V_{(hh \varphi)}(k_1,k_2,k_3;\varepsilon_1,\varepsilon_2) =
	-\frac{i}{2\kappa}\big( 2f_1(k_1,k_2)+(k_1.k_2)\,f_3(k_1,k_2)\big) \times(k_1.\ve_2.\xi)\\
	-\frac{i}{2\kappa}\big(2(k_1.k_2)\,f_2(k_1,k_2)+f_3(k_1,k_2)\big)\times (k_1.\varepsilon_2.k_1)(\xi.k_2).
\end{multline}
Plugging the solution \eqref{eq:vertex_hhd_functions_WRF} into \eqref{eq:lin_gauge_condition_vertex} and setting $k_i^2=0$\footnote{This condition is necessary to impose when working with traceless and transverse gravitons.} we get
\begin{multline}
	\label{eq:lin_gauge_condition_vertex_final}
	\delta V_{(hh \varphi)}(k_1,k_2,k_3;\varepsilon_2) =
	\frac{i\pi^2\kappa \delta\,C_{TT\cO}}{288\sqrt{2}fC_{\cO\cO\cO}} \left(+\frac{5i}{12}(k_1.k_2)^2\right)\times(k_1.\ve_2.\xi)\\
	+0\times (k_1.\varepsilon_2.k_1)(\xi.k_2).
\end{multline}
As we can see, the vertex is non-invariant under the linearized gauge transformation. Remarkably, we obtain the same result as in \eqref{eq:lin_gauge_condition_vertex_alternative}. This means that the violation of the linearized gauge invariance comes purely from the non-conservation of the IR regulator ${\partial_{x_1}}_\mu  f(\rho) \neq 0 $.
Moreover it is contained only in one out of the two  tensor structures in \eqref{eq:lin_gauge_condition_vertex_final}. Comparing \eqref{eq:lin_gauge_condition_vertex} and \eqref{eq:lin_gauge_condition_vertex_final} we see that the functions $f_2(k_1,k_2)$ and $f_3(k_1,k_2)$ must be computed correctly at the order $O(\delta)$ in order to make the coefficient in front of the second tensor structure in \eqref{eq:lin_gauge_condition_vertex} vanish. Instead, the functions $f_1(k_1,k_2)$ is not correct since the coefficient in front of the first tensor structure in \eqref{eq:lin_gauge_condition_vertex} does not vanish.

\section{Discussion}
\label{sec:discussion}

In this section we discuss interesting open problems. We split the discussion into two parts. In the first part we summarize questions about weakly relevant flows which should be understood. In the second part we discuss applications of the technology introduced in this paper in other contexts.

\paragraph{Questions about weakly relevant flows}

\begin{itemize}
	
    \item In section \ref{sec:conformal_perturbation_theory} we have computed scalar two-point functions at the leading order in conformal perturbation. In the future it is very important to understand how to compute scalar three-point functions in conformal perturbation theory. More precisely one needs to learn how to evaluate the integral in equation \eqref{eq:to_compute_3} when the CFT four-point function is expanded in conformal blocks. This result combined with the solution of the Callan-Symanzik equation given in section \ref{eq:CS_3pt} would allow for instance to compute the change in the scalar OPE coefficients between the UV and IR fixed points. We plan to address this problem in the case of 1d weakly relevant flows in \cite{Karateev:2024xxx}.
   
   \item It would be very interesting to compute the correlation function of three stress-tensors
   	\begin{equation}
   	\label{eq:3T}
   	\<T^{\mu\nu}(x_1) T^{\alpha\beta}(x_2) T^{\rho\sigma}(x_3)\>_\text{QFT}
   \end{equation}
   in weakly relevant flows. This would require, however, to learn first how to perform conformal perturbation theory for scalar three-point functions. The expression \eqref{eq:3T} would then allow to directly evaluate $\Delta a$ in 4d. The result must  agree with \eqref{eq:change_trace_anomalies} obtained by the indirect method.

   \item Another interesting correlator to consider is the following three-point function
   	\begin{equation}
   	\label{eq:OTO}
   	\<\cO(x_1) T^{\mu\nu}(x_2) \cO(x_3) \>_\text{QFT}
   \end{equation}
   This correlator can be used to compute the three-point function between two scalars and the Average Null Energy (ANE) operator by integrating the stress-tensor in \eqref{eq:OTO} over the null-direction. In order to integrate the stress-tensor in \eqref{eq:OTO} over the null-direction one would, however, have to carefully address the issue with conservation, which will appear here analogously to equations \eqref{eq:JJO_intro} and \eqref{eq:three-point_function}.
   
   \item The are several sum-rules in the literature which allow to compute $\Delta a$ and $\Delta c$ in 4d as an integral of a certain correlation function. For instance using the stress-tensor two-point function we have computed in section \ref{sec:stress-tensor} one could evaluate $\Delta c$ in an alternative to this paper way by using the integral sum-rule derived in \cite{Karateev:2020axc}. Analogously, in 4d one could use the correlator \eqref{eq:OTO} in order to compute the value of $\Delta a$ and  the correlator \eqref{eq:three-point_function} in order to compute the value of $\Delta c$ by using the sum-rules derived in \cite{Hartman:2023qdn} and \cite{Hartman:2024xkw} respectively. All these sum-rules, however, rely on the exact conservation of the involved correlators. This introduces a challenge for using our expressions of this paper due to subtle non-conservation effects in our approximate expressions.

	\item It would be very useful to consider some explicit realizations of an abstract weakly relevant flow considered in this paper. One could try to construct the UV CFT to be a generalized free field.\footnote{OPE coefficients of scalar operators in generalized free theories have been originally computed in \cite{Fitzpatrick:2011dm}. Various fermion OPE coefficients in 4d were computed in \cite{Elkhidir:2017iov,Karateev:2019pvw}. Generalized free theories received a general treatment in \cite{Karateev:2018oml}.} 
	For instance one could assume the existence of a single fundamental field $\phi(x)$ with the scaling dimension $\Delta_\phi=\frac{d}{2}-\delta/2$, which is parity odd under some discrete symmetry $Z_2$. We can then construct a parity even operator $\cO(x)$, which will be used as the deforming operator, as a product of two fundamental fields $\phi(x)$. The operator $\cO(x)$ will then have the scaling dimension $\Delta=d-\delta$.
	
	This explicit realization of the weakly relevant flow can be studied using Hamiltonian Truncation methods.\footnote{Hamiltonian Truncation is a new promising non-perturbative tool for studying QFTs, see \cite{Fitzpatrick:2022dwq} for a review. It was first introduced in $d=2$ in \cite{Yurov:1989yu}, for the recent developments see \cite{Hogervorst:2014rta,Rychkov:2014eea,Rychkov:2015vap,Elias-Miro:2017xxf,Elias-Miro:2017tup,Anand:2019lkt,Elias-Miro:2020qwz,Anand:2020qnp,EliasMiro:2021aof,Chen:2021bmm,Hogervorst:2021spa,EliasMiro:2022pua,Chen:2022zms,Fitzpatrick:2023aqm,Delouche:2023wsl}.} The obtained results can then be compared to the analytic ones of this paper. This would allow for instance to test Hamiltonian Truncation methods in higher dimensions.

\end{itemize}

\paragraph{Applications of our technology in other contexts}
\begin{itemize}
	
	\item It would be extremely interesting to compute two- and three-point correlation functions along the whole RG flow in the Wilson-Fisher theory. Using them one would be able to compute anomalous dimensions and the change in the OPE coefficients between the UV and IR.

\item There have been numerous studies of RG flows in the presence of defects and boundaries \cite{Affleck:1991tk,Yamaguchi:2002pa,Estes:2014hka,Andrei:2018die,Kobayashi:2018lil,Casini:2018nym,Azeyanagi:2007qj,Lauria:2020emq,Giombi:2020rmc,Wang:2020xkc,Nishioka:2021uef,CarrenoBolla:2023vrv,Padayasi:2021sik,Rodriguez-Gomez:2022gbz,Rodriguez-Gomez:2022gif,Jensen:2015swa,Castiglioni:2022yes,Krishnan:2023cff,Cuomo:2023qvp,Drukker:2023jxp,Shimamori:2024yms,Friedan:2003yc,Casini:2016fgb,Cuomo:2021rkm,Casini:2022bsu,Wang:2021mdq,Shachar:2022fqk,Cuomo:2021kfm,Gaberdiel:2008fn,Shachar:2024ubf,Sato:2021eqo}, where analogous of the $a$-theorem were proposed. It would be interesting to generalize our approach of computing correlation functions and trace anomalies in those scenarios.

\item Finally, it would be interesting to study correlation functions in de Sitter (dS) and Anti-de Sitter (AdS) spaces. For recent interesting related works see  \cite{Meltzer:2021bmb,Lauria:2023uca,Loparco:2023rug,Meineri:2023mps,Loparco:2023akg,Loparco:2024ibp}.

\end{itemize}

\section*{Acknowledgements}
We are grateful to Zohar Komargodski for pointing out to us the model of weakly relevant flows and to Jo\~ao Penedones for emphasising the importance of computing correlation functions along the whole RG flow. We thank Connor Behan, Liam Fitzpatrick, Giovanni Galati, Petr Kravchuk, Hugh Osborn, Jiaxin Qiao, Balt Van Rees, Slava Rychkov, Ritam Sinha, Andy Stergiou, Yannis Ulrich and Matt Walters for very useful conversations. We also thank Connor Behan, Zohar Komargodski and Jo\~ao Penedones for various useful comments on the early versions of this paper.

DK is supported by the SNSF Ambizione grant PZ00P2\_193411. 
The work of BS is supported by the Simons Foundation grant 488649 (Simons Collaboration on the Nonperturbative Bootstrap) and by the Swiss National Science Foundation through the project 200020\_197160 and through the National Centre of Competence in Research SwissMAP. BS is also supported by STFC grant number ST/X000753/1.

\appendix

\section{Basic integrals}
\label{app:integral_3pt}
In this appendix we will be concerned with the following integrals
\begin{equation}
	\label{eq:definition_I_n}
	\bold{I}_n^{(a_1,\ldots ,a_n)}(x_1,\ldots ,x_n)\equiv \int d^d y \f{1}{\left((x_1-y)^2\right)^{a_1}\ldots\left((x_n-y)^2\right)^{a_n}},
\end{equation}
where $a_1$, $a_2$, $\ldots$, $a_n$ are some real parameters.
The integral \eqref{eq:definition_I_n} has $n$ potential (UV) divergences when $y$ coincides with $x_1$, $x_2$, $\ldots$ or $x_n$. These divergences are absent if
\begin{equation}
	\label{eq:range}
	a_1<d/2,\quad
	a_2<d/2,\quad
	\ldots,\quad
	a_n<d/2.
\end{equation}
When $|y|$ is large the integral \eqref{eq:definition_I_n} could have an (IR) divergence. This is absent if
\begin{equation}
a_1+a_2+\ldots+a_n>d/2.
\end{equation}
In what follows we will evaluate the integral \eqref{eq:definition_I_n} focusing on the case of $n=2$ and $n=3$. We will assume the parameter range for $a_1$, $a_2$, $\ldots$, $a_n$ when there are no divergences. The final result, however, can be analytically continued beyond the original parameter range.

In order to write equations in a compact form we will use the following notation
\begin{equation}
	x_{i(n+1)}^2 \equiv (x_i-x_{n+1})^2.
\end{equation}
We will use the $n$-point Feynman parametrization, which reads
\begin{multline}
	\label{eq:Feynman_parametrization}
	 \f{1}{\prod\limits_{i=1}^n \left(x_{i(n+1)}^2\right)^{a_i}}
	 = \f{\Gamma\left(\sum\limits_{i=1}^n a_i\right)}{\prod\limits_{i=1}^n\Gamma(a_i)}\times\\
	  \int_0^1 d\xi_1 \int_0^1 d\xi_2\cdots \int_0^1 d\xi_n
	   \,\delta\left(\sum\limits_{i=1}^n \xi_i-1\right) \f{\prod\limits_{i=1}^n\xi_i^{a_i-1}}{\left(\sum\limits_{i=1}^n\xi_i x_{i(n+1)}^2  \right)^{\sum\limits_{i=1}^n a_i}}.
\end{multline}
In what follows we will also use the following result of integration in $d$-dimensional Euclidean space
\be
\label{eq:volume_integral}
\int d^dx \f{1}{\left((x-y)^2+\Delta\right)^\alpha}=   \f{\pi^{\f{d}{2}}\ \Gamma\left(\alpha -\f{d}{2}\right)}{\Gamma(\alpha) }\ \f{1}{\Delta^{\alpha -\f{d}{2}}},
\ee
where $y$ is $d-$vector and $\Delta$ is some scalar parameter.

\paragraph{Two-points}
Let us consider the $n=2$ case, namely
\begin{equation}
	\label{eq:definition_I2}
	\bold{I}_2^{(a_1,a_2)}(x_1,x_2)\equiv \int d^d x_3 \f{1}{\left(x_{13}^2\right)^{a_1}\left(x_{23}^2\right)^{a_2} }.
\end{equation}
Using the Feynman parametrization \eqref{eq:Feynman_parametrization}, the above integral can be re-written as
\begin{equation}
	\bold{I}_2^{(a_1,a_2)}(x_1,x_2)= \f{\Gamma(a_1+a_2)}{\Gamma(a_1)\Gamma(a_2)}\int_0^1 d\xi\ \xi^{a_1-1}(1-\xi)^{a_2-1} \int d^d x_3 \f{1}{\left(\xi x_{13}^2 +(1-\xi )x_{23}^2\right)^{a_1+a_2} }.
\end{equation}
The denominator of the integrand can be expressed as
\be
\xi x_{13}^2 +(1-\xi )x_{23}^2=\left( x_3-(\xi x_1+(1-\xi)x_2)\right)^2 +\xi(1-\xi)x_{12}^2. 
\ee
Substituting the above expression in $\bold{I}_2$ and using the result of integration \eqref{eq:volume_integral}, we get
\be
\bold{I}_2^{(a_1,a_2)}(x_1,x_2)= \f{\pi^{\f{d}{2}}\Gamma\left(a_1+a_2-\f{d}{2}\right)}{\Gamma(a_1)\Gamma(a_2)}\ \f{1}{(x_{12}^2)^{a_1+a_2-\f{d}{2}}}\int_0^1 d\xi\ \xi^{\f{d}{2}-a_2-1}(1-\xi)^{\f{d}{2}-a_1-1}.
\ee
Now performing the integration over $\xi$ assuming $a_1,a_2\leq \f{d}{2}$, we get
\begin{equation}
	\label{eq:I2}
	\bold{I}_2^{(a_1,a_2)}(x_1,x_2)=
	\f{P_2(a_1,a_2)}{(x_{12}^2)^{a_1+a_2-d/2}},
\end{equation}
where we have
\begin{equation}
	\label{eq:P2}
	P_2(a_1,a_2) \equiv \pi^{d/2}\ 
	\frac{\Gamma \left(d/2-a_1\right) \Gamma \left(d/2-a_2\right)}{\Gamma(a_1)\Gamma(a_2)}\
	\frac{\Gamma\left(a_1+a_2-d/2\right)}{\Gamma (d-a_1-a_2)}.
\end{equation}

\paragraph{Three-points}
Let us now consider the $n=3$ case, namely
\begin{equation}
	\label{eq:definition_I3}
	\bold{I}^{(a_1,a_2,a_3)}_3(x_1,x_2,x_3)\equiv \int d^d x_4 \f{1}{\left(x_{14}^2\right)^{a_1}\left(x_{24}^2\right)^{a_2}\left(x_{34}^2\right)^{a_3} }.
\end{equation}
Using the Feynman parametrization \eqref{eq:Feynman_parametrization}, we can re-write the above integral as
\be
	\bold{I}^{(a_1,a_2,a_3)}_3(x_1,x_2,x_3)&=&\f{\Gamma\left(a_1+a_2+a_3\right)}{\Gamma(a_1)\Gamma(a_2)\Gamma(a_3)}\int_0^1 d\xi_1 \int_0^1 d\xi_2 \int_0^1 d\xi_3\ \delta(\xi_1+\xi_2+\xi_3-1)\nn\\
	&&\times \xi_1^{a_1-1}\xi_2^{a_2-1}\xi_3^{a_3-1} \int d^d x_4 \f{1}{\left(\xi_1x_{14}^2 +\xi_2 x_{24}^2+\xi_3 x_{34}^2 \right)^{a_1+a_2+a_3} }.
\ee
Now we can also re-write the denominator  of the integrand as
\begin{equation}
	\left(\xi_1x_{14}^2 +\xi_2 x_{24}^2+\xi_3 x_{34}^2 \right)=\left( x_4-(\xi_1x_1+\xi_2x_2+\xi_3x_3)\right)^2+\xi_1\xi_2 x_{12}^2+\xi_2\xi_3x_{23}^2+\xi_3\xi_1x_{31}^2,
\end{equation}
under the constraint $\xi_1+\xi_2+\xi_3=1$. Using  \eqref{eq:volume_integral}, we then obtain
\begin{multline}
	\label{eq:I3_generic}
		\bold{I}^{(a_1,a_2,a_3)}_3(x_1,x_2,x_3) = \f{\pi^{\f{d}{2}}\Gamma\left(a_1+a_2+a_3-\f{d}{2}\right)}{\Gamma(a_1)\Gamma(a_2)\Gamma(a_3)}\int_0^1 d\xi_1 \int_0^1 d\xi_2 \int_0^1 d\xi_3\ \delta(\xi_1+\xi_2+\xi_3-1)\\
	\times \xi_1^{a_1-1}\xi_2^{a_2-1}\xi_3^{a_3-1} \left(\xi_1\xi_2 x_{12}^2+\xi_2\xi_3x_{23}^2+\xi_3\xi_1x_{31}^2\right)^{\f{d}{2}-(a_1+a_2+a_3)}.
\end{multline}

A more useful representation of the integral \eqref{eq:definition_I3} for practical purposes is found by using the Schwinger parametrization. This was done in  \cite{Bzowski:2013sza}, their result reads
\begin{align}
&\bold{I}^{(a_1,a_2,a_3)}_3(x_1,x_2,x_3)\nn\\
=& \f{\pi^{\f{d}{2}}\ 2^{a_1+a_2+a_3-d}}{\Gamma(a_1)\Gamma(a_2)\Gamma(a_3)\Gamma(d-a_1-a_2-a_3)} \f{1}{(x_{12}^2)^{\f{a_1+a_2}{2} -\f{d}{4}}}\f{1}{(x_{23}^2)^{\f{a_2+a_3}{2} -\f{d}{4}}}\f{1}{(x_{31}^2)^{\f{a_3+a_1}{2} -\f{d}{4}}} \nn\\
&\times \int_0^\infty dt\ t^{\f{d-4}{4}}\ \mathbf{K}_{\f{d}{2}-a_1-a_2}\left(\sqrt{2tx_{12}^2}\right)\mathbf{K}_{\f{d}{2}-a_2-a_3}\left(\sqrt{2tx_{23}^2}\right)\mathbf{K}_{\f{d}{2}-a_3-a_1}\left(\sqrt{2tx_{31}^2}\right),\label{eq:I3_generic_Schwinger}
\end{align}
where $\mathbf{K}_\nu(z)$ represents the generalized Bessel function and its integral representation in the region $|\text{arg}(z)|< \f{\pi}{4}$ is given by
\begin{equation}
\mathbf{K}_\nu(z)= \f{1}{2}\left(\f{z}{2}\right)^\nu \int_0^\infty du\ u^{-\nu -1} \ e^{-u-\f{z^2}{4u}}.
\end{equation}

It is hard to rewrite our result \eqref{eq:I3_generic_Schwinger} in terms of some known analytic functions. However, in the special case when $a_1+a_2+a_3=d$, the integral becomes conformal and has been evaluate in appendix B of \cite{Dolan:2000uw}. The result reads
\begin{equation}
	\label{eq:I3}
	a_1+a_2+a_3=d:\qquad
	\bold{I}^{(a_1,a_2,a_3)}_3(x_1,x_2,x_3) = \frac{P_3(a_1,a_2,a_3)}{\left(x_{12}^2\right)^{d/2-a_3}\left(x_{13}^2\right)^{d/2-a_2}\left(x_{23}^2\right)^{d/2-a_1}},
\end{equation}
where the coefficient $P_3$ is given by
\begin{equation}
	\label{eq:P3}
P_3(a_1,a_2,a_3) \equiv \pi^{d/2}\  \frac{\Gamma(d/2-a_1)\Gamma(d/2-a_2)\Gamma(d/2-a_3)}{\Gamma(a_1)\Gamma(a_2)\Gamma(a_3)}.
\end{equation}

\section{OPE expansion}
\label{app:OPE_expansion}

In conformal field theories the product of two operators can be written in the following form
\begin{equation}
	\label{eq:OPE}
	\cO_1(x_1)\cO_2(x_2) = \sum_{\cO} C_{\cO_1\cO_2\cO}\textbf{C}_{\cO_1\cO_2\cO}(x_{12},\partial_{x_2})  \cO(x_2)+\ldots,
\end{equation}
known as the OPE expansion. Here $\ldots$ indicate contributions of operators with spin. For simplicity let us only consider the contribution of scalar operators.  Here $C_{\cO_1\cO_2\cO}$ is the OPE coefficient and $\textbf{C}_{\cO_1\cO_2\cO}(x_{12},\partial_{x_2})$ is a function which is completely fixed by conformal invariance.

Let us multiply \eqref{eq:OPE} by $\cO(x_3)$, take vacuum expectation value and use the above expressions for the two- and three-point functions. We get then
\begin{equation}
	\label{eq:condition_C}
	\frac{1}{\left(x_{12}^2\right)^\frac{\Delta_1+\Delta_2-\Delta}{2}\left(x_{13}^2\right)^\frac{\Delta_1-\Delta_2+\Delta}{2}\left(x_{23}^2\right)^\frac{-\Delta_1+\Delta_2+\Delta}{2}} = \textbf{C}_{\cO_1\cO_2\cO}(x_{12},\partial_{x_2})  \frac{1}{\left(x_{23}^2\right)^{\Delta}}.
\end{equation}
From this expression we can systematically compute the function $\textbf{C}_{\cO_1\cO_2\cO}(x_{12},\partial_{x_2})$. We can first write
\begin{equation}
	\label{eq:C_vs_c}
	\textbf{C}_{\cO_1\cO_2\cO}(x_{12},\partial_{x_2}) = \frac{1}{\left(x_{12}^2\right)^\frac{\Delta_1+\Delta_2-\Delta}{2}} \textbf{c}_{\cO_1\cO_2\cO}(x_{12},\partial_{x_2}).
\end{equation}
Then the condition \eqref{eq:condition_C} can be rewritten in terms of the small $\textbf{c}$-function, it reads
\begin{equation}
	\label{eq:condition_c}
\frac{1}{\left(x_{13}^2\right)^\frac{\Delta+\Delta_{12}}{2}\left(x_{23}^2\right)^\frac{\Delta-\Delta_{12}}{2}} = \textbf{c}_{\cO_1\cO_2\cO}(x_{12},\partial_{x_2})  \frac{1}{\left(x_{23}^2\right)^{\Delta}}.
\end{equation}

This condition \eqref{eq:condition_c} can be solved in a series form. Its most general expression reads as
\begin{equation}
	\label{eqC-function}
	\textbf{c}_{\cO_1\cO_2\cO}(x_{12},\partial_{x_2}) =
	1+t^1_{\mu_1}\partial^{\mu_1}_{x_2}
	+t^2_{\mu_1\mu_2}\partial^{\mu_1}_{x_2}\partial^{\mu_2}_{x_2}
	+t^3_{\mu_1\mu_2\mu_3}\partial^{\mu_1}_{x_2}\partial^{\mu_2}_{x_2}\partial^{\mu_3}_{x_2}
	+\ldots,
\end{equation} 
where tensors $t_n$ have the following form
\begin{equation}
	\label{eq:coefficients}
	\begin{aligned}
		t^1_{\mu_1} & = c^1_1 x_{12}^{\mu_1},\\
		t^2_{\mu_1\mu_2} & = c^2_1 \left(x_{12}^{\mu_1}x_{12}^{\mu_2}\right) + c^2_2 \left(x_{12}^2 \delta^{\mu_1\mu_2}\right),\\
		t^3_{\mu_1\mu_2\mu_3} & = c^3_1 \left(x_{12}^{\mu_1}x_{12}^{\mu_2}x_{12}^{\mu_3}\right)+c^3_2 \left(x_{12}^2 \delta^{\mu_1\mu_2}x_{12}^{\mu_3}\right),\\
		t^4_{\mu_1\mu_2\mu_3\mu_4} & = c^4_1 \left(x_{12}^{\mu_1}x_{12}^{\mu_2}x_{12}^{\mu_3}x_{12}^{\mu_4}\right)+ c^4_2 \left(x_{12}^2 \delta^{\mu_1\mu_2}x_{12}^{\mu_3}x_{12}^{\mu_4}\right)+ c^4_3 \left(x_{12}^2 \delta^{\mu_1\mu_2}x_{12}^2 \delta^{\mu_3\mu_4}\right),\\
		&\ldots
	\end{aligned}
\end{equation}
For the special case $\Delta_{12}=0$ the coefficients $c^n_i$ can be written in a compact form as follows
\begin{multline}
	c^n_i = \frac{2^{\Delta+1-2i}(-1)^{i+1}}{\sqrt{\pi}}
	\frac{\Gamma\left(\Delta-d/2+1\right)\Gamma\left(\Delta/2+i-1\right)}
	{\Gamma\left(i\right)\Gamma\left(\Delta/2\right)\Gamma\left(\Delta-d/2+i\right)}\times\\
	\frac{\Gamma\left(\Delta/2+n+1-i\right)\Gamma\left(\Delta/2+1/2\right)}
	{\Gamma\left(n+3-2i\right)\Gamma\left(\Delta+n\right)}.
\end{multline}

\section{Trace of the stress-tensor}
\label{app:trace_ST}
Let us re-write our weakly relevant RG flow action \eqref{eq:theory_definition} placed in curved background
\begin{equation}
	\label{eq:theory_definition_g}
	A =  A_\text{UV CFT}[\Phi,g_{\mu\nu}] + \lambda_0 m^{d-\Delta} \int d^d x\sqrt{g}\, \cO_g(x),
\end{equation}
where $\Phi$ collectively represents the d.o.f of the UV CFT.
Under Weyl transformation with infinitesimal Weyl transformation parameter $\sigma(x)$ we have the following transformation rules
\be
g_{\mu\nu}(x)\rightarrow e^{2\sigma(x)}g_{\mu\nu}(x)\ ,\ \Phi(x)\rightarrow \Phi_\sigma(x)\ ,\  \cO_g(x)\rightarrow e^{-\Delta \sigma(x)} \cO_g(x),
\ee
and $A_\text{UV CFT}$ is Weyl invariant i.e. $A_\text{UV CFT}[\Phi_\sigma, e^{2\sigma}g_{\mu\nu}]=A_\text{UV CFT}[\Phi,g_{\mu\nu}]$. Now the trace of stress tensor for the QFT \eqref{eq:theory_definition_g} is given by
\be
T_{\text{QFT}}{}^{\mu}_{\ \mu}(x)\equiv \f{1}{\sqrt{g}(x)}\f{\delta A}{\delta \sigma(x)}\Bigg{|}_{g_{\mu\nu}=\delta_{\mu\nu}}=(d-\Delta)\lambda_0 m^{d-\Delta}\cO(x).
\ee

On the other hand at scale $\mu$ the effective weakly relevant RG flow action \eqref{eq:theory_alternative} reads
\begin{equation}
	\label{eq:theory_alternative_app}
	A =  A_\text{EFT} + \lambda(\mu) \mu^{d-\Delta} \int d^d x\, \mathbf{O}(x).
\end{equation}
Here $A_\text{EFT}$ is assumed to be scale invariant. The scale transformation is given by
\be
x^\mu\rightarrow e^{\alpha}x^{\mu}\ ,\ \mu\rightarrow e^{-\alpha}\mu\ ,\ \mathbf{O}(x)\rightarrow e^{-\alpha\Delta}\mathbf{O}(x),
\ee
where $\alpha$ is the infinitesimal scale transformation parameter. Hence under the above scale transformation rule, the infinitesimal scale variation of \eqref{eq:theory_alternative_app} turns out
\be
\delta_\alpha A=-\alpha\ \beta(\lambda) \mu^{d-\Delta} \int d^d x\, \mathbf{O}(x),\label{delta_A_1}
\ee
where the expression for $\beta(\lambda)$ is given in \eqref{eq:beta-function_def}. We know that the scale variation of an action  is related to the divergence of the dilatation symmetry current $J^\mu_D(x)=T_{\text{QFT}}^{\mu\nu}(x)x_\nu$ and the precise relation reads
\be
\delta_\alpha A=\alpha\int d^dx \p_\mu J^\mu_D(x)=\alpha \int d^dx\ T_{\text{QFT}}{}^{\mu}_{\ \mu}(x),\label{delta_A_2}
\ee
where to get the expression after second equality we used the conservation of stress tensor. Now comparing \eqref{delta_A_1} and  \eqref{delta_A_2} we get
\be
T_{\text{QFT}}{}^{\mu}_{\ \mu}(x)= -\beta(\lambda) \mu^{d-\Delta}\mathbf{O}(x).
\ee

\section{Solution of the Callan-Symanzik equation: details}
\label{app:CS_details}

By using the definition of the $\beta$-function \eqref{eq:beta-function_def} one can re-write the right-hand side of \eqref{eq:characteristic} in the following way
\begin{equation}
	\label{eq:relation_1}
	\beta\big(\Lambda(s,\lambda)\big) =
	\beta(\lambda)  \frac{\partial \Lambda(s,\lambda)}{\partial \lambda}.
\end{equation}

Let us check explicitly that \eqref{eq:solution_2} indeed satisfies \eqref{eq:CS_2}. Let us first evaluate the derivative in $s$ of the above result. We get
\begin{equation}
	\label{eq:derivative_s}
	s\frac{\partial}{\partial s} f_2(s,\lambda) = -\beta(\lambda)\,\frac{\partial  f_2^\text{boundary}\big(1,\Lambda\big)}{\partial \lambda} \times \exp(\ldots)
	-2 \gamma\big(\Lambda\big)f_2(s,\lambda).
\end{equation}
In order to obtain the first term in the right-hand side we first used \eqref{eq:characteristic} and then \eqref{eq:relation_1}.
In order to proceed let us first re-write the solution \eqref{eq:solution_2} in the following form
\begin{equation}
	\label{eq:solution_2_rewritten}
	f_2(s,\lambda) =   f_2^\text{boundary}\big(1,\Lambda(s,\lambda)\big) \times
	\exp\left(+2 \int_{\Lambda(1,\lambda)}^{\Lambda(s,\lambda)}
	d\Lambda'\,
	\frac{\gamma\big(\Lambda'\big)}{\beta\big(\Lambda'\big)}\right),
\end{equation}
where we simply used \eqref{eq:characteristic}. This expression is more convenient for taking the derivative in $\lambda$, which reads then
\begin{multline}
	\beta(\lambda)\frac{\partial}{\partial \lambda} f_2(s,\lambda) = \beta(\lambda)\,\frac{\partial  f_2^\text{boundary}\big(1,\Lambda\big)}{\partial \lambda} \times \exp(\ldots)\\
	+2\beta(\lambda)f_2(s,\lambda) \left(
	\frac{\gamma\big(\Lambda(s,\lambda)\big)}{\beta(\Lambda(s,\lambda))}\frac{\partial \Lambda(s,\lambda)}{\partial\lambda}-
	\frac{\gamma\big(\Lambda(1,\lambda)\big)}{\beta(\Lambda(1,\lambda))}\frac{\partial \Lambda(1,\lambda)}{\partial\lambda}
	\right).
\end{multline}
Using the relation \eqref{eq:relation_1} and the boundary condition \eqref{eq:boundary_condition} we can bring the above result in its final form
\begin{equation}
	\label{eq:derivative_lambda}
	\beta(\lambda)\frac{\partial}{\partial \lambda} f_2(s,\lambda) = \beta(\lambda)\,\frac{\partial  f_2^\text{boundary}\big(1,\Lambda\big)}{\partial \lambda} \times \exp(\ldots)\\
	+2f_2(s,\lambda) \left(
	\gamma\big(\Lambda\big)-\gamma(\lambda)
	\right).
\end{equation}
Plugging \eqref{eq:derivative_s} and \eqref{eq:derivative_lambda} into the Callan-Symanzik equation \eqref{eq:CS_2} we confirm the equality.

\section{Two-point functions at all orders in conformal perturbation theory}
\label{App:2-point_function_independent_derivation}

Here we review with some modifications an alternative method proposed in \cite{Bzowski:2012ih}, which also allows to compute two-point functions along the full RG flow.

Let us use the action \eqref{eq:theory_alternative}, where $\mu$ is the scale at which the coupling constant $\lambda(\mu)$ is defined. We compute the QFT two-point functions using the following path integral expression
\begin{equation}
	\label{eq:connected_correlation_functions_EFT}
	\langle \mathbf{O}(x_1) \mathbf{O}(x_2) \rangle_\text{QFT} \equiv \frac{ \int [d\Phi] \mathbf{O}(x_1) \mathbf{O}(x_2) e^{-A} }{ \int [d\Phi]\ e^{-A}}. 
\end{equation}
Plugging here \eqref{eq:theory_alternative} and expanding around small $\lambda(\mu)$ we get
\begin{equation}
	\label{eq:expansion}
	\langle\mathbf{O}(x_1)\mathbf{O}(x_2)\rangle_{\text{QFT}}=\sum_{n=0}^\infty \f{(-1)^n}{n!}\lambda(\mu)^n \left(\mu^{\delta}\right)^n\, \mathcal{K}_n(x_1,x_2),
\end{equation}
where we have defined
\begin{equation}
	\label{eq:definition_K_lower}
	\begin{aligned}
		\mathcal{K}_0(x_1,x_2) &\equiv\langle\mathbf{O}(x_1)\mathbf{O}(x_2)\rangle_{\text{EFT}},\\
		\mathcal{K}_1(x_1,x_2) &\equiv\int d^dy\theta\left( |x_1-y|-\mu^{-1}\right) \theta\left( |x_2-y|-\mu^{-1}\right) \langle\mathbf{O}(x_1)\mathbf{O}(x_2)\mathbf{O}(y)\rangle_{\text{EFT}},
	\end{aligned}
\end{equation}
and for $n\geq 2$ we have
\begin{multline}
	\label{eq:definition_K_higher}
		\mathcal{K}_n(x_1,x_2) \equiv  \int d^dy_1 d^dy_2\cdots d^dy_n \left\lbrace \prod_{\substack{i,j=1\\ i>j}}^n \theta\left( |y_{ij}|-\mu^{-1}\right) \right\rbrace\\ \left\lbrace \prod_{\substack{i=1}}^n \theta\left( |x_1-y_{i}|-\mu^{-1}\right) \theta\left( |x_2-y_{i}|-\mu^{-1}\right) \right\rbrace
		\langle\mathbf{O}(x_1)\mathbf{O}(x_2)\mathbf{O}(y_1)\cdots \mathbf{O}(y_n)\rangle_{\text{EFT}},
\end{multline}
All the correlators appearing in \eqref{eq:expansion} are defined only for energies bellow $\mu$ or distances larger than $\mu^{-1}$, namely $|x_{ij}|>\mu^{-1}$. This is the reason we have inserted the $\theta$-functions in the above expressions.

Our normalization condition in section \ref{sec:Callan-Symanzik_correlators} was chosen in such a way that the EFT exhibits conformal invariance. We have for instance
\begin{equation}
	\label{eq:expressions_EFT_CFT}
	\begin{aligned}
			\langle\mathbf{O}(x_1)\mathbf{O}(x_2)\rangle_{\text{EFT}} &=  \langle\cO(x_1)\cO(x_2)\rangle_{\text{UV CFT}} = \frac{1}{\left(x_{12}^2\right)^\Delta},\\
			\langle\mathbf{O}(x_1)\mathbf{O}(x_2)\mathbf{O}(x_3)\rangle_{\text{EFT}} &=  \langle\cO(x_1)\cO(x_2)\cO(x_3)\rangle_{\text{UV CFT}} = \frac{C_{\cO\cO\cO}}{\left(x_{12}^2x_{13}^2x_{23}^2\right)^{\Delta/2}}.
	\end{aligned}
\end{equation}
Inside the EFT correlators appearing in \eqref{eq:definition_K_lower} and \eqref{eq:definition_K_higher} we can take the OPE. As discussed in appendix \ref{app:OPE_expansion} it has the following form
\begin{equation}
	\label{eq:OPE_mu_scale_original}
	\mathbf{O}(z_1)\mathbf{O}(z_2)=\frac{1}{\left(z_{12}^2\right)^\Delta}+\frac{1}{\left(z_{12}^2\right)^{\Delta/2}}C_{\mathcal{O}\mathcal{O}\mathcal{O}}\mathbf{O}(z_2)+\cdots,
\end{equation}
where the first term is the contribution of the identity operator and the second term is the contribution of the operator $\mathbf{O}$ itself. Here and bellow ellipses represent the remaining contributions of primary and descendant operators. We will use the OPE \eqref{eq:OPE_mu_scale_original} inside \eqref{eq:definition_K_lower} and \eqref{eq:definition_K_higher} when $|x_1-y_i|\rightarrow \mu^{-1}$, $|x_2-y_i|\rightarrow \mu^{-1}$ or $|y_i-y_j|\rightarrow \mu^{-1}$. For this purpose instead of \eqref{eq:OPE_mu_scale_original} we can use then
\begin{equation}\label{eq:OPE_mu_scale}
	\mathbf{O}(z_1)\mathbf{O}(z_2)= \mu^{2\Delta}+C_{\mathcal{O}\mathcal{O}\mathcal{O}} \mu^\Delta \mathbf{O}(z_2)+\cdots.
\end{equation}

Let us compute the object $\mathcal{K}_1(x_1,x_2)$ defined in \eqref{eq:definition_K_lower} in the limit $\mu\rightarrow \infty$. We can immediately see that in this limit the $\theta$-functions can be dropped since $\mu^{-1}\rightarrow 0$. Using \eqref{eq:expressions_EFT_CFT} and \eqref{eq:integral_3pt} we simply get
\begin{equation}
	\label{eq:integration_constant}
	\begin{aligned}
		\mathcal{K}_1(x_1,x_2) \Big|_{\mu\rightarrow \infty} &= \int d^d y \langle\cO(x_1)\cO(x_2)\cO(y)\rangle_{\text{UV CFT}} \\
		&=\f{2 C_{\mathcal{O}\mathcal{O}\mathcal{O}}\Omega_{d-1}}{\delta} r^\delta\mathcal{K}_0(x_1,x_2),\quad
		r\equiv |x_{12}|.
	\end{aligned}
\end{equation}

The functions $\mathcal{K}_n(x_1,x_2)$ satisfy a recursion relation. We can derive it by differentiating $\mathcal{K}_n(x_1,x_2)$ with respect to $\mu^{-1}$. When the derivative is acting on a $\theta$-function, it gives us a $\delta$-function. The latter can be used to perform one integral. At the level $n$ there are ${}^nC_2+2n=\f{n(n+3)}{2}$ Heaviside $\theta$-functions, where ${}^nC_2$ is the binomial coefficient.
Using the OPE \eqref{eq:OPE_mu_scale} and only keeping the connected piece proportional to $C_{\mathcal{O}\mathcal{O}\mathcal{O}}$ (ignoring the identity operator) one gets the following recursion relation
\begin{equation}
	\label{eq:K_diff_eq}
	\f{d\mathcal{K}_n(x_1,x_2)}{d\mu^{-1}} = - \f{n(n+3)}{2}C_{\mathcal{O}\mathcal{O}\mathcal{O}}\Omega_{d-1}\mu^{1-\delta} \mathcal{K}_{n-1}(x_1,x_2)+\cdots.
\end{equation}
Our goal now is to solve this recursion relation.

Let us first focus on the case when $n=1$. The recursion relation \eqref{eq:K_diff_eq} then becomes
\begin{equation}
	\label{eq:K_diff_eq_n1}
	\f{d\mathcal{K}_1(x_1,x_2)}{d\mu^{-1}} = - 2C_{\mathcal{O}\mathcal{O}\mathcal{O}}\Omega_{d-1}\mu^{1-\delta} \mathcal{K}_{0}(x_1,x_2)+\cdots.
\end{equation}
In what follows we suppress all the ellipses.
Since $\mathcal{K}_{0}(x_1,x_2)$ is independent of $\mu$ according to our definitions \eqref{eq:definition_K_lower} and \eqref{eq:expressions_EFT_CFT}, we can integrate the above relation. We get
\begin{equation}
	\label{eq:int_n1}
	\mathcal{K}_1(x_1,x_2) = 2C_{\mathcal{O}\mathcal{O}\mathcal{O}}\Omega_{d-1}\frac{\mu^{-\delta}}{\delta} \mathcal{K}_{0}(x_1,x_2)+\text{const},
\end{equation}
where $\text{const}$ is the integration constant independent of $\mu$. We can determine this constant by taking for example the $\mu\rightarrow \infty$ of the above equation. We get then
\begin{equation}
	\text{const} = \mathcal{K}_1(x_1,x_2) \Big|_{\mu\rightarrow \infty}.
\end{equation}
Using \eqref{eq:integration_constant} and plugging this result back into \eqref{eq:int_n1} we get
\begin{equation}\label{eq:K1_final}
	\mathcal{K}_1(x_1,x_2) = 2 C_{\mathcal{O}\mathcal{O}\mathcal{O}}\Omega_{d-1} \mathbf{p}\,\f{r^\delta}{\delta} \, \mathcal{K}_0(x_1,x_2),
\end{equation}
where $\mathbf{p}\equiv 1-(\mu r)^{-\delta}$ and it vanishes in strict $\delta\rightarrow 0$ limit. 
This suggests that $\mathbf{p}$ is the appropriate variable to work with when solving  \eqref{eq:K_diff_eq} in a small $\delta$ expansion.

Let us now rewrite the recursion relation \eqref{eq:K_diff_eq_n1} using instead of $\mu^{-1}$ our new variable $\mathbf{p}$. We simply get
\begin{equation}\label{eq:K_diff_eq_wro_p}
	\f{d\mathcal{K}_n(x_1,x_2) }{d\mathbf{p}} =  \f{n(n+3)}{2}C_{\mathcal{O}\mathcal{O}\mathcal{O}}\Omega_{d-1} \f{r^\delta}{\delta}\mathcal{K}_{n-1}(x_1,x_2) .
\end{equation}
At the level $n=1$ we notice that the solution \eqref{eq:K1_final} can be obtained by trivially integrating \eqref{eq:K_diff_eq_wro_p} and dropping the integration constant. 

We solve now \eqref{eq:K_diff_eq_wro_p} recursively. For instance, at the $n=2$ level we can plug \eqref{eq:K1_final} into \eqref{eq:K_diff_eq_wro_p}. We get then
\begin{equation}
\begin{aligned}
	\label{eq:K_diff_eq_wro_p_n=2}
	\f{d\mathcal{K}_2(x_1,x_2) }{d\mathbf{p}} &=  \f{2(2+3)}{2}C_{\mathcal{O}\mathcal{O}\mathcal{O}}\Omega_{d-1} \f{r^\delta}{\delta}\mathcal{K}_{1}(x_1,x_2)\\
	&=  \f{2(2+3)}{2}C_{\mathcal{O}\mathcal{O}\mathcal{O}}\Omega_{d-1} \f{r^\delta}{\delta}\times  \f{1(1+3)}{2}C_{\mathcal{O}\mathcal{O}\mathcal{O}}\Omega_{d-1}  \mathbf{p} \f{r^\delta}{\delta} \mathcal{K}_0(x_1,x_2)\\
	&=  \f{2!(2+3)!}{3!}\left(C_{\mathcal{O}\mathcal{O}\mathcal{O}}\Omega_{d-1} \f{r^\delta}{2\delta}\right)^2  \mathbf{p}\,\mathcal{K}_0(x_1,x_2).
\end{aligned}
\end{equation}
Integrating this relation and using the fact that, according to \eqref{eq:expressions_EFT_CFT}, $\mathcal{K}_0$ is independent of $\mathbf{p}$, we get
\begin{equation}
	\label{eq:results_K2}
	\mathcal{K}_2 (x_1,x_2)
	=  \f{2!(2+3)!}{3!} \times \f{\mathbf{p}^2}{2!} \left(C_{\mathcal{O}\mathcal{O}\mathcal{O}}\Omega_{d-1} \f{r^\delta}{2\delta}\right)^2\times \mathcal{K}_0(x_1,x_2) +\text{const},
\end{equation}
where $\text{const}$ is the integration constant. In what follows we will always drop its contribution.

We can repeat the above procedure. It is straightforward to see then, that at the $n$-th step we will get the following result
\begin{equation}
	\label{eq:results_Kn}
	\mathcal{K}_n (x_1,x_2) = \f{n!(n+3)!}{3!}\times \f{\mathbf{p}^n}{n!} \left(C_{\mathcal{O}\mathcal{O}\mathcal{O}}\Omega_{d-1} \f{r^\delta}{2\delta}\right)^n\times \mathcal{K}_0(x_1,x_2).
\end{equation}
We obtain the final result by plugging \eqref{eq:results_Kn} and \eqref{eq:expressions_EFT_CFT} into \eqref{eq:expansion}. We get then
\begin{align}\label{eq:QFT_2pt_final}
	\langle\mathbf{O}(x_1)\mathbf{O}(x_2)\rangle_{\text{QFT}}=&  \langle\mathbf{O}(x_1)\mathbf{O}(x_2)\rangle_{\text{EFT}}\nn\\
	&\times \sum_{n=0}^\infty \f{(n+3)!}{3! n!}(-1)^n \times \f{\lambda^n}{\delta^n}  \left(\f{C_{\cO\cO\cO}\Omega_{d-1}}{2}\left((\mu r)^\delta-1\right)\right)^n.
\end{align}
This agree with the result obtained in the main text \eqref{eq:expansions_our_result} - \eqref{eq:definition_q}.

\section{Conservation in the Callan-Symanzik equation}
\label{app:conservation_CS}

Plugging it to \eqref{eq:conservation_JJ_CS} we get a differential equation, which relates the boundary conditions
\begin{multline}
	\frac{\partial}{\partial \Lambda } \left( h_1^\text{boundary}\big(1,\Lambda\big) +  h_2^\text{boundary}\big(1,\Lambda\big)\right) \times s\frac{\partial \Lambda(s,\lambda)}{\partial s} = \\
	(d-1)\,\left(2h_1^\text{boundary}(1,\Lambda)+h_2^\text{boundary}(1,\Lambda)\right).
\end{multline}
Given the boundary function $h_1^\text{boundary}(1,\Lambda)$ we need to solve this equation for $h_2^\text{boundary}(1,\Lambda)$. From \eqref{eq:solution_JJ} we have
\begin{equation}
	h_1(s,\Lambda)  = C_J - \Lambda(s,\lambda) s^{d-\Delta} (\Delta-1)E_{JJ}C^1_{JJ\cO}.
\end{equation}
Also recall from \eqref{eq:lambda-bar_sol} that
\be
\Lambda(s,\lambda)=
\lambda s^\delta\times\frac{1}{1+\frac{\lambda}{\lambda_\star}\,(s^\delta-1)}.
\ee
As a result
\begin{equation}
	s\frac{\partial}{\partial s} \Lambda(s,\lambda) = \frac{\delta}{\lambda_\star} \Lambda\Big(\lambda_\star-\Lambda\Big).
\end{equation}

Using these facts we can write
\begin{multline}
	\left( \frac{\partial}{\partial \Lambda } h_2^\text{boundary}\big(1,\Lambda\big) - (\Delta-1)E_{JJ}C^1_{JJ\cO}\right) \times s\frac{\partial \Lambda(s,\lambda)}{\partial s} = \\
	(d-1)\,\left(2C_J - 2\Lambda (\Delta-1)E_{JJ}C^1_{JJ\cO}+h_2^\text{boundary}(1,\Lambda)\right).
\end{multline}
Performing simple manipulations we can bring this equation into the following form
\begin{multline}
	\label{eq:h2_differential_EQ}
	\frac{\delta}{\lambda_\star} \Lambda\Big(\lambda_\star-\Lambda\Big)\times
	\left( \frac{\partial}{\partial \Lambda } h_2^\text{boundary}\big(1,\Lambda\big) \right)-(d-1)h_2^\text{boundary}(1,\Lambda) = \\
	2(d-1)C_J+(\Delta-1)E_{JJ}C^1_{JJ\cO} \Lambda \left( 2(1-d)+\delta-\frac{\delta}{\lambda_\star}\Lambda\right).
\end{multline}

The differential equation \eqref{eq:h2_differential_EQ} is a first order inhomogeneous differential equation of the following type
	\be
	\f{d h_2^\text{boundary}}{d\Lambda} +p(\Lambda)h_2^\text{boundary}=q(\Lambda),
	\ee
	with
	\begin{align}
		p(\Lambda)&= -(d-1)\f{\lambda_\star}{\delta} \f{1}{\Lambda (\lambda_\star -\Lambda)},\\
		q(\Lambda)&= \left[2(d-1)C_J+(\Delta-1)E_{JJ}C^1_{JJ\cO} \Lambda \left( 2(1-d)+\delta-\frac{\delta}{\lambda_\star}\Lambda\right)\right]\f{\lambda_\star}{\delta} \f{1}{\Lambda (\lambda_\star -\Lambda)}.
	\end{align}
	The general solution of the differential equation reads
	\begin{align}
		h_2^\text{boundary}(1,\Lambda)&= e^{-\int^\Lambda dt \ p(t)}\left( C\ + \int^\Lambda ds\ q(s)\  e^{\int^s dt \ p(t)}\right),
	\end{align}
	where $C$ is an integration constant to be fixed. Substituting $P(t)$ and $q(s)$ the solution simplifies to
	\begin{align}
		h_2^\text{boundary}(1,\Lambda)&= \left(\f{\Lambda}{\lambda_\star -\Lambda}\right)^{\f{d-1}{\delta}}\left( C\ + \int^\Lambda ds\ q(s)\  \left(\f{\lambda_\star -s}{s}\right)^{\f{d-1}{\delta}}\right).
	\end{align}
	We then substituting $q(s)$ in the above integrand and perform the integration explicitly. In order to match the boundary condition \eqref{eq:solution_JJ} at the leading order in $\lambda$ expansion we are also forced to set $C=0$. We obtain the following final result then
\begin{multline}
	h^\text{boundary}_2(1,\Lambda)  = -2C_J +
	(\Delta-1)E_{JJ}C^1_{JJ\cO}\\ \times\left((\Lambda+\lambda_\star)+(\Lambda-\lambda_\star)\,{}_2F_1\left(1,1,\frac{1-d+\delta}{\delta},\frac{\Lambda}{\lambda_\star}\right)\right).\label{eq:h2_solution_conservation}
\end{multline}

\section{Conformal correlator $\<TT\cO\>$}
\label{sec:Osborn_Petkou}
Let us study in this section the $\<TT\cO\>$ correlator. In what follows we will only consider the case when $\cO$ is a scalar operator. It has been first addressed by Osborn and Petkou in \cite{Osborn:1993cr}. We will also present its structure in the index-free formalism.

\paragraph{Index-free formalism}
The UV and IR fixed points of weakly relevant flows are described by two different CFTs. These CFTs have each its own stress-tensor. Let us provide the explicit expressions for the two- and three-point functions of the stress-tensor in any CFT. For doing that it is convenient to use the index-free notation of \cite{Costa:2011mg}. We define
\begin{equation}
	T(x,z) \equiv
	T^{\mu\nu}(x)\,z_{\mu} z_{\nu},
\end{equation}
where $z$ is an auxiliary vector with complex components. The expressions for the stress-tensor correlators in CFTs were found by Osborn and Petkou in \cite{Osborn:1993cr}. In index-free form they can be written as 
\begin{equation}
	\label{eq:TTO_IF}
	\begin{aligned}
		\< T(x_1,z_1) T(x_2,z_2)\>_\text{CFT} &= C_{T}\times \frac{H_{12}^2}{\left(x_{12}^2\right)^{d+2}},\\
		\< T(x_1,z_1) T(x_2,z_2) \cO(x_3) \>_\text{CFT} &= C_{TT\cO}\times \mathbf{T}_{TT\cO}(x_i,z_i),
	\end{aligned}
\end{equation}
Here $C_T$ and $C_{TT\cO}$ are real non-negative coefficients. The $c$-anomaly in $d=4$ is defined as
\begin{equation}
	c \equiv \frac{\pi^2}{640} \times C_T.
\end{equation}
The tensor structure $\mathbf{T}_{TT\cO}$ reads as
\begin{equation}
	\label{eq:tensor_structure_index_free}
	\mathbf{T}_{TT\cO}(x_i,z_i) \equiv \frac{a_1 \left(H_{12}\right)^2+a_2H_{12}V_{1,23}V_{2,13}+a_3\left(V_{1,23}V_{2,13}\right)^2
	}{\left(x_{12}^2\right)^{d+2-\Delta/2}\left(x_{13}^2\right)^{\Delta/2}\left(x_{23}^2\right)^{\Delta/2}}.
\end{equation}
Finally, the coefficients $a_i$ read as
\begin{equation}
	\label{eq:coefficeints_a}
	a_1 = 2\mathbb{C}, \qquad
	a_2 = -4(\mathbb{B}+2\mathbb{C}), \qquad
	a_3 = \mathbb{A}+8\mathbb{B}+8\mathbb{C},
\end{equation}
where
\begin{equation}
	\label{eq:epsilon-expansion}
	\begin{aligned}
		\mathbb{A} &= 1,\\
		\mathbb{B} &= -\frac{d^2-d \Delta -d+\Delta -2}{2 (d-2) (2 d-\Delta +2)},\\
		\mathbb{C} &=-\frac{2 d^2 \Delta -d^3+d^2-d \Delta ^2-2 d \Delta +2 d+\Delta ^2}{2 (d-2) (2 d-\Delta ) (2 d-\Delta +2)} .
	\end{aligned}
\end{equation}

The index-full form of the stress-tensor correlators can be obtained by using the following equality
\begin{equation}
	T^{\mu\nu}(x)= D^\mu D^\nu T(x,z),
\end{equation}
where $D^\mu$ is the Todorov differential operator defined as
\begin{equation}
	\label{eq:todorov}
	D^\mu\equiv \left(d/2-1+z\cdot \partial_z\right)\partial_z^\mu
	-\frac{1}{2}\,z^\mu\,\partial^2_z.
\end{equation}
We summarize the result of Osborn and Petkou in index-full form in appendix \ref{app:conventions_CFT}.
Using the results of appendix \ref{app:conventions_CFT} it is easy to find for example that
\begin{equation}
	\label{eq:contraction_2pt}
	\delta_{\mu\rho}\delta_{\nu\sigma}\langle T^{\mu\nu}(x_1)T^{\rho\sigma}(x_2)\rangle= \f{1}{2}(d-1)(d-2)\ \f{C_T}{\left(x_{12}^2\right)^{d}}.
\end{equation}

\paragraph{Result of Osborn and Petkou}
\label{app:conventions_CFT}

The two- and three-point functions of the stress-tensor in CFTs were derived \cite{Osborn:1993cr}. Their result reads as
\be 
\langle T^{\mu\nu}(x_1)T^{\rho\sigma}(x_2)\rangle&= C_T\times \f{\mathcal{I}^{\mu\nu,\rho\sigma}(x_{12})}{\left(x_{12}^2\right)^d},
\label{eq:TTUV}\\
\langle T^{\mu\nu}(x_1)T^{\rho\sigma}(x_2)\mathcal{O}(x_3)\rangle
&= C_{TT\cO}\times\f{\mathcal{I}^{\mu\nu,\alpha\beta}(x_{13})\mathcal{I}^{\rho\sigma, \gamma\delta}(x_{23})t_{\alpha\beta,\gamma\delta}(X_{12})}{\left(x_{12}^2\right)^\frac{2d-\Delta}{2}\left(x_{23}^2\right)^\frac{\Delta}{2}\ \left(x_{31}^2\right)^\frac{\Delta}{2}} , \label{eq:TTO}
\ee
where we have defined 
\begin{equation}
	x_{ij}^\mu \equiv x_i^\mu -x_j^\mu,\qquad
	X_{12}^{\mu}\equiv \f{x_{13}^{\mu}}{x_{13}^2}-\f{x_{23}^{\mu}}{x_{23}^2},
\end{equation}
together with 
\begin{equation}
	I^{\mu\nu}(x) \equiv \delta^{\mu\nu} - 2\,\frac{x^\mu x^\nu}{x^2},\quad
	\mathcal{I}^{\mu\nu,\rho\sigma}(x)\ \equiv \f{1}{2}I^{\mu\rho}(x)I^{\nu\sigma}(x)+\f{1}{2}I^{\mu\sigma}(x)I^{\nu\rho}(x)-\f{1}{d}\delta^{\mu\nu}\delta^{\rho\sigma}.\label{eq:I_definitions}
\end{equation}
Finally, the object $t$ reads as
\begin{equation}
	t_{\alpha\beta ,\gamma\delta}(X)\equiv \mathbb{A}h^1_{\alpha\beta}(X) h^1_{\gamma\delta}(X)+\mathbb{B}h^2_{\alpha\beta ,\gamma\delta}(X)+\mathbb{C}h^3_{\alpha\beta ,\gamma\delta}(X).
\end{equation}
The coefficients $\mathbb{A}$, $\mathbb{B}$ and $\mathbb{C}$ were given in \eqref{eq:epsilon-expansion}.
The functions $h^i$ are defined as
\be
h^1_{\mu\nu}(X) &\equiv \f{X_\mu X_\nu}{X^2} - \frac{1}{d} \delta_{\mu\nu},\nn\\
h^2_{\mu\nu ,\rho\sigma}(X)&\equiv\f{1}{X^2}\Big[X_\mu X_\rho \delta_{\nu\sigma}+X_{\mu}X_{\sigma}\delta_{\nu\rho}+X_{\nu}X_{\rho}\delta_{\mu\sigma}+X_{\nu}X_{\sigma}\delta_{\mu\rho}-\f{4}{d}X_\mu X_\nu \delta_{\rho\sigma}\nn\\
&\ -\f{4}{d}X_\rho X_\sigma \delta_{\mu\nu}\Big]+\f{4}{d^2}\delta_{\mu\nu}\delta_{\rho\sigma},\nn\\
h^3_{\mu\nu ,\rho\sigma}(X)&\equiv\ \delta_{\mu\rho}\delta_{\nu\sigma}+\delta_{\mu\sigma}\delta_{\nu\rho}-\f{2}{d}\delta_{\mu\nu}\delta_{\rho\sigma}.
\ee

It is straightforward to show that by contracting \eqref{eq:TTUV} and \eqref{eq:TTO} with the auxiliary vectors $z_1$ and $z_2$ we obtain exactly \eqref{eq:TTO_IF}.

\section{Double integral evaluation}
\label{app:double_integral_ebaluation}

In this appendix we compute the integral \eqref{eq:integral_J}. Let us write it here again for convenience
\begin{equation}
	\label{eq:integral_J_app}
	\bold{J}(a,b,c)\equiv 
	w\, C_{TT\cO}\times
	\int d^dz_1\int d^dz_2\ \f{\big(\chi\left(\mu \rho,\lambda\right)\big)^2}{\left(z^2_{12}\right)^\frac{a+\delta}{2} \left(z_1^2\right)^\frac{b-\delta}{2} \left(z_2^2\right)^\frac{c-\delta}{2}},
\end{equation}
where $\chi\left(s,\lambda\right)$ is defined in \eqref{eq:running_coupling}, the coefficient $w$ is defined in \eqref{eq:def_w} and $\rho$ is given by
\begin{equation}
\label{eq:z_def}
	\rho= \left(z_1^2 z_2^2 (z_1-z_2)^2\right)^\frac{1}{6}.
\end{equation}

The integral \eqref{eq:integral_J_app} is divergent for very particular points
\begin{equation}
	z_1^\mu=0,\qquad
	z_2^\mu=0,\qquad
	z_1^\mu=z_2^\mu.
\end{equation}
We refer to these as the UV divergences. In order to deal with UV divergences we will work with some particular values of $a$, $b$ and $c$ for which these divergences are absent. We will then analytically continue the final answer to generic values.
Using the change of variables we can deduce from \eqref{eq:integral_J} the following relations
\begin{equation}
	\bold{J}(a,b,c) = \bold{J}(a,c,b),\qquad
	\bold{J}(a,b,c) = \bold{J}(c-2\delta,b,a+2\delta). \label{eq:J_identities}
\end{equation}
In what follows we will also restrict our attention to the range of parameters obeying \eqref{eq:condition}, namely
\begin{equation}
	\label{eq:condition_app}
	a+b+c=3d-4.
\end{equation}

\paragraph{Integral evaluation. Step 1}
Let us now set $z_2^\mu= r e_2^\mu$, where $e_2$ is a unit vector. The integral \eqref{eq:integral_J_app} takes the following form then
\begin{equation}
	\bold{J}(a,b,c)= w C_{TT\cO} \Omega_{d-1}\int d^dz_1\int_0^{\infty} r^{d-1}dr\ \f{\big(\chi\left(\mu \rho,\lambda\right)\big)^2}{|z_1-r e_2|^{a+\delta} z_1^{b-\delta} r^{c-\delta}}.
\end{equation}
Let us now move to spherical coordinated for $y_1$
\begin{equation}
	z_1^\mu = R\, e_1^\mu,\qquad
	d^d z_1 =R^{d-1} dR \times d \Omega_{d-2}\times \sin^{d-2}\theta d \theta,
\end{equation}
where $e_1^\mu$ is some unit vector and $\theta$ is the angle between $e_1$ and $e_2$. We have then
\begin{equation}
	z_1^2=R^2,\qquad
	(z_1-r e_1)^2 = R^2 + r^2-2 r R \cos\theta.
\end{equation}
Our integral then reduces to
\begin{multline}
	\bold{J}(a,b,c)=w C_{TT\cO} \Omega_{d-1}\Omega_{d-2}\int_0^{\infty} dR R^{d-1-b+\delta} 
	\int_0^{\infty}  dr r^{d-1-c+\delta} \\
	\int_0^\pi d \theta \sin^{d-2}\theta\f{\big(\chi\left(\mu \rho,\lambda\right)\big)^2}{(R^2 + r^2-2 r R  \cos\theta)^\frac{a+\delta}{2}}.
\end{multline}
Now to perform the integration over $r$ and $R$ we can make the following change of variables
\begin{equation}
	r=z\cos\phi\ ,\hspace{0.5cm} R=z\sin\phi \ ,\hspace{0.5cm} \hbox{with} \hspace{0.3cm} 0\leq \phi\leq \f{\pi}{2}\ \hbox{and} \ 0\leq z<\infty\ ,
\end{equation}
and get
\begin{multline}
	\bold{J}(a,b,c)=  w C_{TT\cO}\Omega_{d-1}\Omega_{d-2}\int_0^{\infty} dz   \int_0^{\f{\pi}{2}} d\phi\   \int_0^\pi d \theta z^{2d-1-(a+b+c)+\delta }\\
	\times (\sin\phi ) ^{d-1-b+\delta } (\cos\phi ) ^{d-1-c+\delta }  \sin^{d-2}\theta \f{\big(\chi\left(\mu \rho,\lambda\right)\big)^2}{(1-\cos (\theta ) \sin (2 \phi ))^{\frac{a+\delta}{2}}}
\end{multline}

\paragraph{Integral evaluation. Step 2}
Let us now do another set of change of variables from $(\phi, z)$ to $(s, \rho)$ with $s=\cot\phi, \rho=z(\sin\phi\cos\phi)^{\f{1}{3}}\left(1-\sin(2\phi)\cos\theta\right)^{\f{1}{6}}$ in the ranges $0\leq s<\infty\ ,\ 0\leq \rho<\infty$ and substitute $a+b+c=3d-4$ to simplify and get
\begin{multline}
	\bold{J}(a,b,c)=  w C_{TT\cO}\Omega_{d-1}\Omega_{d-2}\,
	\int_{0}^{\infty} d\rho\ \rho^{-d+3+\delta}\big(\chi\left(\mu \rho,\lambda\right)\big)^2
	\int_0^\pi d \theta \sin^{d-2}\theta \\
	\times \int_{0}^{\infty} ds\ s^{d-1-c+\delta+\f{d-4-\delta}{3}}
	\f{1}{(1 + s^2-2 s  \cos\theta)^{\frac{a+\delta}{2}-\f{d-4-\delta}{6}}}
\end{multline}
Let us denote the integral over $z$ as follows 
\be
h(\delta)\equiv  wC_{TT\cO}\, \int_0^\infty d\rho\ \rho^{-d+3+\delta} \big(\chi\left(\mu \rho,\lambda\right)\big)^2\label{L_definition}
\ee
Hence in terms of $h(\delta)$, our $\bold{J}$ integral becomes
\begin{multline}
	\bold{J}(a,b,c)=  \Omega_{d-1}\Omega_{d-2}\,
	h(\delta)	\int_0^\pi d \theta \sin^{d-2}\theta \\
	\times \int_{0}^{\infty} ds\ s^{d-1-c+\delta+\f{d-4-\delta}{3}}
	\f{1}{(1 + s^2-2 s  \cos\theta)^{\frac{a+\delta}{2}-\f{d-4-\delta}{6}}}
\end{multline}
Now we can perform Newton's generalized binomial expansion of the denominator in the integrand in the following way
\begin{multline}
	\f{1}{(1 + s^2-2 s  \cos\theta)^{\frac{a+\delta}{2}-\f{d-4-\delta}{6}}}
	=\f{1}{(1+s^2)^{\f{a+\delta}{2}-\f{d-4-\delta}{6}}}\\
	\times \ \sum_{n=0}^{\infty} \f{\left( \f{a+\delta}{2}-\f{d-4-\delta}{6}\right)_n}{n!}\ \left(\f{2s}{1+s^2}\right)^n\ (\cos\theta)^n
\end{multline}
Above $(x)_n\equiv \f{\Gamma(x+n)}{\Gamma(x)}$ represents the Pochhammer symbol. Now we can substitute the above series expansion in the  $\bold{J}$ integral and commute the integrals and series summation to first perform the integration over $\theta$. Using the following integration result
\be
\int_0^\pi d\theta\ \sin^{d-2}\theta\ \cos^{n} \theta=\frac{\left((-1)^{n}+1\right) \Gamma \left(\frac{d-1}{2}\right) \Gamma \left(\frac{n+1}{2}\right)}{2 \Gamma \left(\frac{d+n}{2}\right)}
\ee
the $\bold{J}$ integral simplifies to
\begin{multline}
	\bold{J}(a,b,c)=  \Omega_{d-1}\Omega_{d-2}\,
	h(\delta)\sum_{n=0}^{\infty} \f{\left( \f{a+\delta}{2}-\f{d-4-\delta}{6}\right)_n}{n!} \frac{\left((-1)^{n}+1\right) \Gamma \left(\frac{d-1}{2}\right) \Gamma \left(\frac{n+1}{2}\right)}{2 \Gamma \left(\frac{d+n}{2}\right)}\\
	\int_{0}^{\infty} ds\ s^{d-1-c+\delta+\f{d-4-\delta}{3}}\  \f{1}{(1+s^2)^{\f{a+\delta}{2}-\f{d-4-\delta}{6}}} \left(\f{2s}{1+s^2}\right)^{n} 
\end{multline}
After performing the integration over $s$ we get

\begin{multline}
	\bold{J}(a,b,c)=  \Omega_{d-1}\Omega_{d-2}\,
	h(\delta)\sum_{n=0}^{\infty} \f{\left( \f{a+\delta}{2}-\f{d-4-\delta}{6}\right)_n}{n!} \frac{\left((-1)^{n}+1\right) \Gamma \left(\frac{d-1}{2}\right) \Gamma \left(\frac{n+1}{2}\right)}{2 \Gamma \left(\frac{d+n}{2}\right)}\\
	\times\frac{ \Gamma \left(\frac{1}{6} (-3 c+4 d+3 n+2 \delta -4)\right) \Gamma \left(\frac{1}{6} (3 a+3 c-5 d+3 n+2 \delta +8)\right)}{2\Gamma \left(\frac{1}{6} (3 a-d+6 n+4 \delta +4)\right)}\times 2^n
\end{multline}
Summing the infinite series in the above expression we get
\begin{multline}
	\bold{J}(a,b,c)=  \Omega_{d-1}\Omega_{d-2}\,
	h(\delta) \f{\sqrt{\pi } \Gamma \left(\frac{d-1}{2}\right) \Gamma \left(\frac{1}{6} (-3 c+4 d+2 \delta -4)\right) \Gamma \left(\frac{1}{6} (3 a+3 c-5 d+2 \delta +8)\right)}{2 \Gamma \left(\frac{d}{2}\right) \Gamma \left(\frac{1}{6} (3 a-d+4 \delta +4)\right)}\\
	\times \, \, _2F_1\left(\frac{a}{2}+\frac{c}{2}+\frac{\delta }{3}-\frac{5 d}{6}+\frac{4}{3},-\frac{c}{2}+\frac{2 d}{3}+\frac{\delta }{3}-\frac{2}{3};\frac{d}{2};1\right).
\end{multline}
The above expression is convergent only when $a<d-1-\delta$. In this region we can use Gauss' summation theorem to simplify the above result as
\begin{multline}
	\label{eq:result_integral_J}
	\bold{J}(a,b,c)=  \Omega_{d-1}\Omega_{d-2}\,
	h(\delta) \frac{\sqrt{\pi } \Gamma \left(\frac{d-1}{2}\right) \Gamma \left(-\frac{a}{2}+\frac{2 d}{3}-\frac{2 \delta }{3}-\frac{2}{3}\right) }{2 \Gamma \left(\frac{1}{6} (3 a-d+4 \delta +4)\right) \Gamma \left(\frac{c}{2}-\frac{\delta }{3}+\frac{2}{3}-\frac{d}{6}\right) }\\
	\times \f{\Gamma \left(\frac{1}{6} (-3 c+4 d+2 \delta -4)\right) \Gamma \left(\frac{1}{6} (3 a+3 c-5 d+2 \delta +8)\right)}{\Gamma \left(-\frac{a}{2}+\frac{4 d}{3}-\frac{c}{2}-\frac{\delta }{3}-\frac{4}{3}\right)}.
\end{multline}

\paragraph{Integral evaluation. Step 3}
By evaluating the integral $h(\delta)$ defined in \eqref{L_definition} in $d=4$, we simply get
\begin{equation}
	h(\delta) = \frac{wC_{TT\cO}}{\delta\mu^\delta} \frac{\lambda_\star}{\lambda} \frac{1}{1-\frac{\lambda}{\lambda_\star}}.
\end{equation}
Plugging here the explicit expression for the coefficient $w$ given in \eqref{eq:def_w} and taking into account \eqref{eq:lambda(mu)}, \eqref{eq:renormalization_Z} and \eqref{eq:lambda_star} we can write
\begin{equation}
	\label{eq:h-integral_result}
		h(\delta) =  \lambda_\star C_{TT\cO}  \frac{1-\frac{2\lambda}{\lambda_\star}}{1-\frac{\lambda}{\lambda_\star}} = \lambda_\star C_{TT\cO} +O\left(\lambda\right).
\end{equation}
The final result for the integral $\bold{J}(a,b,c)$ can be gotten from \eqref{eq:result_integral_J} after setting $d=4$ together with \eqref{eq:h-integral_result}.

\bibliographystyle{JHEP}
\bibliography{refs}

\end{document}